\documentclass[a4paper,11pt]{article}
\pdfoutput=1
\usepackage{jheppub}
\usepackage{lineno}
\usepackage{upgreek}
\usepackage{float, extarrows, tikz-cd}
\usepackage{graphicx}
\usepackage{slashed}
\usepackage{tabularx,ragged2e}
\usepackage{amssymb}
\usepackage{subcaption}
\usepackage{amsmath,amssymb}
\usepackage{slashed}
\usepackage{caption}
\usepackage{xcolor}
\usepackage{dsfont}
\usepackage{verbatim}
\usepackage{mathtools, xcolor,ytableau, amsfonts,tikz}
\usepackage{graphicx}
\usepackage{physics}
\usepackage{simpler-wick}
\usepackage{microtype}
\usepackage{graphicx}
\usepackage[nobottomtitles*]{titlesec}
\newcolumntype{C}{>{\Centering\arraybackslash}X}

\usepackage{placeins}

\numberwithin{equation}{section}

\def\le{\left}
\def\ri{\right}
\newcommand\ov{\over}
\newcommand\p{\ensuremath{\partial}}
\newcommand{\es}[2] {\begin{equation} \label{#1} \begin{split} #2 \end{split} \end{equation}}

\def\<{\langle}
\def\>{\rangle}

\newcommand\lam{\lambda}
\newcommand\Lam{\Lambda}

\newcommand\Om{\Omega}
\newcommand\ga{{\ensuremath{{\gamma}}}}

\newcommand\de{{\ensuremath{{\delta}}}}

\def\mop#1{\mathop{\rm #1}\nolimits}
\def\vol{\mop{vol}}

\title{The entanglement membrane in 2d CFT: \\
 \Large reflected entropy, RG flow, and information velocity}

\author[a]{Hanzhi Jiang,}        
\author[b]{ M\'ark Mezei,}        
\author[c]{ and Julio Virrueta}

                \affiliation[a]{Rudolf Peierls Centre for Theoretical Physics, University of Oxford, Oxford OX1 3PU, U.K.}                                                              
               \affiliation[b]{Mathematical Institute, University of Oxford, Woodstock Road, Oxford, OX2 6GG, U.K.} 
               \affiliation[c]{Theoretisch-Physikalisches Institut, Friedrich-Schiller-Universit\"at Jena, Max-Wien-Platz 1, D-07743 Jena, Germany} 

\emailAdd{hanzhi.jiang@physics.ox.ac.uk}     
\emailAdd{mezei@maths.ox.ac.uk}
\emailAdd{julio.virrueta@uni-jena.de}   

\abstract{The time evolution of entanglement entropy in generic chaotic many-body systems has an effective description in terms of a minimal membrane, characterised by a tension function. For 2d CFTs, a degenerate tension function reproduces several results regarding the dynamics of the entropy; this stands in contrast to higher dimensions where the tension is non-degenerate. 
In this paper we use holography to show that, in order to correctly capture the reflected entropy in 2d CFT, one needs to add an additional degree of freedom to the membrane description. Furthermore, we show that the conventional non-degenerate membrane tension function emerges upon introducing a relevant deformation of the CFT, dual to a planar BTZ black hole with scalar hair and with an interior Kasner universe. Finally, we also study the membrane description for reflected entropy and information velocity~\cite{Couch:2019zni} in higher dimensions.
}

\graphicspath{{./Figures/}}

\begin{document}
\maketitle
\flushbottom

\section{Introduction}

\begin{figure}[htbp]
\centering
\includegraphics[width=.5\textwidth]{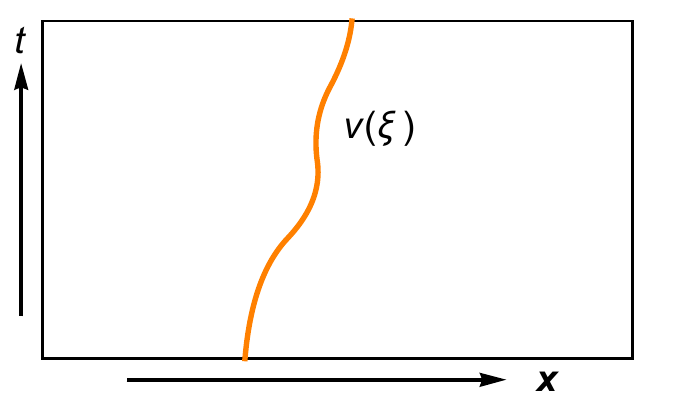}
\qquad
\caption{Cartoon of a minimal membrane (orange) extending in spacetime in $(1+1)$-dimensions. The figure can either represent entropy dynamics in a quench, where $A(t)$ is on the top time slice and the bottom time slice represents the initial state, or the operator entanglement of the time evolution operator $e^{-(\beta/2+it)H}$, where the top region is $A_\text{ket}$ and the bottom one is $A_\text{bra}$. \label{fig:MembraneCartoon}}
\end{figure}

The dynamics of entanglement entropy in systems out-of-equilibrium is one of the most fundamental probes of thermalisation. In chaotic many-body quantum systems, there is mounting evidence that the dynamics of entanglement and R\'enyi entropies in the limit of large region sizes and long times can be captured by an effective description in terms of a minimal membrane~\cite{Nahum:2016muy,Jonay:2018yei,Mezei:2018jco}. This effective description states that the entanglement entropy is computed by the action
\begin{equation}\label{MinMemb}
S=\min_{x(\xi)} \int d^{d-1}\xi\, \sqrt{\abs{\gamma(\xi)}}\, \frac{s_\text{th}\, {\cal E}(v(\xi))}{\sqrt{1-v^2(\xi)}}\,.
\end{equation}
In the above formula, $\xi$ are the membrane worldvolume coordinates,\footnote{In $d=1+1$, the membrane is a curve in spacetime.} $x(\xi)$ is its embedding into spacetime, $\gamma$ is the induced metric on the membrane, $s_\text{th}$ is the coarse grained entropy density, $v(\xi)$ is local transverse velocity of the membrane, and ${\cal E}(v)$ is the membrane tension function that depends on the microscopic details of the many-body system. The minimisation is performed over all membrane shapes satisfying boundary conditions determined by the situation of interest. If we wish to study a quench, the membrane connects the time slice of the boundary of the subregion $A(t)$ and the time slice of the initial state; similarly, if we are interested in operator entanglement, the membrane connects the $A_\text{bra}$ and $A_\text{ket}$ subregions of the operator, see Fig.~\ref{fig:MembraneCartoon}.

The membrane effective description can be derived for random quantum circuits~\cite{Nahum:2016muy,Jonay:2018yei,Zhou:2018myl}, Floquet circuits (up to a plausible assumption about the convergence of an RG scheme)~\cite{Zhou:2019pob}, in generalised dual-unitary circuits~\cite{Rampp:2023vah}, Brownian Hamiltonians~\cite{Vardhan:2024wxb}, and in holography~\cite{Mezei:2018jco,Mezei:2019zyt}. There is also convincing numerical evidence for this description in chaotic spin chains~\cite{,Jonay:2018yei}. 

Even though two-dimensional CFTs were some of the first studied examples of entanglement dynamics~\cite{Calabrese:2005in,Calabrese:2007rg}, and that they are generically chaotic systems for central charge above some threshold value,\footnote{If they only have Virasoro symmetry, the criterion is $c>1$. If they have extended chiral symmetry characterised by an effective central charge $c_\text{currents}$  for the chiral sector of the theory, then the criterion is $c>c_\text{currents}$~\cite{Asplund:2015eha}. Below these threshold values the CFTs are integrable and their entanglement dynamics are described by the quasiparticle effective description~\cite{Calabrese:2005in,Calabrese:2007rg}.} the holographic derivation of the entanglement membrane breaks down for them. One could dismiss this issue by attributing it to some peculiarity of the dual 3d gravity.
Indeed, the entanglement entropy dynamics in quenches and in the operator entanglement setup obtained using 2d CFT technology is reproduced by the membrane effective description, provided we choose the degenerate membrane tension function ${\cal E}(v)=1$. However, there are two indications that there is a fundamental issue at play: first, the R\'enyi entropy dynamics in 2d CFT for two intervals or more~\cite{Asplund:2015eha} is incompatible with the membrane theory~\cite{Mezei:2018jco}. Second, in quantum circuits the entropy dynamics is translated to the dynamics of a domain wall in a lattice spin model. The diffusion constant of the domain wall is proportional to $1/{\cal E}''(v)$~\cite{Zhou:2019pob}, which diverges for the degenerate membrane tension; indeed, just like 2d CFTs, dual unitary circuits formally give rise to the degenerate membrane tension ${\cal E}(v)=1$~\cite{Zhou:2019pob}, and in the equivalent lattice model the domain wall delocalises.

In this paper we present holographic studies of the entanglement dynamics of 2d CFTs and their deformations. We derive a generalised membrane with an additional scalar degree of freedom, that we will represent in figures as a colour, on its worldline as an 
effective theory of entanglement dynamics in 2d CFT. We show that upon adding a relevant deformation to the CFT, this additional degree of freedom gets ``gapped out" and we recover the entanglement membrane with non-degenerate membrane tension. This dichotomy is analogous to how hydrodynamics emerges from holomorphic stress tensor dynamics once a 2d CFT is deformed by a relevant operator~\cite{Davison:2024msq}. These observations make us speculate that the additional degree of freedom on the generalised membrane worldline is capturing the infinite-dimensional symmetry charge dynamics of 2d CFT.\footnote{An additional evidence in this direction is the aforementioned failure of the ordinary membrane to capture the R\'enyi entropy for multiple intervals, since R\'enyi entropy should be more sensitive to the conserved charge dynamics. See evidence in this direction in Refs.~\cite{Rakovszky:2019oht,Huang:2019hts,Znidaric:2020czi,Rakovszky:2020idb}.}

We also show that the reflected entropy, that is computed holographically by the entanglement wedge cross section (EWCS), is also captured by the generalised membrane in 2d CFT, and that this quantity, in contrast to entanglement entropy, is not captured by the ordinary membrane with degenerate membrane tension. To complete our investigations, we also study the reflected entropy in higher dimensional CFTs, and demonstrate that the ordinary membrane correctly captures them. 

Additionally, to further illustrate the versatility and utility of the entanglement membrane, we show that the information velocity defined in~\cite{Couch:2019zni} through entanglement dynamics can also be efficiently translated to and computed through the membrane description.

\section{Entanglement membrane and 2d CFT}\label{Sec:EEMembrane}

\subsection{The entanglement membrane for a half space subregion}

The holographic derivation of the entanglement membrane~\cite{Mezei:2018jco,Mezei:2019zyt} starts with the study of Hubeny-Rangamani-Ryu-Takayanagi (RT/HRT) surfaces~\cite{Ryu:2006bv,Ryu:2006ef,Hubeny:2007xt} in black brane spacetimes. In the quench setup, at early times we have some non-equilibrium features in the spacetime that get encoded in boundary conditions for the membrane. At later times local thermal equilibrium is reached, which is dual to the black brane. In the operator entanglement setup, we consider an eternal black brane and put the $A_\text{bra}$ and $A_\text{ket}$ subregions on the left and right asymptotic boundaries~\cite{Hartman:2013qma}.

Hence, we consider an 
 asymptotically AdS$_{d+1}$  black brane in infalling coordinates:
\begin{equation}
ds^2 = \frac{1}{z^2}\left(-a(z)du^2 -{2\ov b(z)}dudz+dx^2+ d\vec{y}_{d-2}^2\right)\,,\label{blackBraneInf}
\end{equation}
where the conformal boundary is at $z=0$, and the horizon is at $z=1$; therefore $a(1)=0$. The  AdS$_{d+1}$ asymptotics requires $a(0)=b(0)=1$. For a neutral black brane $a(z)=1-z^d,\, b(z)=1$. 

\begin{figure}[htbp]
\centering
\includegraphics[width=.45\textwidth]{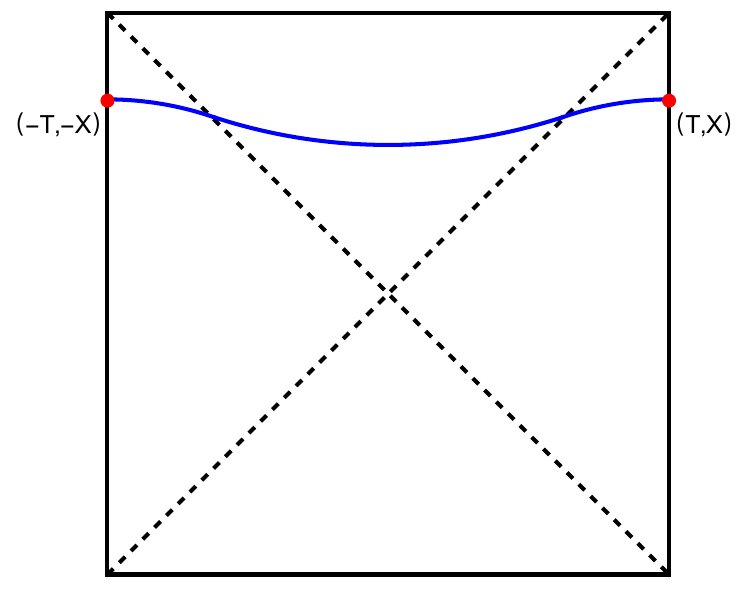}
\includegraphics[width=.365\textwidth]{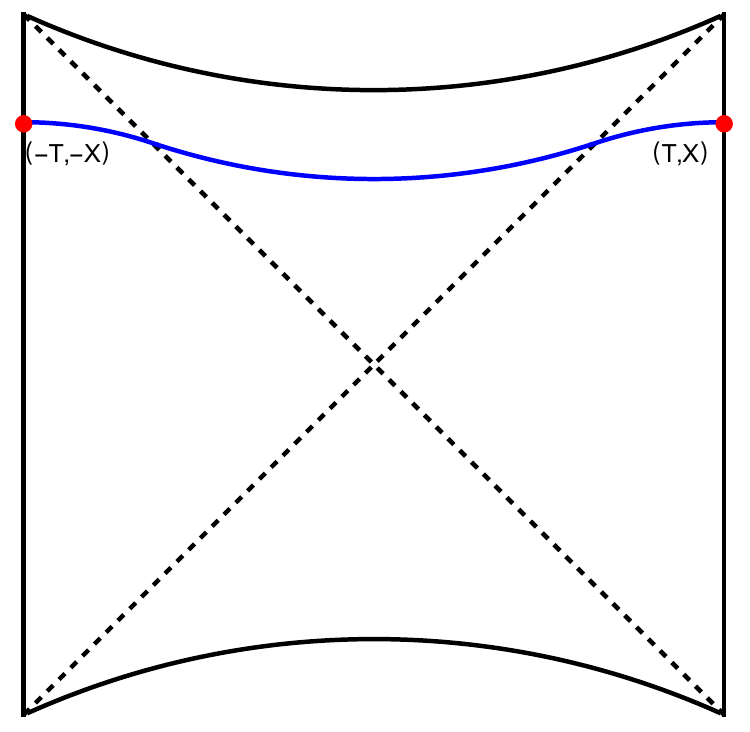}
\qquad
\caption{Penrose diagram of the HRT surface computing the time-dependent entanglement entropy in planar BTZ (left) and AdS$_{d+1}$-Schwarzschild (right, $d>2$) with entangling subregion taken to be half the space on both sides of the thermofield double. The $\pm X$ displacement in the $x$ direction is perpendicular to the plane of the paper. \label{fig:DispHalfSpaceBTZ}}
\end{figure}

We would like to determine the two-sided entanglement entropy for half spaces extending in the $\vec{y}$ directions at late-time, see Fig.~\ref{fig:DispHalfSpaceBTZ}. We parametrise the HRT surface as $x(u),\, z(u)$, and obtain the area functional for the metric~\eqref{blackBraneInf}:
\begin{equation}
\label{AreaFunct}
\begin{aligned}
S &= s_\text{th} \vol(\p A) \int du  \frac{\sqrt{Q}}{z^{d-1}}\,, \qquad Q\equiv \dot x^2-a(z)-{2\ov b(z)}\dot z\,,
\end{aligned}
\end{equation}
where $\dot f\equiv\frac{df}{du}$. 

It was shown in~\cite{Hartman:2013qma,Liu:2013qca,Liu:2013iza,Mezei:2016zxg,Mezei:2018jco} that for $d>2$ black branes in the leading behaviour of entanglement entropy in the late time and large entangling subregion limit is determined by the portion of the HRT surfaces that lie inside the horizon; at late time, the HRT surfaces get stuck at some special extremal slice $z=z_*>1$ in the interior instead of moving deeper towards the singularity.\footnote{As we will see in section~\ref{KasnerSec}, these conclusions also hold for $d=2$ hairy BTZ black branes (strings). } Therefore, to extract the leading behaviour of entanglement entropy in the late time and large entangling subregion limit, we can impose the following self-consistent scaling Ansatz~\cite{Mezei:2016zxg,Mezei:2018jco}:
\es{CoordScaling}{
u\rightarrow \Lambda u, \hspace{0.5cm} x \rightarrow \Lambda x, \hspace{0.5cm} z\rightarrow z\,,
}
for $\Lambda\gg1$. In this scaling the $\dot z$ term can be dropped from $Q$ in \eqref{AreaFunct}. Then the $z$ equation of motion becomes algebraic in $z$
\es{projectionz}{
v^2\equiv \dot x^2  = a(z) - \frac{z a'(z)}{2(d-1)} \equiv c(z)\,;
}
e.g.~for neutral black branes $c(z)=1-{d-2\ov 2(d-1) }\, z^d$. The function $c(z)$ is monotonically decreasing and positive in the range $1\leq z \leq z_*$, where $z_*$ is the first zero of $c(z)$. Then $z$ in this range can be solved for in terms of $v^2$ and we have managed to reduce the problem in dimension. The resulting reduced action is the membrane theory introduced around \eqref{MinMemb}:
\begin{equation}
\label{AreaFunct2}
\begin{aligned}
S &= s_\text{th} \vol(\p A) \int du \,  \mathcal{E}(v^2)\,, \qquad
\mathcal{E}(v^2) =\left. \sqrt{\frac{-a'(z)}{2(d-1)z^{2d-3}}}\right|_{z=c^{-1}(v^2)}\,.
\end{aligned}
\end{equation}
The bulk infalling time $u$ assumes the role of boundary theory time. (Recall that at the AdS boundary infalling and boundary times agree.) 

The function $\mathcal{E}(v^2)$ obeys a series of properties. To state them, we first introduce two important velocities that characterise the spread of quantum information: 
\begin{itemize}
    \item The entanglement velocity $v_E$~\cite{Hartman:2013qma,Liu:2013iza,Liu:2013qca} determines the speed at which entanglement grows linearly
\begin{align}
    v_E=\sqrt{-\frac{a(z)}{z^{2(d-1)}}}\Big|_{z=z_*}\,,\label{vE}
\end{align}
where $z_*$ maximises the expression under the square root. $z_*$ is the deepest point in the black hole interior the HRT surface can reach in the scaling regime~\eqref{CoordScaling}.\footnote{For planar AdS-Schwarzschild black branes, $z_*=\Big(\frac{2(d-1)}{d-2}\Big)^{1/d}$~\cite{Hartman:2013qma,Liu:2013iza,Liu:2013qca}. When $d=2$, i.e. in the BTZ black brane, $z_*=\infty$. } 
    \item The butterfly velocity $v_B$~\cite{Roberts:2014isa} is related to the out-of-time order correlator (OTOC) 
\begin{align}
    v_B=\sqrt{-\frac{a'(1)}{2(d-1)}}\,.\label{vB}
\end{align}
$c(z)$ maps $v_B$ to the horizon at $z=1$. In previous work~\cite{Mezei:2018jco} the near-horizon region around $z=1$ was not treated with enough care. In this work we fix the resulting imprecisions that involve membrane with regions, where $v=v_B$.
\end{itemize}
In general, $v_E\leq v_B$ due to the constraint from null energy condition~\cite{Mezei:2016zxg}.\footnote{This inequality also follows from general quantum many-body considerations that are independent from holography~\cite{Mezei:2016wfz}.} Notice that when the geometry \eqref{blackBraneInf} is the BTZ black brane, $a(z)=1-z^2,\,b(z)=1$, we get $v_E=v_B= \mathcal{E}(v)=1$.

We are now ready to state the constraints obeyed by the membrane tension function: it is an even function, monotonically increasing and convex for $0\leq v <1$. It diverges as $v\rightarrow 1$ and certain important values are~\cite{Mezei:2018jco}:\footnote{The behaviour around $v_B$ follows from consistency conditions related to operator spreading~\cite{Jonay:2018yei}. }
\begin{align}
    \mathcal{E}(0)=v_E\,, && \mathcal{E}'(0)=0\,, && \mathcal{E}(v_B)=v_B\,, && \mathcal{E}'(v_B)=1\,, && \mathcal{E}'(v)>0\,, && \mathcal{E}''(v)>0\,. \label{constraints}
\end{align}

For the $d=2$ BTZ black hole, however, we get $c(z)=1$, which either enforces  $\dot x^2 =1$ or signals the breakdown of the scaling Ansatz. Indeed, two-sided geodesics reach $e^\Lam$ depths inside the BTZ black hole~\cite{Hartman:2013qma}. Then a different scaling can be applied: let us define $z\equiv e^\xi$ and scale $\xi, u, x$ linearly with $\Lam$. The resulting action in the $d=2$ BTZ black hole is 
\es{AreaFunct3}{
S &= s_\text{th} \int du\,  \sqrt{1+{\dot x^2-1-2\dot z\ov z^2}}\\
&=s_\text{th} \int du \le[1+{d(e^{-\xi})\ov du}+ \frac12\,e^{-2\xi}(\dot x^2-1-\dot \xi^2)+\dots \ri]\,.
}
We note that to leading order the action is simply given by the difference in infalling times $u$. All curves that obey the scaling assumptions and stretch between the same endpoints have the same leading order action. The shape of the curves is determined by the $e^{-2\xi}$ term (since the $e^{-\xi}$ term is a total derivative). We can think of \eqref{AreaFunct3} as a generalised membrane theory, where besides the membrane shape $x(u)$ there is an extra degree of freedom $\xi(u)$ that lives on the membrane. Of course, we have not done much compared to the HRT prescription: $\xi(u)=\log z(u)$ is the logarithm of the bulk depth, and we expanded for large $\xi, u, x$.

Since irrespective of the value of $\xi$ or the membrane shape we get the same value for the entropy, one may question the utility of the above exercise: to obtain the entropy we could just as well use the ordinary membrane theory with ${\cal E}(v)=1$. However, we will show how $\xi$ plays an essential role in computing the reflected entropy in 2d CFT. We will also analyse how the membrane theory arises in an RG flow from a 2d CFT, by gapping out the $\xi$ degree of freedom, which then can be integrated out, leaving us with a local theory for $x(u)$, the membrane shape.

\subsection{The entanglement membrane shape for a half space}\label{ApproxApproxLag}

The membrane shape is determined from the effective Lagrangian 
\es{Leff}{
\mathcal{L}_\text{eff}=\frac12\,e^{-2\xi}(\dot x^2-1-\dot \xi^2)\,.
}
derived from \eqref{AreaFunct3}. The momentum $p=e^{-2\xi} \, \dot x$ is conserved and the $\xi$ equation of motion is
\es{xiEOM}{
\ddot \xi=-1+\dot \xi^2+p^2 e^{4\xi}\,,
}
which motivates us to introduce $p\equiv e^{-2\xi_p}$. We can understand the gross features of the solutions that stay in the large $\xi$ region for long $u$ as follows: For $\xi<\xi_p$ the $p^2 e^{4\xi}$ term is negligible, and the solution is  $\xi(u)= u$ with exponentially small corrections. Once this curve reaches the vicinity of $\xi_p$ it transitions to the solution $\xi(u)= \xi_p$ with exponentially small corrections. The transition between the two behaviours is abrupt and can be neglected in the scaling regime. All of these statements can be verified by straightforward numerical solutions of \eqref{xiEOM}. It turns out that \eqref{xiEOM} can also be solved in closed form, see Appendix~\ref{ApproxLagDispHalf}; the exact solution also confirms the above approximations in the large $T$, large $X$ limit. Finally, since the BTZ black brane is AdS$_3$ with some identifications, we are studying AdS$_3$ geodesics in disguise: this perspective is developed in Appendix~\ref{Displaced half spaces}.

\begin{figure}[htbp]
\centering
\includegraphics[width=.38\textwidth]{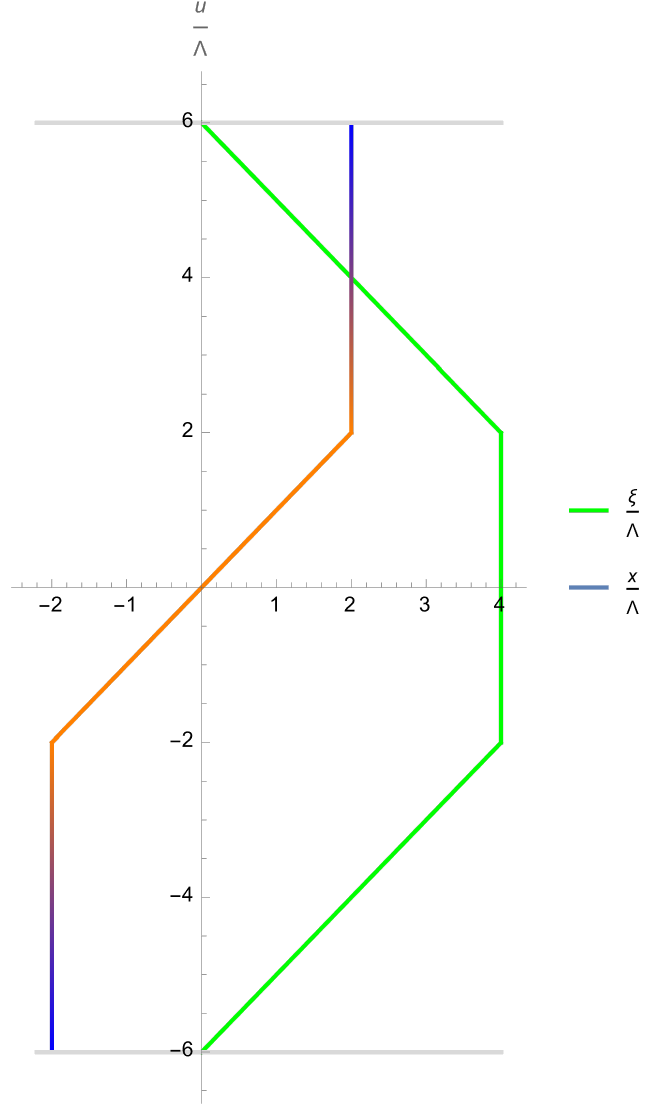}\hspace{2cm}
\includegraphics[width=.22\textwidth]{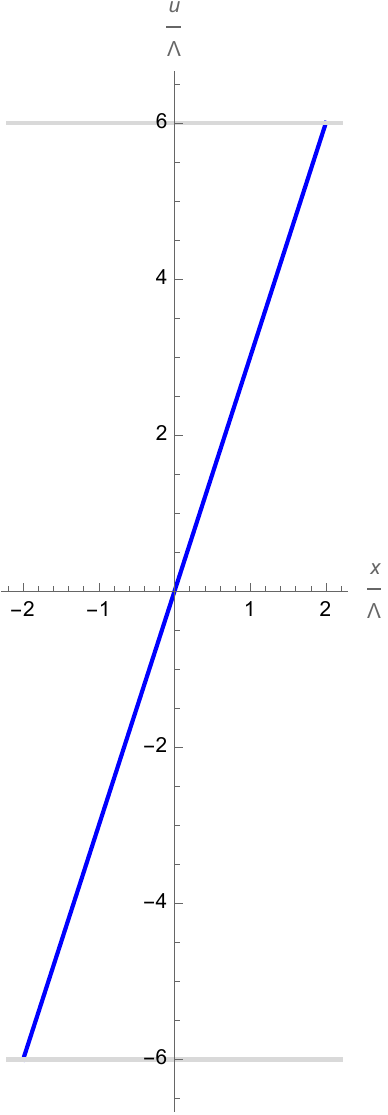}
\qquad
\caption{$\emph{Left}$: A plot of $x(u)$ and $\xi (u)$ (green) in \eqref{FinalXi} in the scaling, i.e. large $\Lambda$ limit in generalised membrane theory. The light gray horizontal lines denote $u=\pm T\Lambda$. As for $x(u)$, we use blue $\to$ orange to denote the increase in the bulk depth in $\xi$. $\emph{Right}$: A plot of the membrane shape function \eqref{MembraneShapeHigherD} in planar AdS$_{d+1}$-Schwarzschild black brane ($d\geq 2$), with $v=\frac{1}{3}$. The membrane is of uniform blue color as it stays at a constant $z\in O(1)$ at late time. \label{fig:zxApproxLagrangian}}
\end{figure}

In summary we find (when $T>X$)
\es{FinalXi}{
(x(u),\xi(u))=\begin{cases}
\le(-X,T+u\ri)\qquad &-T<u<-X\,,\\
\le(u,T-X\ri)\qquad &-X<u<X\,,\\
\le(X,T-u\ri)\qquad &X<u<T\,,
\end{cases}
} 
where we used the result $\xi_p=T-X$ obtained from matching to boundary conditions. See Fig.~\ref{fig:zxApproxLagrangian} (left) for a plot of \eqref{FinalXi}. In this and subsequent figures we plot the membrane shape $u(x)$ and visualise the extra degree of freedom $\xi$ on the membrane worldline with a colour interpolating between blue ($\xi=0$) to orange ($\xi=\xi_p$). In Fig.~\ref{fig:zxApproxLagrangian} we also plotted $\xi(x)$ with green, which is redundant information. Also in plots we will always take the time and space extents to be $T\Lam$ and $X\Lam$ with $T,X=O(1)$ and $\Lam\to\infty$.

\begin{figure}[htbp]
\centering
\includegraphics[width=.28\textwidth]{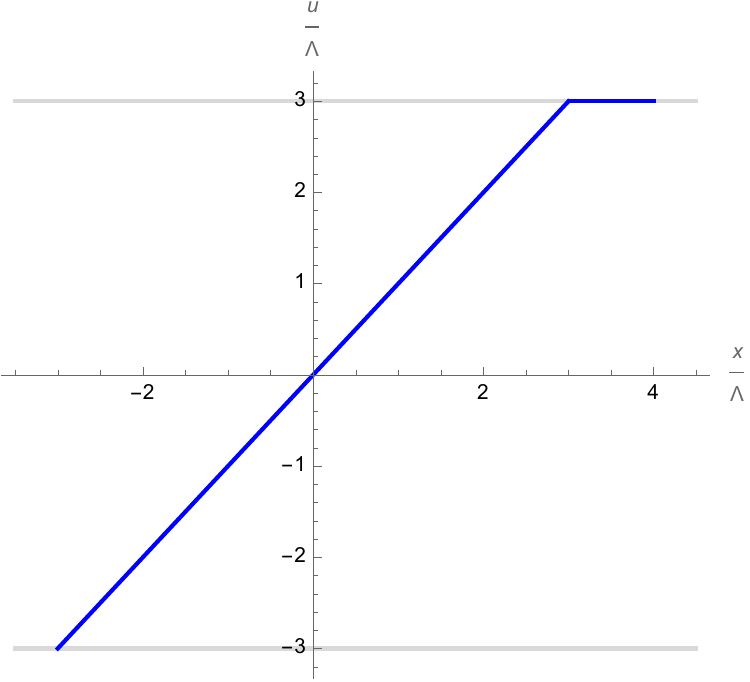}\hspace{1cm}
\includegraphics[width=.28\textwidth]{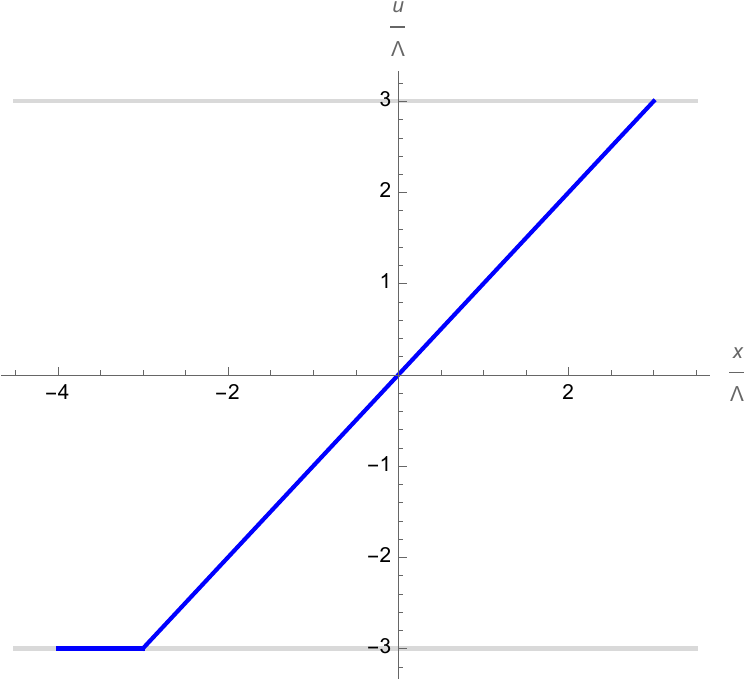}\hspace{1cm}
\includegraphics[width=.28\textwidth]{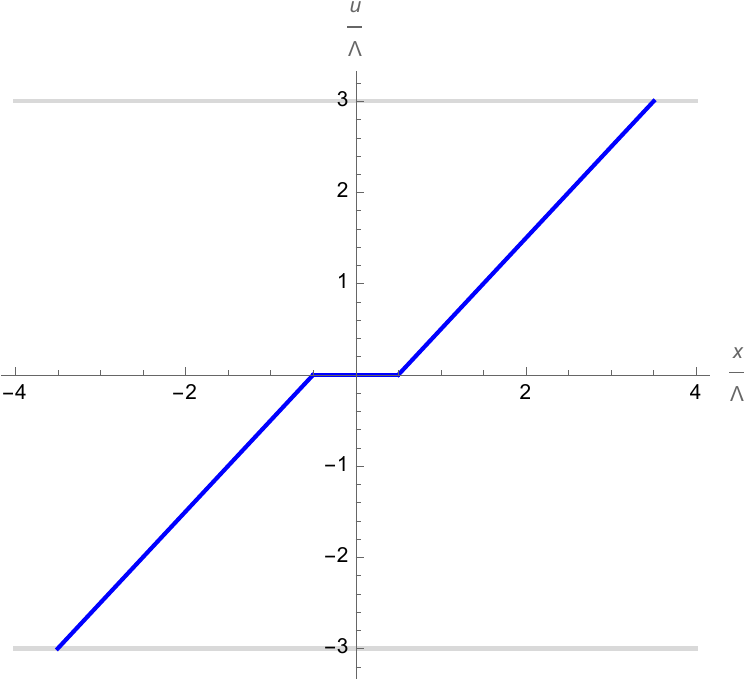}
\caption{Equivalent membranes computing the operator entanglement for $T<X$ in the scaling limit. Here we plotted for $T=6\Lambda$ and $X=7\Lambda$. The membranes are in blue, as the corresponding HRT surface is in the near-horizon region. \label{fig:XgTmem}}
\end{figure}

The corresponding operator entanglement entropy is
\es{OpEntFinal}{
S_{\text{op}}=s_\text{th}\,2T\,,
}
which simply follows from the extent of the generalised membrane in $u$, see the discussion below \eqref{AreaFunct3}. In mapping the HRT surface to the boundary generalised membrane, we have to decide when to switch from the right to the left infalling time. It does not matter at what exact point we switch, as long as the HRT surface is not at or very close to the horizon, see a detailed analysis in~\cite{Mezei:2016zxg}.
The right and left time extents of the membrane then add up to~$2T$.\footnote{A good choice is to switch between the two sides, when the HRT surface reaches the deepest point behind the horizon $z_*$. If we parametrise the membrane in boundary polar coordinates as $r(u,\Om)$, then generically the switch will happen at different locations as we move in polar angle, $u_\text{switch}(\Om)$, so the mapping will be dependent on the subsystem through the membrane shape. A simpler choice would be to switch at $u=0$, but this choice is not time translation invariant, and under large time shifts (corresponding to boosts in the near-horizon region) it leads to the switch happening in the problematic near-horizon region.

There is one extra subtlety: there exist HRT surfaces that live entirely or partly (for some region of $\Om$) in the near-horizon region. Hence they never reach $z_*$ and the above switch procedure is not well-defined for them. These membranes or membrane sections have $v=v_B$, which maps to the near-horizon region $z=1$. In these cases we use the $u_\text{switch}=\text{const}$ switch, with the constant set by continuity in $\Om$ or if the entire membrane has $v=v_B$ we set  $u_\text{switch}=-T,0,T$ depending on what looks most natural, see Fig.~\ref{fig:XgTmem}. This choice however cuts out some portion of the HRT surface, which we need to account for to correctly compute the entropy. We glue in a horizontal membrane for this portion. This rule is consistent, because ${\cal E}(v_B)/v_B=1$. It is also natural from the circuit perspective, where it corresponds to a horizontal cut of legs~\cite{Jonay:2018yei}.  See the first membrane in Fig.~3 in~\cite{Mezei:2018jco} for illustration. Unfortunately, the last membrane on that figure, corresponding to saturation is misleading: we explain below that saturated membranes are more naturally $v_B$ cones. \label{matching_footnote}}    

It is instructive to compare the generalised membrane theory equations of motion \eqref{FinalXi} to that in the ordinary membrane theory~\cite{Mezei:2018jco}, where the HRT surface is described solely by the membrane shape function $x(u)$, since the bulk $z$ degree of freedom can be integrated out. For the two-sided entanglement entropy in half spaces offset by $2X$ $(v_BT>X)$, the membrane shape is a straight line of slope $v$~\cite{Jonay:2018yei,Mezei:2018jco}
\begin{align}
    x(u)=v u\,, && v=\frac{X}{T}\in[0,v_B)\,. \label{MembraneShapeHigherD}
\end{align}
The comparison between ordinary \eqref{MembraneShapeHigherD} and generalised \eqref{FinalXi} membrane theory is illustrated in Fig.~\ref{fig:zxApproxLagrangian}. A natural question to ask, then, is how to interpolate between them. We will answer this question in section~\ref{The transition between the generalised and the ordinary membrane theory}. 
The corresponding operator entanglement entropy however should not differ between the generalised and ordinary membrane prediction, which is~\cite{Jonay:2018yei,Mezei:2018jco}
\es{OpEntFinal2}{
S_{\text{op}}=s_\text{th}\,2T\,{\cal E}(v)\,.
}
For the degenerate membrane tension ${\cal E}(v)=1$ this indeed agrees with \eqref{OpEntFinal}.


So far we have focused on the $T>X$ case. When $T<X$, one can show that $\xi$ is small, i.e. the geodesic goes just inside the horizon. As $z$ is no longer large, the effective action \eqref{AreaFunct3} is no longer valid. In the discussion above \eqref{AreaFunct3} we encountered the possibility that $z=O(1)$ and $\dot x^2=1$, and this is exactly the HRT surface we need for  $T<X$. 
This conclusion is also supported by the exact BTZ geodesics as shown in Appendix~\ref{Displaced half spaces}. The membrane mapping of this HRT surface is a bit unnatural; see Fig.~\ref{fig:XgTmem}, as we need to include a horizontal membrane section between sections of $v=v_B=1$ ones. We can add this section at arbitrary positions, and all of these shapes map to the same HRT surface in the bulk. See footnote~\ref{matching_footnote} for an explanation.\footnote{For this setup, it may seem more natural from the bulk point of view to extend the time extent of spacetime from $2T$ to $2X$, and then the membrane would be a straight $v=1$ line. However, such a choice would lead to a very convoluted membrane mapping for more complicated shapes in higher dimensions. To us, the addition of horizontal membrane segments produces the most intuitive membrane theory. }
The corresponding operator entanglement is:
\es{OpEntFinal3}{
S_{\text{op}}=s_\text{th}\,2X\,,
}
in agreement with earlier results in~\cite{Jonay:2018yei,Mezei:2018jco}.

\subsection{Membranes for saturated strips}\label{MembraneStatic2d}

The displaced half space discussion is easily generalised for displaced intervals of width $D$. In this case there are two candidate HRT surfaces or equivalently generalised membranes: the first is the doubling of the HRT surface corresponding to displaced half spaces and it connects the interval endpoints on the two sides, the second is a static HRT surface that connects the interval endpoints on the same sides. For early times $T<D$, the first is minimal, and for late times $T>D$ the second becomes dominant, leading to the saturation of operator entanglement entropy. In this section we discuss the second HRT surface and its membrane equivalent.

We show the derivation in detail, as the mapping of the near-horizon region to the membrane was not considered with enough care in prior literature. We consider static RT surfaces outside the horizon on a constant Schwarzschild time slice $t=T$.\footnote{The Schwarzschild time $t$ in the geometry \eqref{blackBraneInf} is defined as
\begin{equation}
t = u+\int_0^z\frac{dz'}{a(z')b(z')}
\end{equation}} We can choose a coordinate system such that our two boundary endpoints are symmetric at $\Big(\pm\frac{D}{2},T\Big)$.\footnote{To study the static geodesic only, it is easiest to set $T=0$ as its geometry does not change with time. }
The static geodesic on the $t=T$ slice in BTZ is depicted in Fig.~\ref{fig:StaticGeod} (see Appendix~\ref{Static geodesic} for the explicit expressions). We observe that the geodesic roughly consists of two parts: a $\emph{plateau}$,\footnote{Throughout this paper, we use the word ``plateau" to refer to the portion of the HRT surface that stays at constant $z$ (either inside or outside the horizon) in the scaling limit.} where it lies close to the horizon $z=1$ and moves transversely; and two regions where it $\emph{shoots}$ from the near-horizon region towards the conformal boundary $z=0$. 

\begin{figure}[htbp]
\centering
\includegraphics[width=.55\textwidth]{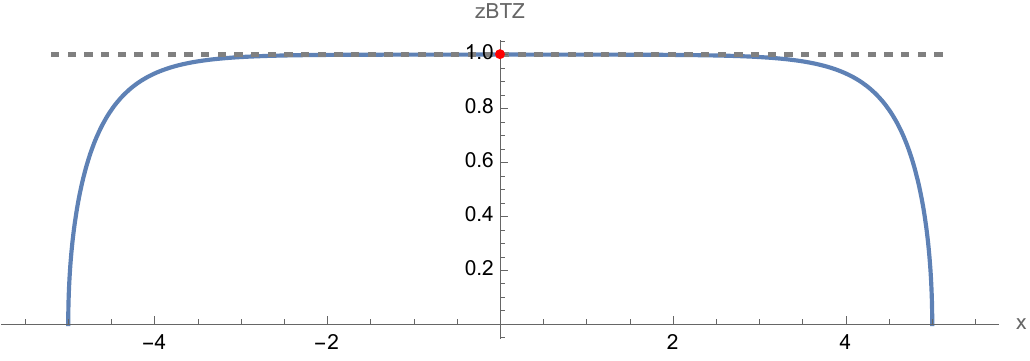}
\qquad
\caption{Plot of the static spacelike geodesic on the $t=T$ slice. The dashed gray line is the black hole horizon. The geodesic consists of a plateau and two parts that shoot towards the conformal boundary. The red dot denotes the deepest point $\tanh \frac{D}{2}$ the geodesic can reach in the bulk. \label{fig:StaticGeod}}
\end{figure}

The static curve shown in Fig.~\ref{fig:StaticGeod} is on the constant Schwarzschild time slice $t=T$. In (generalised) membrane theory, however, we would like to understand the projection of this RT surface to the boundary along constant infalling time. In $d=2$, the infalling time $u$ along the static geodesic on the $t=T$ slice is given by (see Appendix~\ref{SchVSInfallingSec})
\es{u(z)2dExpansion}{
    u(z)&=T-\text{arctanh}z\\
    &\to T+\frac{1}{2}\log\frac{1-z}{2},\ \ (z\to 1^-)
}
Notice that $u(z)\to -\infty$ as $z\to 1^-$. Therefore, close to the horizon, a small shift in $z$ leads to a large shift in $u$. This shift then leads to a finite difference between infalling and Schwarzschild time. See Fig.~\ref{fig:RTInfalling} in Appendix~\ref{Static geodesic} for an illustration of the geodesic in Fig.~\ref{fig:StaticGeod} in infalling coordinates. 

\begin{figure}[htbp]
\centering
\includegraphics[width=.55\textwidth]{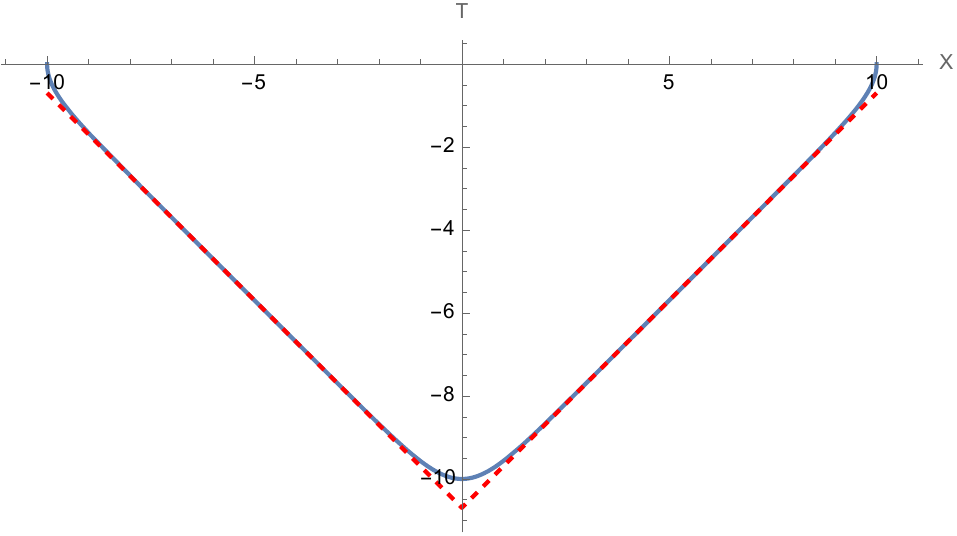}
\qquad
\caption{Projection of the static RT surface (see $\eqref{zLambdaStatic}$ and $\eqref{xLambdaStatic}$ for explicit expressions) to the boundary along constant infalling time. In the scaling limit, the projection is a cone of slope $\pm 1$ \eqref{vBcone2dApprox}, which are depicted here as dashed red lines. Notice that its two corners and tip are all blunt.
 \label{fig:vBcone2d}}
\end{figure}

The explicit projection of the static geodesic shown in Fig.~\ref{fig:StaticGeod} to the boundary along constant infalling time is given by (see Appendix~\ref{Static geodesic} for details)
\begin{equation}
\label{vBcone2dApprox}
    u(x)\approx 
    \left\{
    \begin{aligned}
        &x+T-\frac{D}{2}-\log 2,&&x>0\,,\\
        &-x+T-\frac{D}{2}-\log 2,&&x<0\,,
    \end{aligned}
    \right.
\end{equation}
where we kept the subleading $\log 2$ term to get a more accurate match on Fig.~\ref{fig:vBcone2d}.
We see that the projection is a $\emph{cone}$ with slope $\pm 1$, as is depicted in Fig.~\ref{fig:vBcone2d}. Notice that the majority of the cone comes from the projection of the plateau part of the geodesic; the parts that shoot exponentially towards the boundary are projected to the two blunt corners. The latter is because when $z\to 0$, $\text{arctanh}z$ is small, so $u(z)$ is close to $T$. Notice also that the tip of the cone is blunt in a region vanishingly small in the scaling limit, and hence is not captured by \eqref{vBcone2dApprox}.  

\section{Reflected entropy from the entanglement wedge cross section in 2d CFT}\label{sec:2dEWCS}

The generalised membrane theory and the ordinary membrane theory results with degenerate membrane tension ${\cal E}(v)=1$ predict the same entropy evolution in quenches and for operator entanglement. However, in this and the next section we show that the two give drastically different predictions for reflected entropy: it is the generalised membrane theory that computes it correctly for 2d CFTs. 

\subsection{Reflected entropy and the entanglement wedge cross section}
For a bipartite system with density matrix $\rho_{AB}$, the reflected entropy is given by
\begin{equation}\label{eq:RefEnt}
S_{R}(A:B) = S(\rho_{AA^*})\,, \quad \rho_{AA^*} = \Tr_{BB^*}\ket{\rho^{1/2}_{AB}}\bra{\rho^{1/2}_{AB}}\,,
\end{equation}
where $\ket{\rho^{1/2}_{AB}}\in \mathcal{H}_{ABA^*B^*}$ is the canonical purification of  $\rho_{AB}$, and $S(\rho)$ is the von Neumann entropy~\cite{Dutta:2019gen}. The extended Hilbert space $\mathcal{H}_{ABA^*B^*}$ is constructed out of mirror copies of the original system.
In applications $A$ and  $B$ are often disjoint regions and are parts of a larger system. Then $\rho_{AB}$ is obtained by reducing the state of this larger system onto the region $A\cup B$.

The holographic entanglement entropy for the density matrix $\rho_{AB}$ is:
\begin{equation}\label{eq:HRT}
S(AB) = \frac{1}{4G_N}\text{Area}(\gamma_{AB})\,,
\end{equation}
where $\gamma_{AB}$ is the codimension-two extremal surface homologous to the entangling region $A\cup B$ (HRT surface). Given the HRT surface $\gamma_{AB}$, the cross section $\Sigma_{AB}$ is the minimal area surface among the (potentially multiple) extremal surfaces bisecting the entanglement wedge, defined as the causal development of $\gamma_{AB}\cup A \cup B$~\cite{Takayanagi:2017knl,Akers:2022max}. 

It was shown in~\cite{Dutta:2019gen} that the holographic dual of the reflected entropy is the entanglement  wedge cross section (EWCS):
\begin{equation}\label{eq:holoRefEntCov}
S_{R}(A:B) = \min_{\Sigma_{AB}}\frac{1}{2G_N}\text{Area}(\Sigma_{AB})\,,
\end{equation}
where the minimisation is done over all extremal surfaces.

\subsection{Setup of the problem}
In order to study the time evolution of the entanglement wedge cross-section, we consider a bipartite system consisting of the union of two strips: $A\cup B = [0,\ell]\cup [\ell+D,\infty)$. As initial state, we consider $\ket{\psi_0}=e^{-\frac{\beta}{4}H}\ket{B}$, where $\ket{B}$ is a conformal boundary state. The time evolution of such a system is described holographically by a one-sided black hole with an end-of-the-world (EoW) brane~\cite{Hartman:2013qma}\footnote{Another well-known set-up is that of a black hole form from the collapse of null dust, described by the Vaidya geometry. The evolution of the EWCS in this scenario has been studied in \cite{BabaeiVelni:2020wfl,BabaeiVelni:2023cge}.}. The location of the EoW brane in infalling time is
\begin{equation}\label{eq:EoWBrane}
u_{\text{EoW}}(z) = \frac{i\pi}{a'(1)b(1)} - \int_0^z \frac{dz'}{a(z')b(z')}\,.
\end{equation}
The reflected entropy was first studied in this setup in~\cite{Kudler-Flam:2020url}, where three computations were presented: a 2d CFT and the holographic calculation using BTZ geodesics gave agreement, and these were reproduced by a heuristic membrane picture. In what follows, we will reproduce the results of~\cite{Kudler-Flam:2020url}  using the generalised membrane theory, and explain how it is related to their heuristic membrane picture. The main novelty in our treatment is the introduction of a generalised membrane characterising the equilibrium EWCS (section~\ref{EqEWCS}), which describes a spacelike geodesic in BTZ that contains a portion that is null in the scaling limit.\footnote{If extended beyond the HRT surface it ends on, it reaches the BTZ singularity (Appendix~\ref{SatEWCS}).} 
The computations will build on the results on half-space entanglement discussed in section~\ref{Sec:EEMembrane}.

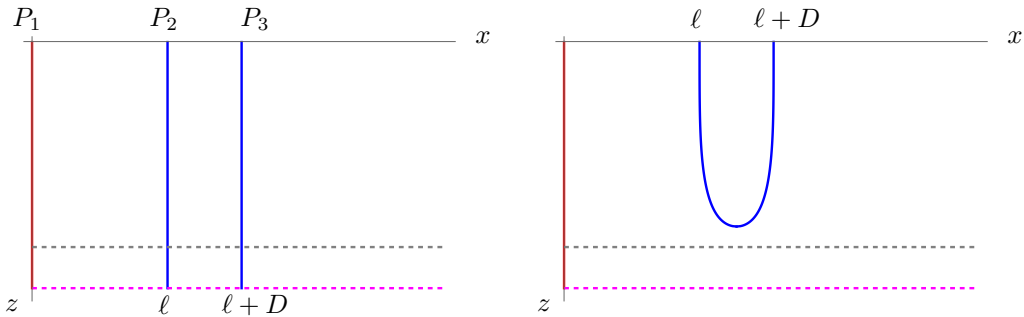
\begin{figure}[ht]
\centering
\begin{tikzpicture}
\node (hrt1) at (0,0)
    {\includegraphics[scale=0.3]{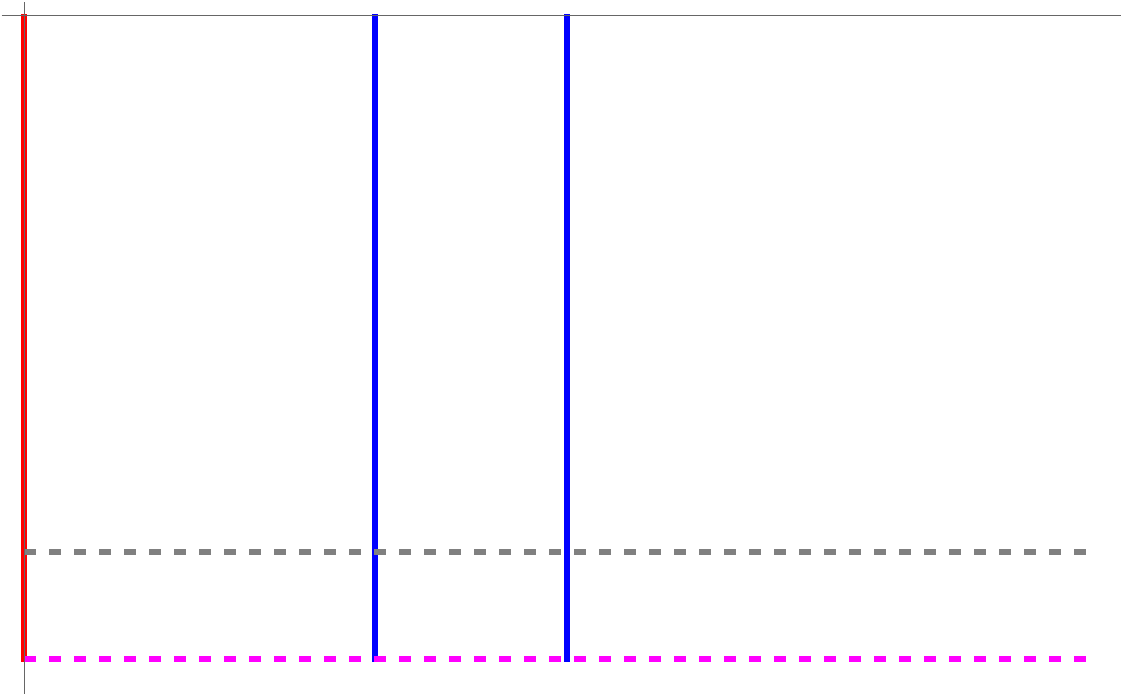}};
\node at (-3,-1.8) {\small $z$};
\node at (4,-1.8) {\small $z$};
\node at (3.2,1.75) {\small $x$};
\node at (10.2,1.75) {\small $x$};
\node at (-2.8,2) {\small $P_1$};
\node at (-1,2) {\small $P_2$};
\node at (0.2,2) {\small $P_3$};
\node at (-1,-1.8) {\small $\ell$};
\node at (0.2,-1.8) {\small $\ell+D$};
\node at (6,2) {\small $\ell$};
\node at (7.2,2) {\small $\ell+D$};
\node (hrt2) at (7,0)
    {\includegraphics[scale=0.3]{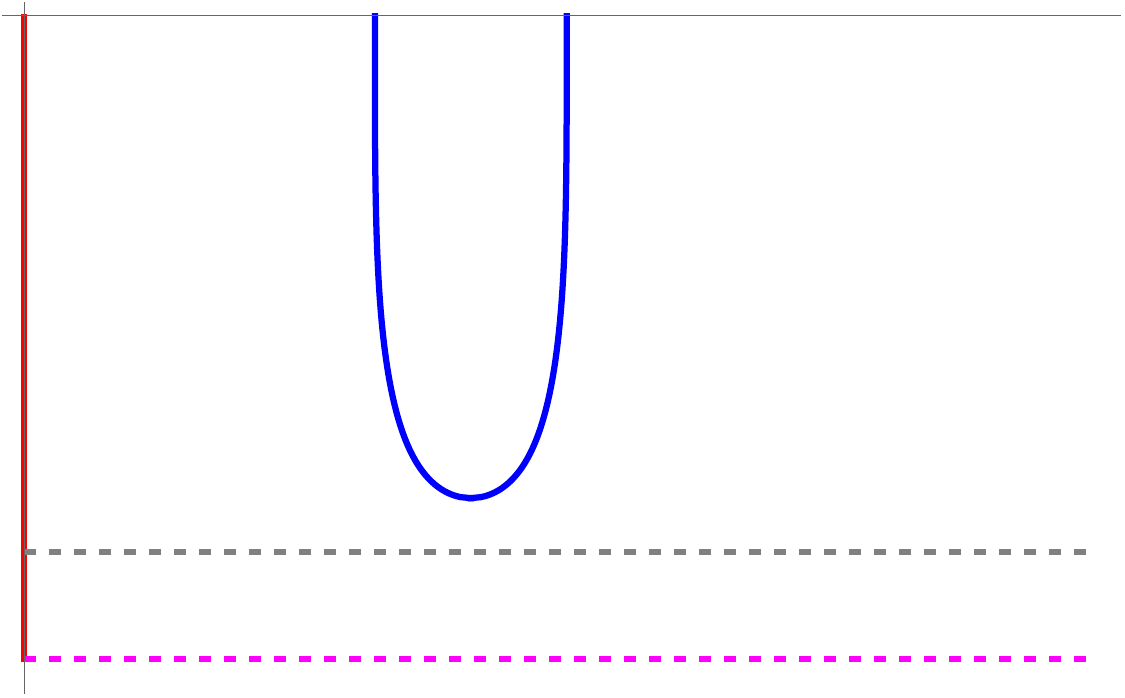}};
\end{tikzpicture}
\caption{HRT surfaces for the interval $[0,\ell]\cup [\ell+D,\infty)$. The black dashed line indicates the horizon, while the magenta line corresponds to the end-of-the-world brane. In the right graph, the non-equilibrium and equilibrium HRTs are plotted in red and blue, respectively. }
\label{fig:hrts}
\end{figure}

The case $\ell<D$ is uninteresting for our purposes since its entanglement wedge is always disconnected. Hence we will focus on the $\ell>D$ case. The possible extremal surfaces computing the entanglement entropy are shown in Fig.~\ref{fig:hrts}. At early times, the entanglement wedge is disconnected, and so the cross section simply vanishes. At time $t=\frac{D}{2}$, there is a transition to a connected entanglement wedge. Our main goal here will be to characterise the dynamics of the EWCS in this scenario.

\begin{figure}[H]
\centering
\begin{tikzpicture}
\node at (0,0) {\includegraphics[scale=0.4]{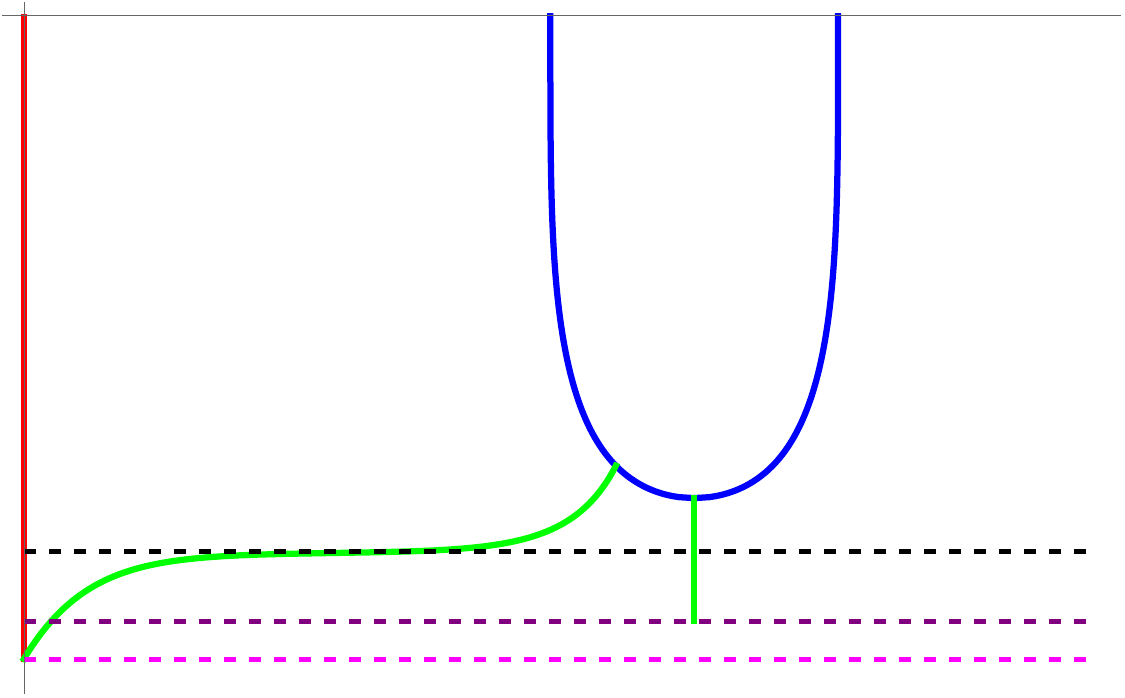}};
\node at (4,2.25) {$x$};
\node at (-3.7,-2.5) {$z$};
\node at (-0.1,2.5) {$\ell$};
\node at (2,2.5) {$\ell+D$};
\node at (4,-1.3) {$z_{h}$};
\node at (4,-1.7) {$z_{0}'$};
\node at (4,-2.1) {$z_{0}$};
\end{tikzpicture}
\caption{Possible non-vanishing entanglement wedge cross sections (green). Since these curves move in time, they may encounter the EoW brane at different $z$'s, which itself is moving according to~\eqref{eq:EoWBrane}.  }
\label{fig:ECSPlot}
\end{figure}
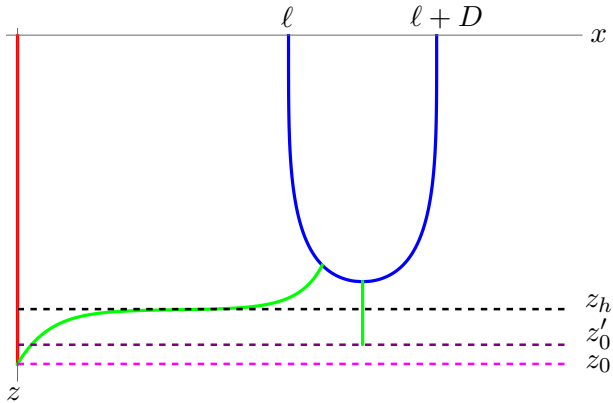

In time-dependent scenarios there are multiple extremal surfaces bisecting the entanglement wedge and, in order to correctly capture the EWCS, one must select the surface with minimal area. For our configuration, the possible surfaces are depicted in Fig.~\ref{fig:ECSPlot}. 

\subsection{Non-equilibrium EWCS}
\begin{figure}[htbp]
\centering
\includegraphics[width=.6\textwidth]{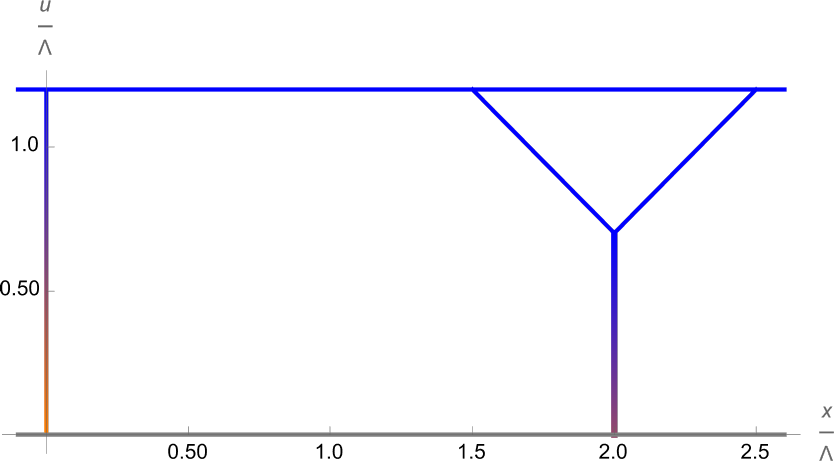}
\qquad
\caption{Projection of non-equilibrium EWCS in $d=2$ along constant infalling time in the scaling limit, for $T=1.2\Lambda$, $\ell=1.5\Lambda$, and $D=\Lambda$. We use blue $\to$ orange to denote the increase in the bulk depth in $\xi$, and plot the EWCS as thick to contrast with HRT. The non-equilibrium EWCS starts from the top tip of the projection of the static RT surface, and then enters the interior where its length grows linearly in time. The non-equilibrium EWCS ends perpendicularly on the EoW brane, whose location is given by $u=0$ for large $z$. We plotted the EoW brane as gray, since HRT and EWCS can end on it at any depth. From this projection, one can easily read off the length of the EWCS to be $T-\frac{D}{2}$. See Fig.~\ref{fig:NeqEWCS} for the full 3d plot. \label{fig:NeqEWCSProj}}
\end{figure}

At $t\geq\frac{D}{2}$, the RT surface connecting $P_2$ and $P_3$ saturates, see Fig.~\ref{fig:hrts}. The projection of this RT surface is therefore a cone of height $\frac{D}{2}$. The EWCS starts from this RT surface, enters the black hole horizon, and ends on the EoW brane in the interior. The stretch of the spacelike interior time direction~\cite{Hartman:2013qma} then leads the EWCS to grow in time. The EWCS can also be described by the effective Lagrangian method described in section~\ref{ApproxApproxLag}. The symmetry of the problem indicates the EWCS to be a vertical line starting from the tip of the RT surface, where $u=T-\frac{D}{2}$. Thus, we have $X=0$ and $T\to T-\frac{D}{2}$ in \eqref{FinalXi}. The time-dependent EWCS is described by 
\begin{align}
    (x(u),\xi(u))=\left(\ell+\frac{D}{2},T-\frac{D}{2}-u\right)
\end{align}
See Fig.~\ref{fig:NeqEWCSProj}. Therefore, the length of the EWCS is $T-\frac{D}{2}$, which starts from 0 at $T=\frac{D}{2}$ and grows linearly in time with slope $v_E=1$.

\subsection{Equilibrium EWCS}\label{EqEWCS}
To describe the entanglement wedge cross section, we introduce a section of a generalised membrane that moves negligibly in $u$ with a bit of hindsight.\footnote{See Appendix~\ref{SatEWCS} for the exact solution of this problem.} Correspondingly, its contribution to the entropy is zero in the scaling limit. It is describing a portion of an HRT surface that becomes null in the scaling limit. Nevertheless it is important in piecing together the generalised membrane that computes the EWCS. 

\begin{figure}[htbp]
\centering
\includegraphics[width=.5\textwidth]{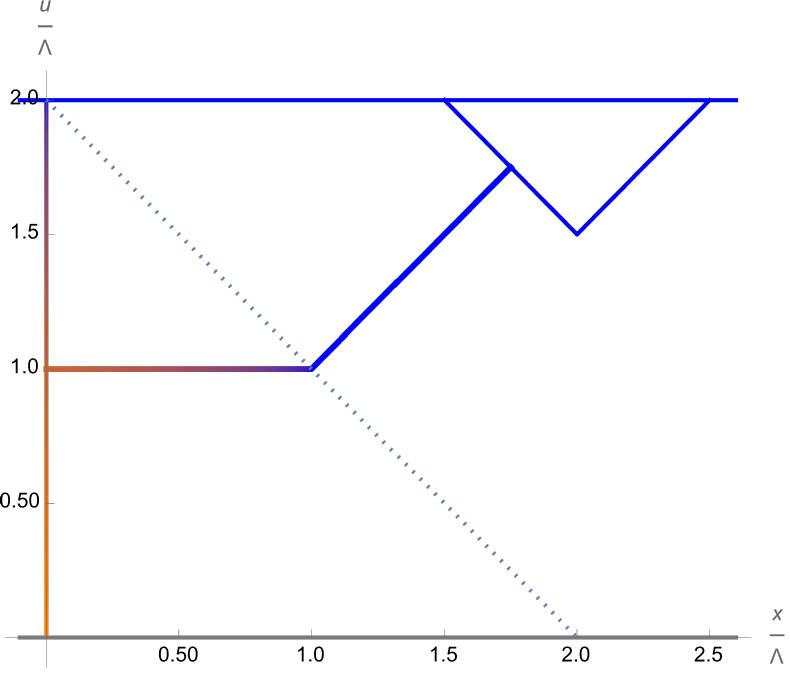}
\qquad
\caption{Projection of equilibrium EWCS in $d=2$ along constant infalling time, for $T=2\Lambda$, $\ell=1.5\Lambda$, and $D=\Lambda$. We use blue $\to$ orange to denote the increase in the bulk depth in $\xi$, and plot the EWCS as thick to contrast it with the HRTs. The equilibrium EWCS starts from the mid-point of the projection of static RT surface, enters the horizon, and arrives at a plateau where it moves in $x$ for a distance $\ell/2$. The EWCS then becomes null and shoots deep into the interior, where it meets the HRT surface. Note that at the connection points the colours of the EWCS match the colours of the HRT surfaces, as required by the boundary conditions. The dotted 45$^\circ$ denotes an alternative projection of the HRT surface along the null generalised membrane, which allows us to match our results with the heuristic picture of~\cite{Kudler-Flam:2020url}. See Fig.~\ref{fig:SatEWCS} for the full 3d plot. \label{fig:EqEWCSProj}}
\end{figure}

Since this generalised membrane moves a large amount in the $x, \xi$ directions but only a little in the $u$ direction, it is easier to describe it by changing variable from $u$ to $x$. Implementing this change of variables in \eqref{xiEOM} and using momentum conservation $\dot x=e^{2(\xi-\xi_p)}$, we get:
\es{xiEOMx}{
 \xi''&=1-\xi'^2 - e^{4(\xi_p-\xi)}\,,\\
 u'&=e^{-2(\xi-\xi_p)}\,.
}
where $f'=\frac{df}{dx}$. This set of equations has the following solution that will be relevant for us:\footnote{Generically, there is another kink at some $x=x_1$, but this additional feature will not be needed. One can think of the solution presented below as the $x_1\to \infty$ special case of the general solution.}
\begin{equation}
    \label{XiEWCS}
    (u(x),\xi(x))=
    \left\{
    \begin{aligned}
        &(u_p,-x+x_0+\xi_p),&& 0<x<x_0\\
        &(x+u_p-x_0,\xi_p),&& x_0<x
    \end{aligned}
    \right.
\end{equation}
where the constants are chosen for later convenience, see Fig.~\ref{fig:EqEWCSProj}. To match the two HRT surfaces or generalised membranes, we require that
\begin{align}
    &\xi_p+x_0=T-u_p\,&&\text{to match the non-equilibrium HRT,}\label{matchingConds1}\\
    &(u_p+x_c-x_0, \xi_p)=(T+\ell-x_c, 0)\,&&\text{to match the equilibrium HRT,}\label{matchingConds2}
\end{align}
where $x_c$ is the connection point between the EWCS and the equilibrium HRT in Fig.~\ref{fig:EqEWCSProj}. In \eqref{matchingConds1}, the non-equilibrium HRT is described by half of \eqref{FinalXi} with $X=0$, hence $x(u)=0$; in \eqref{matchingConds2}, the equilibrium HRT is given by $x=-u+T+\ell$ and $\xi(u)=0$ (as it is outside the horizon). Only three out of four parameters get determined:
\begin{align}
 \xi_p=0\,,&&   u_p=T+{\ell\ov 2}-x_c\,, && x_0=x_c-{\ell\ov 2}\,.\label{matchingConds}
\end{align}
The EWCS is given by the change of $u$ for these generalised membranes
\es{ECWS}{
\ell(\text{EWCS})=x_c-x_0={\ell\ov 2}\,.
}
$\ell(\text{EWCS})$ is independent of $x_c$, so each of the generalised membranes correctly computes the EWCS. This degeneracy does not get resolved even if we solve the approximate equations of motion \eqref{xiEOMx} exactly, as is done in Appendix~\ref{Plateau+Null geodesic}. Once we back off from the large $z$ approximation, subleading terms fix $x_c$ to be $x_c=\ell+{D\ov 4}$, see Appendix~\ref{SatEWCS} for details.

Our generalised membrane description shown in Fig.~\ref{fig:EqEWCSProj} is different from the heuristic membrane picture of~\cite{Kudler-Flam:2020url}. Recall that in getting Fig.~\ref{fig:EqEWCSProj}, we project the bulk HRT surfaces to the boundary along constant infalling time. We can find the connection between our result and that in~\cite{Kudler-Flam:2020url} by projecting instead along the null generalised membrane in~\eqref{XiEWCS}, which amounts to a shift of $\xi$ in the $x$ coordinate compared to projecting along constant infalling time.\footnote{Recall that projecting the HRT surfaces to the boundary along constant infalling time is to plot $(x,u)$ on flat spacetime. When projecting along the null generalised membrane, we plot instead $(x+\xi,u)$.}  Under such a prescription, the horizontal generalised membrane in Fig.~\ref{fig:EqEWCSProj} is projected to a $\emph{point}$, and the non-equilibrium HRT in the left is projected to the dotted 45$^\circ$ line in Fig.~\ref{fig:EqEWCSProj}.\footnote{See Fig.~\ref{fig:SatEWCS} from the exact solution of equilibrium EWCS for a better visualisation.} The resulting projection then matches with the picture of~\cite{Kudler-Flam:2020url}.\footnote{We find it somewhat unsatisfactory that to make the heuristic picture of~\cite{Kudler-Flam:2020url} work for the reflected setup $B'\cup A' =(-\infty,-(\ell+D)]\cup [-\ell,0]$, we would also have to reflect the non-equilibrium membrane, even though it computes the entropy of the complement half-space $(-\infty,0]$ in a pure state. Relatedly, the mapping from the holographic bulk to the membrane picture would need to be dependent on the EWCS we are interested in computing.}

To sum up, in 2d CFT the generalised membrane description of the reflected entropy is given by 
\begin{equation}
\label{GenMembraneEWCS2d}
\frac{S_R}{2} = 
\begin{cases}
0 \ &T < \frac{D}{2}\,, \\ 
T - \frac{D}{2}\ \hspace{0.5cm} &\frac{D}{2} < T < \frac{\ell+D}{2}\,, \\ 
\frac{\ell}{2}  \  &T > \frac{\ell+D}{2}\,.
\end{cases}
\end{equation}
One can compare \eqref{GenMembraneEWCS2d} to the exact solution to this problem detailed in Appendix~\ref{ExactEWCS}, see Figure~\ref{fig:EWCSDyn}. 

\begin{figure}[htbp]
\centering
\includegraphics[width=.55\textwidth]{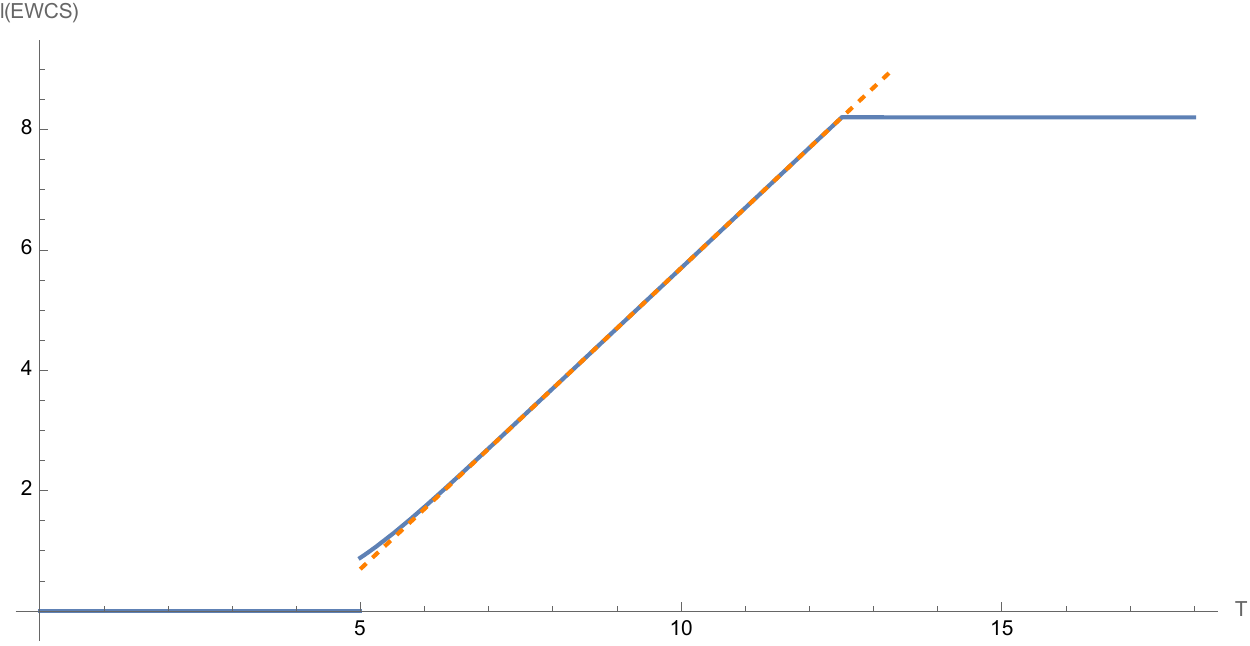}
\qquad
\caption{Time evolution of EWCS, for $\ell=15$, and $D=10$. At early time, the EWCS is 0, as the entanglement wedge is disconnected (Fig.~\ref{fig:hrts} left). After $T=\frac{D}{2}$, the RT surface connecting $P_2$ and $P_3$ saturates (Fig.~\ref{fig:hrts} right), and the EWCS starts to grow linearly with slope 1 in the scaling limit (dashed orange line). Notice that there is a small jump of EWCS length at $T=\frac{D}{2}$; this jump is negligible in the scaling limit. The EWCS then saturates at $\frac{\ell}{2}$ after $T=\frac{\ell+D}{2}$. In the scaling limit, the EWCS dynamics is described by~\eqref{GenMembraneEWCS2d}.\label{fig:EWCSDyn}}
\end{figure}

\section{Reflected entropy in the membrane effective theory}\label{sec:EWCS}
In section~\ref{sec:2dEWCS} we explored the dynamics of the reflected entropy in the context of $2d$ CFTs. Now, we will show how to describe the corresponding dynamics for the case of $d>2$. The setup is the same as that in Fig.~\ref{fig:hrts}, but the equations describing the HRT surfaces and EWCS are different due to the increase in dimensionality. While many of the features explored earlier remain true in higher dimensions, we also observe some important differences. 

While in the preceding sections we only showed computations in the scaling limit that is described by the (generalised) membrane description and relegated finite size computations to appendices, in this section we include finite size numerical computations in the main text, as we believe they may be of methodological interest.

\subsection{Time evolution of entanglement cross-section}
As indicated before, we consider the same configuration as in Fig.~\ref{fig:hrts}. The evolution of entanglement entropy in this setting was first studied in \cite{Hartman:2013qma}, and further analysed in \cite{Liu:2013iza,Liu:2013qca,Mezei:2016zxg,Mezei:2019sla}.

Following the aforementioned references, we parameterise the equilibrium HRT as $(u_1(z),x_1(z) )$. While there is no closed-form solution to the extremisation problem in $d>2$, we can use reparameterisation invariance and translational symmetry to express the solutions in terms of integrals as
\begin{equation}
\label{EqHRTEqn}
\begin{gathered}
u_1(z) = T - \int_0^z \frac{dz'}{a(z')b(z')},\\
x_1(z) = \frac{2\ell+D}{2} + \int_{\Tilde{z}}^{z} \frac{dz'}{b(z')\sqrt{a(z')}}\frac{z'^{d-1}}{\sqrt{\Tilde{z}^{2(d-1)}-z'^{2(d-1)}}}\,.\\
\end{gathered}
\end{equation}
The parameter \(\tilde{z} < 1\) denotes the deepest point reached by the HRT surface in the bulk. This parameter can, in turn, be related to the size of the entangling region by evaluating the integral near the asymptotic boundary, $z=0$. The boundary time at which the entanglement entropy is evaluated is denoted as $T$.

The non-equilibrium HRT surface is simply $(u_2(z),0)$. In this case, reparameterization invariance suffices to solve the extremisation equations as
\begin{equation}
\begin{gathered}
u_2(z) = u_{\text{EoW}}(z_0) - \int_{z_0}^{z} \frac{dz'}{a(z')b(z')}\left(1 - \frac{\sqrt{-a(z_0)}z'^{d-1}}{\sqrt{z_0^{2(d-1)}a(z')-z'^{2(d-1)}a(z_0)}}\right)\,,\\
x_2(z) = 0\,,
\end{gathered}
\end{equation}
where $u_{\text{EoW}}(z)$ is as in eq. \eqref{eq:EoWBrane} and $z_0$ denotes the point at which the surface intersects the end-of-the-world brane. In this case, the boundary time $T$ is related to the parameter $z_0$ by evaluating the integral at $z=0$.

The EWCS that intersects the end-of-the-world brane corresponds to a portion of the non-equilibrium HRT described above. Concretely, it is parameterised as
\begin{equation}\label{intForm}
\begin{gathered}
u_{\text{EWCS}}^{(1)}(z) = u_{\text{EoW}}(z_0') - \int_{z_0'}^z \frac{dz'}{a(z')b(z')}\left(1 - \frac{\sqrt{-a(z_0')}z'^{d-1}}{\sqrt{z_0'^{2(d-1)}a(z')-z'^{2(d-1)}a(z_0')}}\right)\,,\\
x_{\text{EWCS}}^{(1)}(z) = \frac{2\ell+D}{2},
\end{gathered}
\end{equation}
but now the parameter $z_0'$ is determined by the intersection condition
\begin{equation}
\label{Intersection1}
u_{\text{EWCS}}^{(1)}(\Tilde{z}) = u_1(\Tilde{z})\,,
\end{equation}
where $\tilde{z}$ is the same parameter as in \eqref{EqHRTEqn}.

The area of the surface is
\begin{equation}
\label{ECS1}
\mathcal{A}_{\text{EWCS}}^{(1)}(z_0') = \int_{\Tilde{z}}^{z_0'} \frac{dz}{z^{d-1}b(z)}\frac{z_0'^{d-1}}{\sqrt{z_0'^{2(d-1)}a(z)-z^{2(d-1)}a(z_0')}}\,.
\end{equation}
By solving the intersection condition \eqref{Intersection1}, and evaluating the integral \eqref{ECS1} at different values of boundary time $T$, one can obtain the time-dependence of the surface. The result is a monotonic increasing function that is approximately linear for the parameter choices we explored.

The other possible EWCS is a surface connecting the two HRTs, without reaching the EoW brane. Unlike the previous case, this does not correspond to a portion of a known HRT. Fortunately, we can employ the symmetries of the problem to solve the extremisation equations in a similar fashion.

We choose to parameterise this surface as $(u_{\text{EWCS}}^{(2)}(z),x_{\text{EWCS}}^{(2)}(z))$. The area of this surface is given by the functional
\begin{equation}\label{eq:EWCS2Func}
\mathcal{A} = \int_{z_1}^{z_2}\frac{dz}{z^{d-1}}\sqrt{-a(z)\left(\frac{du}{dz}\right)^2- \frac{2}{b(z)}\frac{du}{dz} + \left(\frac{dx}{dz}\right)^2}\,.
\end{equation}
The equations of motion extremising this expression are both reparameterisation and translation invariant, hence we have a couple of conserved quantities denoted as $(j,e)$. In terms of these constants, the solution can be written as
\es{SatEWCSHigherD}{
u_{\text{EWCS}}^{(2)}(z) &= u_2(z_2) - \int_{z_2}^{z} \frac{dz'}{a(z')b(z')}\left(1-\frac{ez'^{d-1}}{\sqrt{H(z')}}\right)\,,\\
x_{\text{EWCS}}^{(2)}(z) &= - \int_{z_2}^{z} dz' \frac{z'^{d-1}}{b(z')\sqrt{H(z')}}\,,\\
H(z) &= \left(e^2 - a(z)\right)z^{2(d-1)} + j^2 a(z)\,,
}
where $(z_1,z_2)$ are the values of $z$ at which the surface intersects the corresponding HRT surface. The integration constants were chosen to guarantee intersection with the non-equilibrium surface.

Up to this point, the calculation is not dissimilar to those of usual HRT surfaces. The main difference arises because this new surface does not obey boundary conditions at $z=0$, instead we demand intersection with the equilibrium surface:
\begin{equation}
\label{Intersection2}
\begin{gathered}
u_{\text{EWCS}}^{(2)}(z_1) = u_1(z_1)\,,\\
x_{\text{EWCS}}^{(2)}(z_1) = x_1(z_1)\,.
\end{gathered}
\end{equation}
The solutions to these conditions provide a family of extremal curves. 

For HRT surfaces, we are able to fix the integration constants, $\tilde{z}$ and $z_0$, by symmetry considerations or by imposing intersection with the end-of-the-world brane. For the new surface associated with the EWCS, we require the intersection with both HRT surfaces to occur perpendicularly; this guarantees that the surface is not only extremal but also minimal. This perpendicularity condition can be used to determine the integration constants $(j,e)$. We then solve the intersection conditions numerically in order to fix $(z_1,z_2)$.

The area of the surface can be written as
\begin{equation}
\mathcal{A}(z_1,z_2) = \int_{z_1}^{z_2} dz \frac{j}{z^{d-1}b(z)\sqrt{H(z)}}\,,
\end{equation}
which we evaluate numerically after solving the intersection conditions.

\begin{figure}[H]
\centering
\begin{tikzpicture}
\node at (0,0) {\includegraphics[scale=0.4]{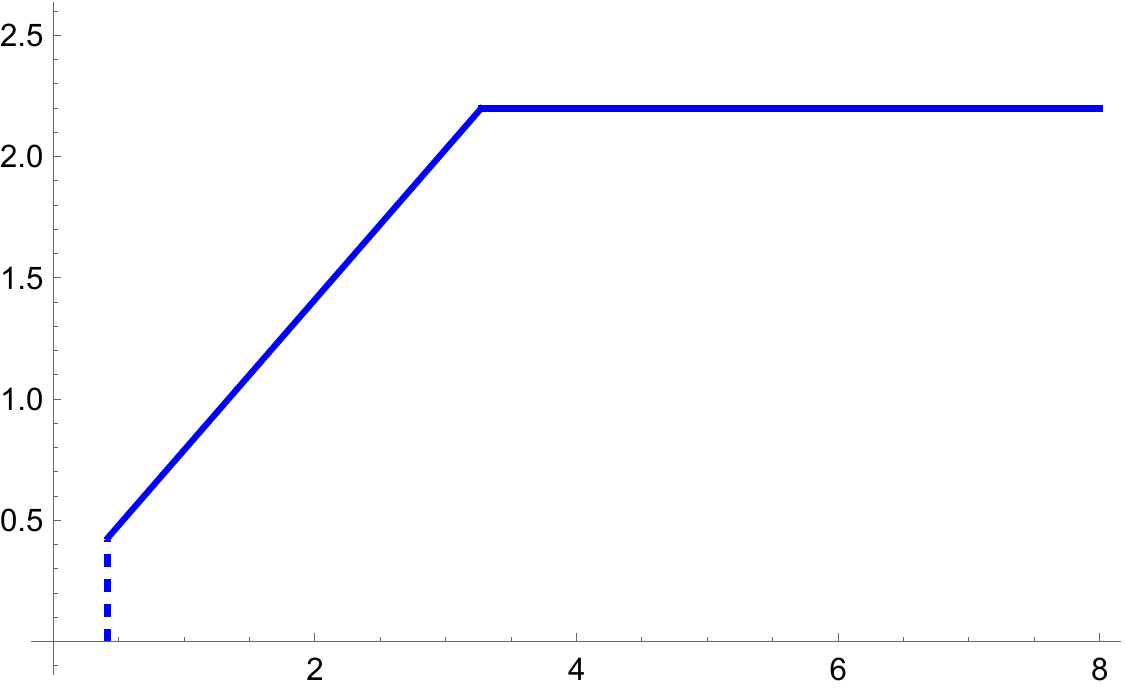}};
\node at (-3.8,2.5) {$\frac{S_R}{2}$};
\node at (4,-2.3) {$T$};
\end{tikzpicture}
\caption{Time evolution of EWCS for $D=1.07$ and $\ell=1.95$ in $d=4$.}
\label{fig:ECSTimeDependence}
\end{figure}

The first candidate surface for the EWCS directly depends on the position of the EoW brane. The second candidate only depends on the position of the EoW brane through the change in shape of the non-equilibrium HRT $u_2(z)$; it approaches a constant as $T\to\infty$. The total time evolution of the EWCS is depicted in Fig.~\ref{fig:ECSTimeDependence}. After the linear growth phase, the time dependence becomes negligible for the choice of parameters considered.

\subsection{Membrane theory for reflected entropy}\label{sec:MemRef}

Let us now develop a description for the general behaviour of the reflected entropy as a function of time, assuming the membrane theory for entanglement entropy is valid. We expect the membrane description to be accurate only in the scaling limit. In order to ease the comparison between the exact numerical results and the membrane prediction, we parameterise the size of the finite interval, $\ell$, as $\ell=\eta D$, with $\eta>1$. We can then take $D\gg1$.

First, we have to determine the scaling limit of the HRT surfaces. The non-equilibrium HRT does not move in $x$, hence it corresponds to the $v=0$ membrane (and most of it sits at $z=z_*$ in the bulk). 
We have seen in section~\ref{MembraneStatic2d} that in $d=2$, the saturated (H)RT surfaces become cones
of slope $\pm 1$ under the membrane mapping. In $d\geq 2$ dimensions this result generalises as follows:
we can approximate \eqref{EqHRTEqn} by expanding the integrand in $z$ around $\tilde z \to 1^-$, giving
\es{ExpandedInts}{
u(z)&=T-\frac{1}{b(1)a'(1)}\log (1-z)+\dots\,,\\
x(z)&={2\ell+D\over 2}-\frac{\log (1-z)}{b(1)\, \sqrt{-2(d-1)a'(1)}}\,,
}
and eliminating $z$ we get
\es{ExpandedInts2}{
x(u)&=\sqrt{-{a'(1)\over 2(d-1)}}u+\text{const}=v_B u+\text{const}\,,
}
where we used \eqref{vB}.
Therefore, projections of static (H)RT surfaces to the boundary along constant infalling time are $\emph{cones}$ of slope $v_B$. These conclusions remain true for arbitrary shapes as can be inferred from the results of~\cite{Mezei:2016zxg}, see also~\cite{Mezei:2016wfz,Mezei:2020knv}.

\begin{figure}[htbp]
\centering
\includegraphics[width=.4\textwidth]{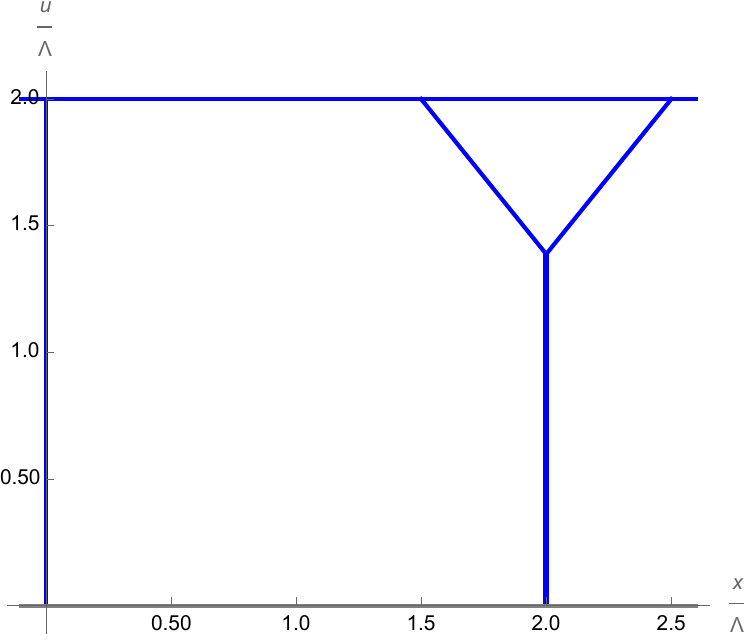}\hspace{0.5cm}
\includegraphics[width=.4\textwidth]{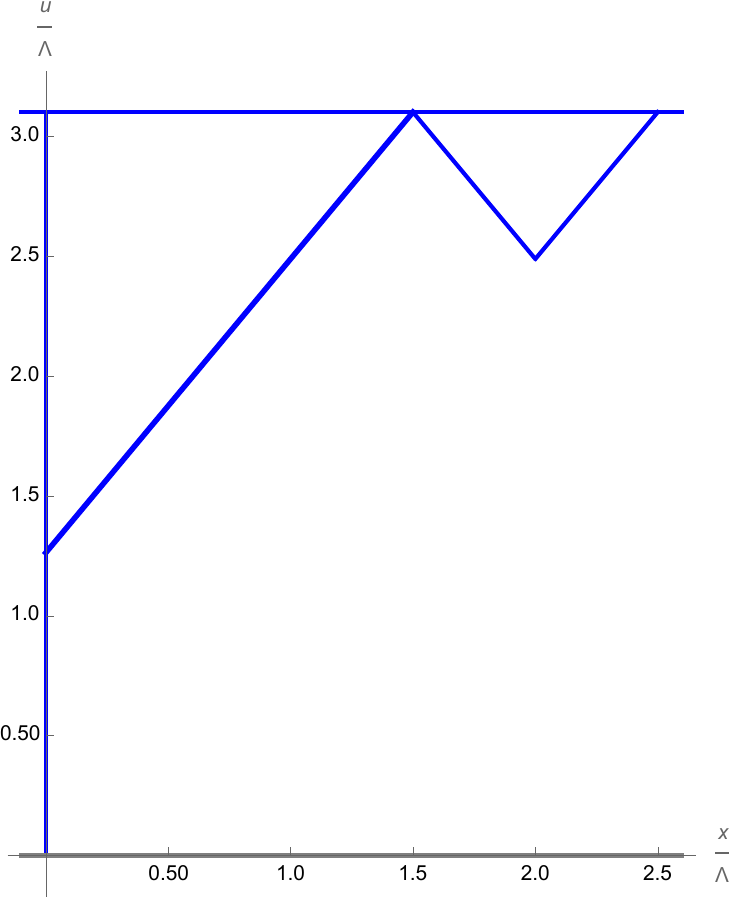}
\qquad
\caption{Projection of EWCS in $d=4$ along constant infalling time in the scaling limit, for $\ell=1.5\Lambda$ and $D=\Lambda$. The entire plots are in blue, as all surfaces are at $z=O(1)$. We plot the EWCS as thick to contrast with HRT. $\emph{Left}:$ Plot at $T=2\Lambda$. The non-equilibrium EWCS starts from the tip of the static RT surface, and then enters the interior and ends on the EoW brane (plotted in gray, since HRT and EWCS can end on it at any depth). Its length grows linearly in time with slope $v_E$. $\emph{Right}:$ Plot at $T=3.1\Lambda$. The equilibrium EWCS skims the horizon from the interior, its projection is therefore a line with slope $v_B$. \label{fig:MembraneEWCS}}
\end{figure}

We can now propose a membrane description for the reflected entropy for the same configuration above. As observed, at early times the entanglement wedge is disconnected and so its cross section vanishes. This continues until the smaller HRT transitions from linear growth to equilibrium, according to the membrane description of entanglement, this occurs at time $t=\frac{D}{2v_E}$. After this time, the cross section is described by a surface connecting the equilibrium HRT to the end-of-the-world brane. Such a surface is described by the same type of membrane as the linear growth of entanglement entropy. Finally, this linear growth continues until the dominant surface is the one connecting the two HRTs. In the scaling limit it is a straight line with slope $v_B$ and its  contribution is equal to $\ell$. See Fig.~\ref{fig:MembraneEWCS}.  A check of the consistency of the membrane description with the full HRT surface is that the EWCS connecting the two HRTs~\eqref{SatEWCSHigherD} is dominated by the part on the horizon in the large $D$ limit, see Figure~\ref{fig:ECSPlot} which is plotted using~\eqref{SatEWCSHigherD}. A calculation similar to~\eqref{ExpandedInts2} then shows that the projection of this EWCS to the boundary along constant $u$ is indeed a line of $v_B$ slope.\footnote{Note that the membrane theory only characterises this surface under the scaling limit and does not encode its local geometry near the connection points to the two HRTs. To investigate these details, one needs to invoke the full equations of motion~\eqref{SatEWCSHigherD}.}  

In summary, the membrane description of the reflected entropy is
\begin{equation}
\label{MembraneEWCS}
\frac{S_R}{2} = 
\begin{cases}
0 \ &T < \frac{D}{2v_E}\,, \\ 
v_E \left(T - \frac{D}{2v_B}\right) \ \hspace{0.5cm} &\frac{D}{2v_E} < T < \frac{\ell}{v_E}+\frac{D}{2v_B}\,, \\ 
\ell  \  &T > \frac{\ell}{v_E}+\frac{D}{2v_B}\,.
\end{cases}
\end{equation}

\begin{figure}[h!]
\centering
\begin{tikzpicture}
\node at (0,0) {\includegraphics[scale=0.4]{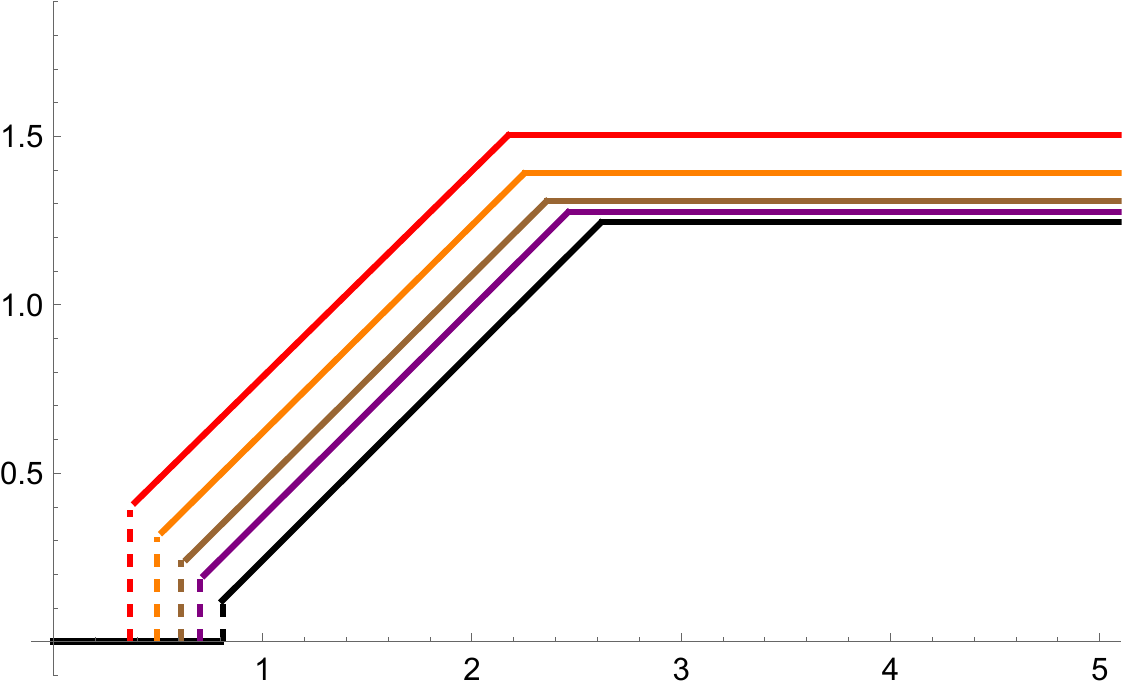}};
\node at (4.5,-2) {$T/D$};
\node at (-3.5,2.5) {$S_R/2D$};
\end{tikzpicture}
\caption{Dynamics of entanglement cross section for $D=1.07$ (red), $D=1.3$ (orange), $D=1.95$ (brown), and $D=3.3$ (purple). This is compared against the membrane prediction (black). As noted in the main text, we emphasise the numerical limitation in solving the intersection problem, which permits us to use only relatively small values of $D$.}
\label{fig:MembraneComp}
\end{figure}

We can compare the membrane theory prediction against the holographic calculation outlined earlier, see Fig.~\ref{fig:MembraneComp}.

As noted earlier, we expect the membrane theory to be accurate in the limit $D\gg 1$. This seems to be supported by the numerical results depicted in Fig.~\ref{fig:MembraneComp}, however we are somewhat constrained from a perfect agreement due to numerical limitations. In Appendix~\ref{App:ScalingECS}, we further explore this point.

\subsection{Comparison between the generalised and ordinary membrane results for reflected entropy}\label{sec:diffs}

\begin{figure}[htbp]
\centering
\includegraphics[width=.45\textwidth]{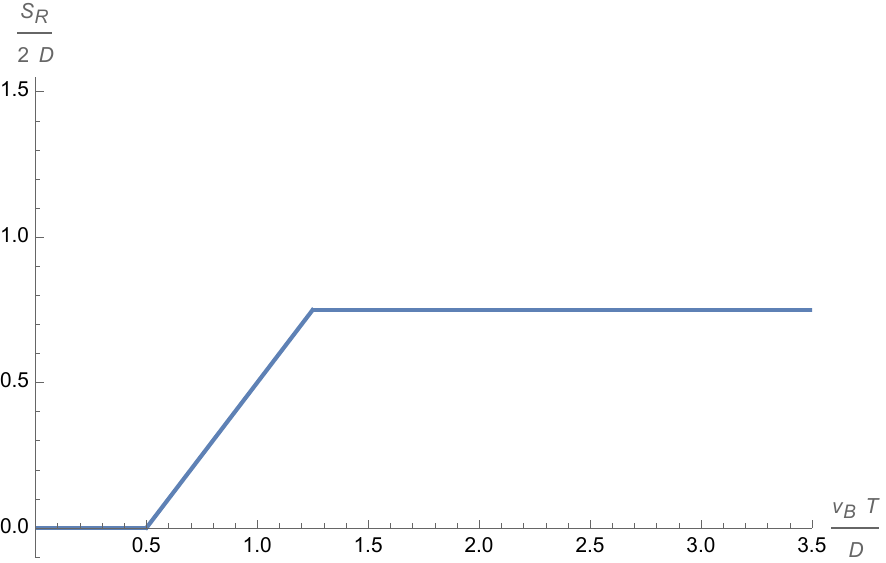}\hspace{0.5cm}
\includegraphics[width=.45\textwidth]{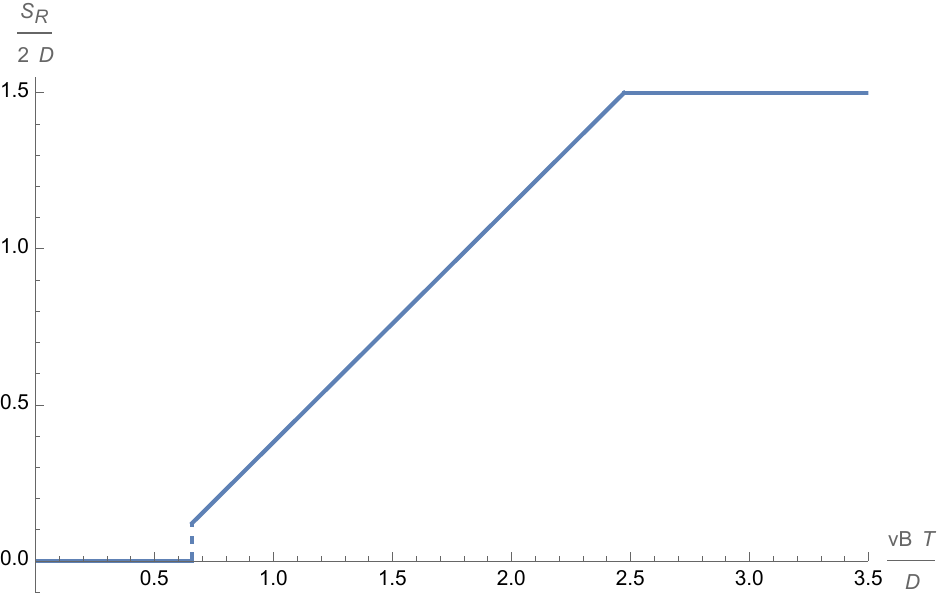}
\qquad
\caption{Time evolution of reflected entropy in $d=2$ (left) and $d=4$ (right) for $\ell/D=3/2$. We use the same ranges on the horizontal and vertical axes to facilitate the comparison. \label{fig:EWCS2d4d}}
\end{figure}

Let us compare the results for reflected entropy from the generalised \eqref{GenMembraneEWCS2d} and ordinary membrane results  \eqref{MembraneEWCS} applicable to 2d and higher-dimensional CFTs (and other generic chaotic systems). There are key differences that we illustrate on Fig.~\ref{fig:EWCS2d4d}. On the horizontal axis we put $v_B T/D$ to make them look as similar as possible. The first difference that we notice is that the membrane theory result is not continuous, it starts with a jump of size 
\es{JumpSize}{
{\Delta S_R\ov 2D}=\frac{1}{2}\left(1-\frac{v_E}{v_B}\right)D\,, \qquad \text{at ${v_B T\ov D}={v_B\ov 2 v_E}$.}
}
Note however that this difference would vanish if $v_E=v_B$ were to hold, i.e.~in the limit of the degenerate membrane tension ${\cal E}(v)=1$. The slope of the two curves are also different, by a factor $v_E/v_B$, which again would go away for $v_E=v_B$. However, the plateau values of the curves $\ell/2D$ and $\ell/D$ do not match even for  the degenerate membrane tension: \emph{this is the prediction that genuinely distinguishes the two effective theories}.

\section{Entanglement membrane and 2d RG flows}\label{KasnerSec}
In section~\ref{Sec:EEMembrane}, we have explored the generalised entanglement membrane in 2d CFT with a degenerate membrane tension, and compared the resulting equation of motion with that of ordinary membrane theory~\cite{Mezei:2018jco}. In this section, we show that deforming a 2d CFT with a relevant operator leads to a
renormalisation group (RG) flow between the generalised and ordinary membrane theories and the emergence of a  
 non-degenerate membrane tension.  In the small deformation limit we maintain full analytic control over this RG interpolation. 

\subsection{Holographic RG flow from BTZ to AdS$_3$-Kasner}
\label{sec:Basic Setup}
We consider a two-sided planar BTZ black hole with a relevant deformation. Such a deformation breaks the infinite-dimensional conformal symmetry of the dual (1+1)-dimensional field theory. The action is given by
\begin{align}
    \mathcal{L}=\frac{1}{2}(R+2)-\frac{1}{2}(g^{ab}\partial_a\phi \partial_b\phi+m^2\phi^2)\,.\label{action}
\end{align}
Here we choose $m^2=-\frac{3}{4}$, which is above the Breitenlohner-Freedman bound $m^2_\text{BF}=-1$~\cite{Breitenlohner:1982jf}. It is convenient to rewrite the metric \eqref{blackBraneInf} in the following form: 
\begin{align}
    ds^2=\frac{1}{z^2}\Big(-f(z)e^{-\chi(z)}du^2-2e^{-\chi(z)/2}dudz+dx^2\Big)\,,\label{metricInfalling}
\end{align}
where the horizon is at $z_+=1$, hence $f(1)=0$. Without the relevant deformation, the metric \eqref{metricInfalling} becomes the planar BTZ metric, where $f(z)=1-z^2$ and $\chi(z)=0$. Unlike BTZ, the metric \eqref{metricInfalling} does contain a curvature singularity at $z=\infty$. 
The Einstein-scalar equations of motion can be obtained by varying the action \eqref{action}:\footnote{See~\cite{Caceres:2021fuw} for the Einstein-scalar equations of motion in general $d\geq 2$ dimensions with arbitrary $\Delta$. }
\begin{align}
    \phi''+\Big(\frac{f'}{f}-\frac{1}{z}-\frac{\chi'}{2}\Big)\phi'+\frac{3}{4}\frac{\phi}{z^2f}&=0\label{EoM1}\\
    \chi'-2\frac{f'}{f}-\frac{3}{4}\frac{\phi^2}{zf}-\frac{4}{zf}+\frac{4}{z}&=0\label{EoM2}\\
    \chi'-z\phi'^2&=0\label{EoM3}
\end{align}

The scalar field and the geometry behave asymptotically ($z\to 0$) as~\cite{Witten:1998qj,Gubser:1998bc,deHaro:2000vlm}: 
\begin{align}
    \phi=\phi_0 z^{\frac{1}{2}}+\langle O\rangle z^{\frac{3}{2}}+..., && \chi=\frac{\phi_0^2}{4}z+\frac{3}{4}\phi_0\langle O\rangle z^2+..., && e^{-\chi}f=1-\langle T_{tt}\rangle z^2+...\label{falloffs},
\end{align}
where $\phi_0$ serves as the source for the dual boundary operator whose expectation value is $\langle O\rangle$. As $-\frac{d^2}{4}<m^2<-\frac{d^2}{4}+1$, both modes are normalizable~\cite{Klebanov:1999tb}. Here we implement standard quantisation, with the dimension of $O$ given by $\Delta=\frac{d}{2}+\sqrt{\frac{d^2}{4}+m^2}=\frac{3}{2}$.
The temperature of the black hole is: 
\begin{align}
    T=\frac{|f_+'|e^{-\frac{\chi_+}{2}}}{4\pi}\label{temperature}
\end{align}
where $\chi_+$ and $f_+'$ are the horizon value of $\chi$ and $f'$, respectively. 

From the equations of motion \eqref{EoM1}-\eqref{EoM3}, one can read off the following near-singularity ($z\to\infty$) behaviours:  
\begin{align}
    \phi=\sqrt{2}c\log{z}+..., && \chi=2c^2\log{ z}+\chi_{\infty}+...,&& f=-f_{\infty}z^{2+c^2}+...\,,\label{Kasner}
\end{align}
where the constant $c=0$ for BTZ. Let $z^{2+c^2}=\frac{1}{\tau^2}$, the metric \eqref{metricInfalling} then takes the Kasner form~\cite{Frenkel:2020ysx,Kasner:1921zz,Belinski:1973zz} in the interior 
\begin{align}
    ds^2\sim -d\tau^2+\tau^{2p_{t}}dt^2+\tau^{2p_{x}}dx^2\,, && \phi\sim -\sqrt{2}p_{\phi}\log{\tau}\,,
\end{align}
where the Kasner exponents are given by 
\begin{align}
    p_t=\frac{c^2}{2+c^2}\,, && p_x=\frac{2}{2+c^2}\,, && p_{\phi}=\frac{2c}{2+c^2}\,.
\end{align}
We have $p_t+p_x=1$, $p_t^2+p_x^2+p_{\phi}^2=1$.  For the BTZ black hole, $p_t=p_{\phi}=0$ and $p_x=1$. One can think of $p_t$ as measuring the deformation of the geometry away from BTZ and therefore characterizing the relevant deformation. 

\begin{figure}[htbp]
\centering
\includegraphics[width=.48\textwidth]{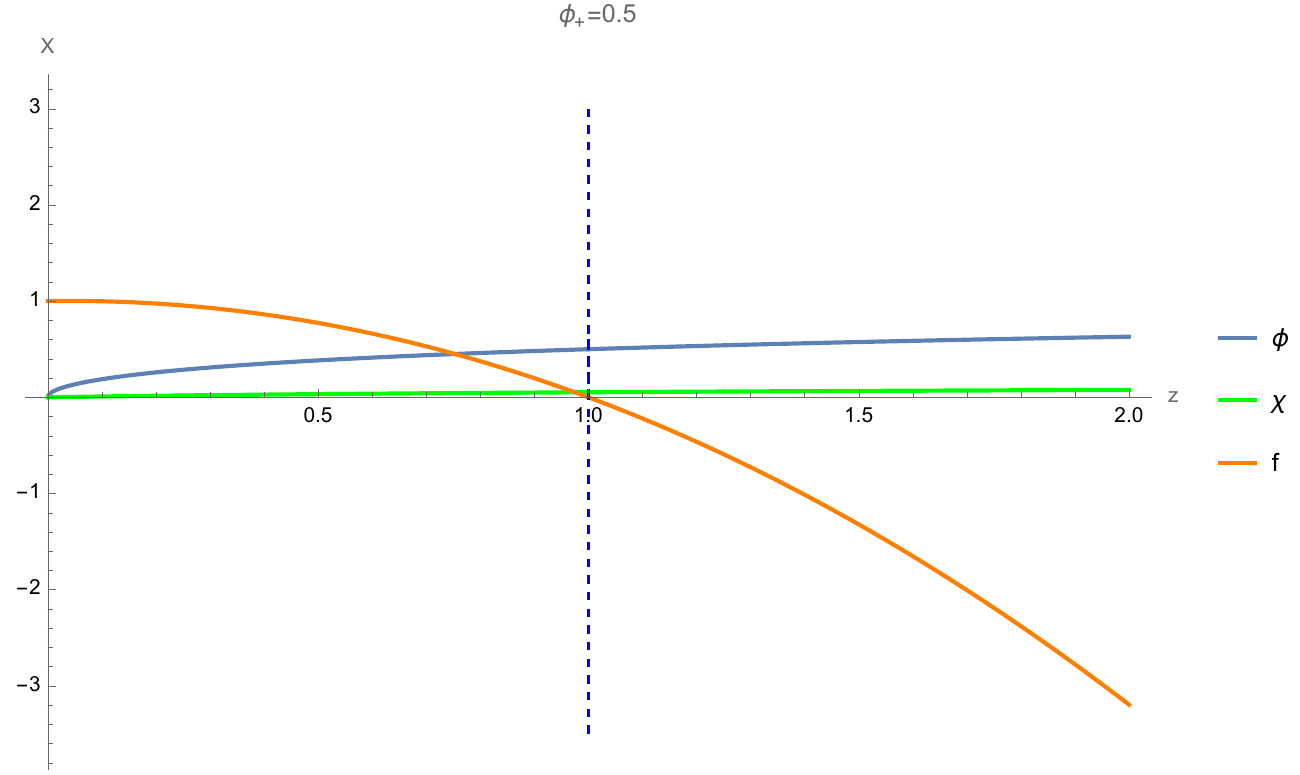}
\includegraphics[width=.45\textwidth]{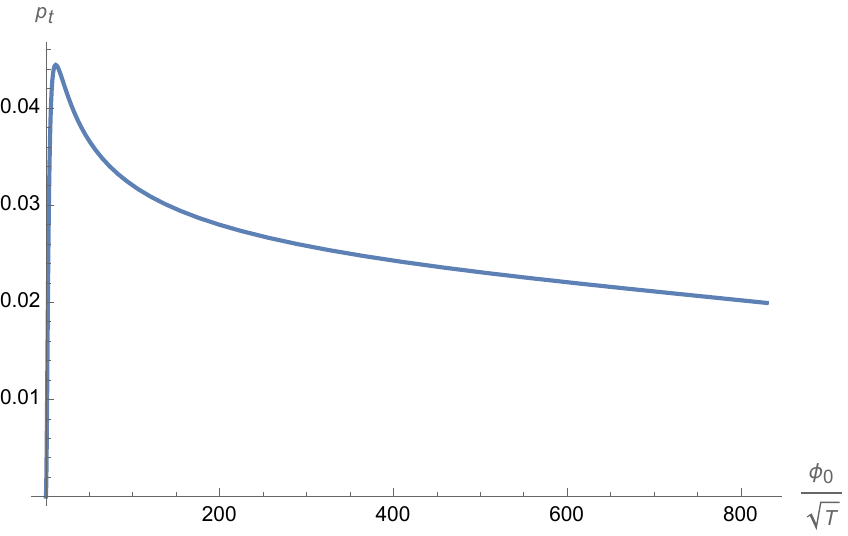}
\qquad
\caption{A demonstration of the shooting solutions. $\emph{Left}$: Plot of $\phi$ (blue), $\chi$ (green) and $f$ (orange) when $\phi_+=0.5$. The vertical dashed blue line is the black hole horizon $z=1$. $\emph{Right}$: Plot of the Kasner exponent $p_t$ with respect to the dimensionless parameter $\frac{\phi_0}{\sqrt{T}}$. Notice that for larger values of $\frac{\phi_0}{\sqrt{T}}$, $p_t$ returns to its BTZ value $p_t=0$. \label{fig:ShootingEgl}}
\end{figure}

\begin{figure}[htbp]
\centering
\includegraphics[width=.65\textwidth]{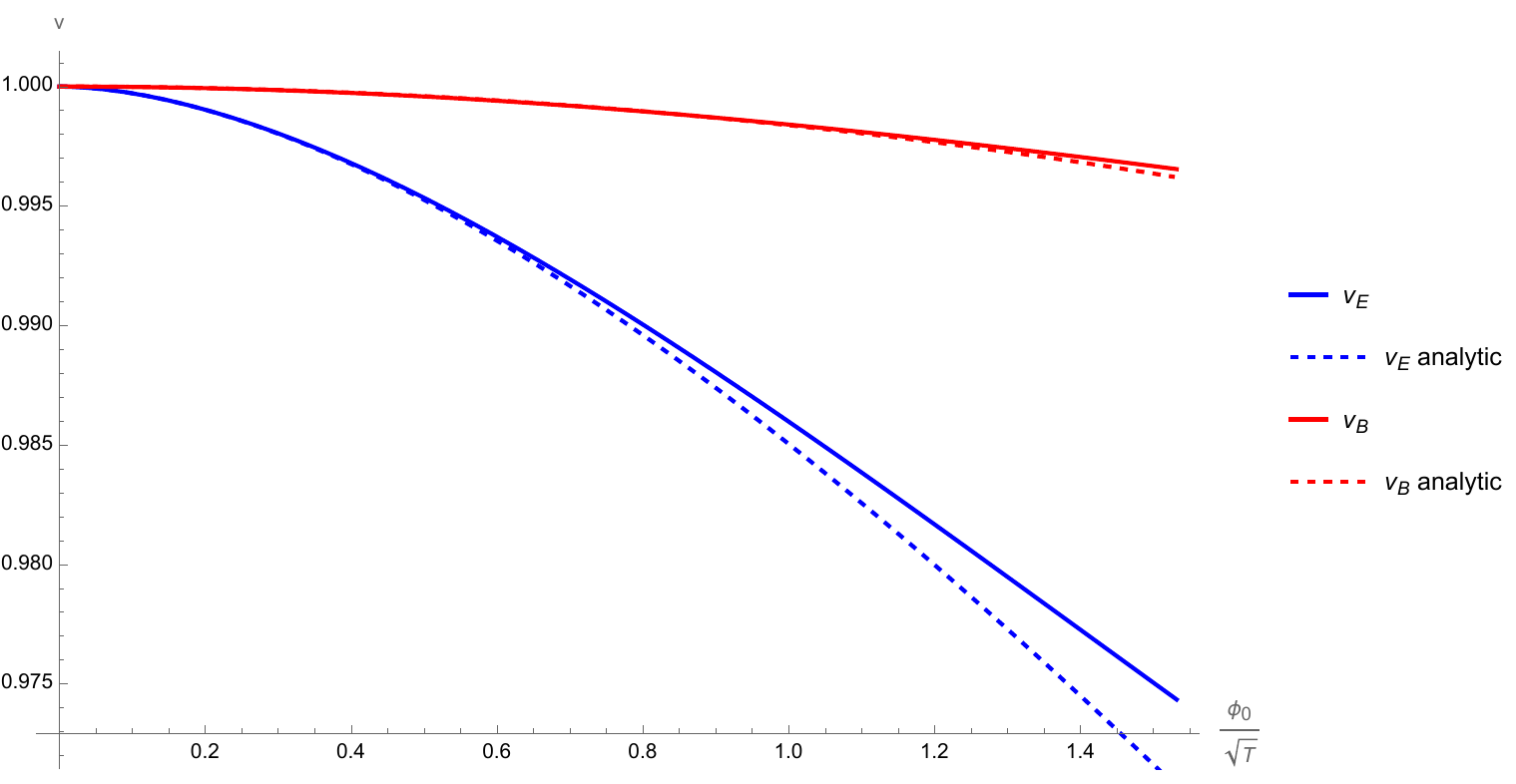}
\qquad
\caption{Plot of $v_E$ (blue) and $v_B$ (red) with respect to the dimensionless parameter $\frac{\phi_0}{\sqrt{T}}$. The solid lines are numerical solutions, whereas the dashed lines are analytic expansions in the small-$\phi$ limit. At $\phi_0=0$, i.e. when the geometry is planar BTZ black hole, $v_E=v_B=1$. When $\phi_0>0$, we have $v_E<v_B$ due to the constraint from null energy condition~\cite{Mezei:2016zxg}. When $\frac{\phi_0}{\sqrt{T}}$ is small, the numerical solutions of both $v_E$ and $v_B$ are well approximated by small-$\phi$ perturbation theory. \label{fig:vEvB}}
\end{figure}

\begin{figure}[htbp]
\centering
\includegraphics[width=.5\textwidth]{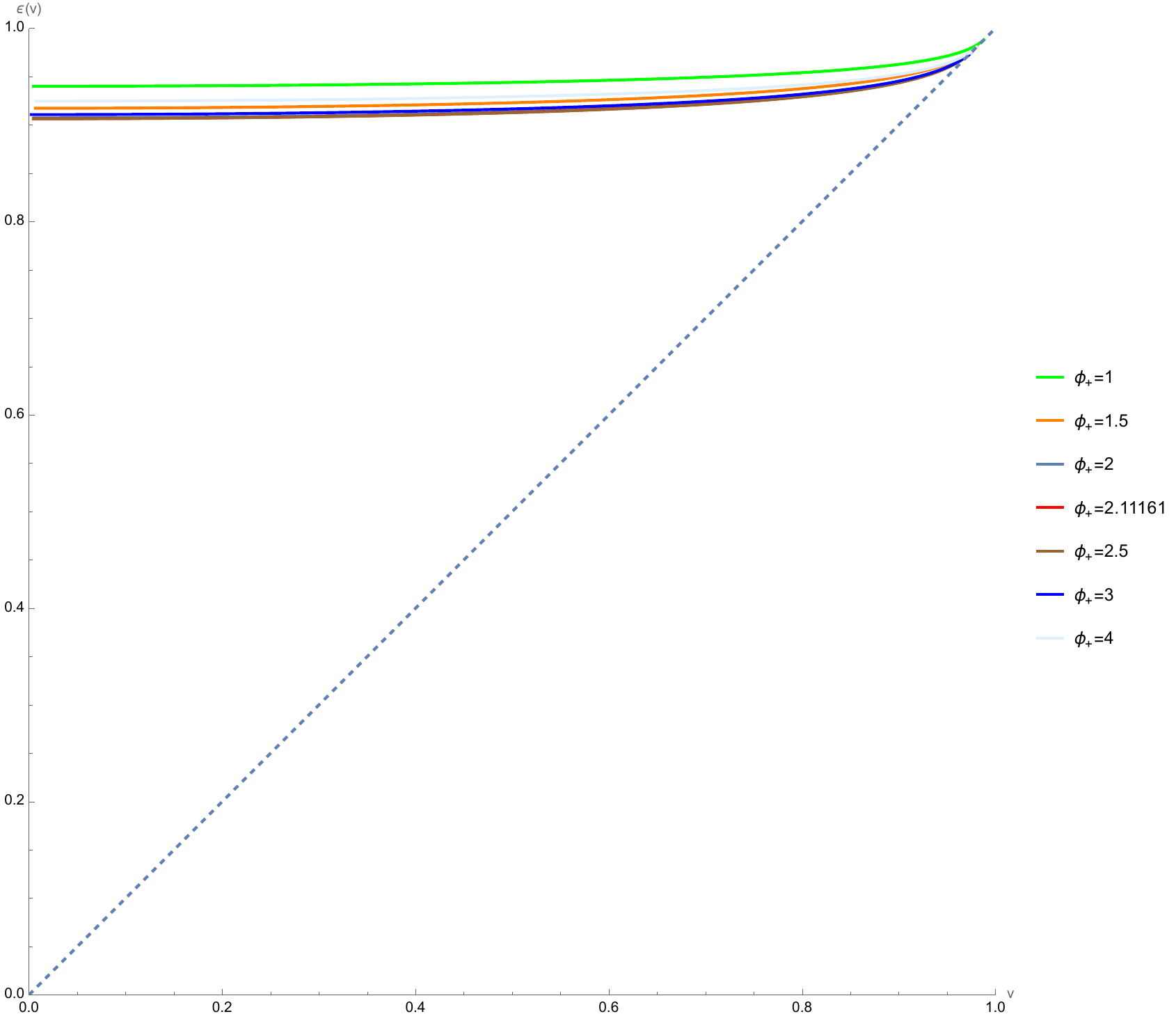}
\qquad
\caption{Plot of $\mathcal{E}(v)$ in AdS$_3$-Kasner geometry, for $\phi_+=1,1.5,2,2.11,2.5,3,4$. The dashed line is at $45^{\circ}$. As $\phi_+$ gets large, the membrane tension returns to its undeformed degenerate BTZ value $\mathcal{E}(v)=1$, see Fig.~\ref{fig:ShootingEgl} (right). \label{fig:Ev}}
\end{figure}

The equations of motion \eqref{EoM1}-\eqref{EoM3} can be solved numerically using the so-called ``shooting" method~\cite{Hartnoll:2008vx,Hartnoll:2008kx}. First, expand $\phi$, $\chi$ and $f$ around the horizon $z_+=1$:
\begin{align}
    \phi(z)=\sum_{n=0}^N\phi_{(n)}(z-1)^n\,, && \chi(z)=\sum_{n=0}^N\chi_{(n)}(z-1)^n\,, && f(z)=\sum_{n=1}^N f_{(n)}(z-1)^n \,\label{shootingExpansion}
\end{align}
where $N$ is some positive integer,  and we have used $f(1)=0$. Then, plugging \eqref{shootingExpansion} into the equations of motion \eqref{EoM1}-\eqref{EoM3} allows one to express all the expansion parameters in \eqref{shootingExpansion} in terms of the horizon value of the scalar field $\phi_{(0)}=\phi_+$. One can therefore integrate the field equations numerically in terms of the dimensionless parameter $\frac{\phi_0}{\sqrt{T}}$ from an $\epsilon$-ball centred at the horizon $z=1$. Here, in addition to integrating to the boundary $z=0$~\cite{Hartnoll:2008vx,Hartnoll:2008kx}, we also integrate towards the singularity $z=\infty$~\cite{Frenkel:2020ysx}. A demonstration of the shooting solution is presented in Fig.~\ref{fig:ShootingEgl}, where we show the plots of $\phi$, $\chi$, and $f$, as well as the Kasner exponent $p_t$.\footnote{Notice that $p_t>0$ in $d=2$, whereas for $d\geq 3$, $p_t<0$~\cite{Frenkel:2020ysx}. } Notice that as $\frac{\phi_0}{\sqrt{T}}$ increases from 0 to $\infty$, $p_t$ first increases and then decreases after reaching its maximal value, indicating that the geometry returns to BTZ in the large $\phi_+$ limit. The translation between the metric functions featuring in \eqref{blackBraneInf} and \eqref{metricInfalling} is $a(z)=f(z)e^{-\chi(z)}$ and $b(z)=e^{\chi(z)/2}$, hence $a(z),\, b(z)$ move away from their BTZ values under the relevant deformation, and $v_E$ \eqref{vE} and $v_B$ \eqref{vB} also become non-trivial functions of $\frac{\phi_0}{\sqrt{T}}$, see Fig.~\ref{fig:vEvB}. For the same reason, the membrane tension $\mathcal{E}(v)$ \eqref{AreaFunct2} is no longer degenerate. See Fig.~\ref{fig:Ev} for a plot of numerical membrane tensions in AdS$_3$-Kasner geometry. It is found that as $\phi_+$ increases from 0, $\mathcal{E}(v)$ starts to deflect from $\mathcal{E}(v)=1$ and becomes non-trivial; for large $\phi_+$, however, we observe that $\mathcal{E}(v)$ asymptotes to $\mathcal{E}(v)=1$. This is likely because as $\phi_+\to\infty$ the Kasner exponent $p_t\to 0$, the BTZ value.

\subsection{Small $\phi$ perturbation theory}
In the previous subsection, we have solved the equations of motion \eqref{EoM1}--\eqref{EoM3} numerically for generic perturbations. For small values of $\phi$, however, one can establish a perturbation theory and find analytic solutions order by order. Let us consider the leading order perturbation in $\phi$
\begin{align}
    \phi(z)=\lambda \phi_1(z)\label{phiLambdaExp}
\end{align}
where  $\lambda$ is a small constant. From the equations of motion \eqref{EoM3} and the action \eqref{action}, both $\chi(z)$ and $f(z)$ are second order in $\lambda$, 
\begin{align}
    \chi(z)&=\lambda^2 \chi_1(z)\label{chiLambdaExp}\\
    f(z)&=1-z^2+\lambda^2 f_1(z)\label{fLambdaExp}
\end{align}
Plugging \eqref{phiLambdaExp}-\eqref{fLambdaExp} into the equations of motion \eqref{EoM1}-\eqref{EoM3}, one can obtain   $\phi_1(z)$ in closed form, and obtain integral expressions for $\chi_1(z)$ and $f_1(z)$,
\begin{align}
    \phi_1(z)&=\frac{2}{\pi}\sqrt{z}K\Big(\frac{1-z}{2}\Big)\,,\label{phi1}\\
    \chi_1(z)&=\int_0^{z}dz' \frac{\left((z'+1) K\left(\frac{1-z'}{2}\right)-2 z'
   E\left(\frac{1-z'}{2}\right)\right)^2}{\pi ^2 \left(z'^2-1\right)^2}\,, \label{chi1IntExpr}\\
   f_1(z)&=z^2 \int_1^{z}dz' \frac{-(2 z'^2+z'-1) K\left(\frac{1-z'}{2}\right)^2+2 z' (z'+1)
   K\left(\frac{1-z'}{2}\right) E\left(\frac{1-z'}{2}\right)-2 z'^2
   E\left(\frac{1-z'}{2}\right)^2}{\pi ^2 z'^2 \left(z'^2-1\right)}\,,  \label{f1IntExpr}
\end{align}
where $K(z)$ and $E(z)$ denote the complete elliptic integral of the first and second kind, respectively. Integrating \eqref{EoM3} from 0 to $z$ in \eqref{chi1IntExpr} ensures $\chi(0)=0$. When $\phi_+$ is small, one can check that~\eqref{phi1} matches well with numerical solution. According to \eqref{phi1}, $\phi(1)=\lambda$ at the horizon $z=1$. From \eqref{phi1}, one learns that $\phi(z)$  behaves near the boundary $z\to 0$ as 
\begin{align}
    \phi(z)\to\lambda\frac{ \Gamma \left(1/4\right)}{\sqrt{2 \pi }\  \Gamma
   \left(3/4\right)}z^{\frac{1}{2}}+\lambda\frac{ \Gamma \left(-1/4\right)}{8 \sqrt{2
   \pi }\ \Gamma \left(5/4\right)}z^{\frac{3}{2}}+...\label{phi1bdy}
\end{align}
and near the singularity $z\to \infty$ as
\begin{align}
    \phi(z)\to\lambda\frac{\sqrt{2}}{\pi }\log{z}+...\label{phi1sing}
\end{align}
Comparing \eqref{phi1bdy} to \eqref{falloffs}, and \eqref{phi1sing} to \eqref{Kasner} then yields a relation among the small perturbation parameter $\lambda$, the value of $\phi$ at the horizon $\phi_+$, the falloff $\phi_0$, as well as the Kasner parameter $c$, 
\begin{align}
    \lambda=\phi_+=\pi c= \sqrt{2 \pi }\ \frac{\Gamma
   \left(3/4\right)}{\Gamma \left(1/4\right)} \phi_0\,.
\end{align} 


The integral expressions \eqref{chi1IntExpr} and \eqref{f1IntExpr} make it hard to derive a closed form expression for $\mathcal{E}(v)$ \eqref{AreaFunct2} even in perturbation theory. To get more insight into its behaviour, in the next subsection we expand $\mathcal{E}(v)$ around $v=0$ and $v=v_B$ to leading non-trivial orders in $\lambda$. As $v^2=c(z)$ \eqref{projectionz} is monotonically decreasing in $z$, these expansions take place around $z=z_*$ and $z=1$, respectively. 

\subsubsection{Expansion near the singularity}\label{Expansion near the singularity}
Let us first consider the expansion of $\mathcal{E}(v)$ at $v=0$ or equivalently $z=z_*$, where $\mathcal{E}(0)=v_E$. Intuitively, $z_*$ is large for small $\lambda$, as $z_*=\infty$ in the undeformed BTZ black hole. Therefore, to find the expansion of $\mathcal{E}(v)$ at $v=0$, one can start by studying the behaviors of $\chi$ and $f$ at large $z$:
\begin{align}
    \chi_1(z)&\approx \frac{2}{\pi^2}\log{z}+\frac{-2-\pi+12\log{2}}{2\pi^2}\,,\label{chi1C}\\
    f_1(z)&\approx -\frac{z^2}{\pi^2}(\log{z}+3\log{2})\,.\label{f1C}
\end{align}
\eqref{chi1C} and \eqref{f1C} then allow us to find the small $\lambda$ expansion of $v^2(z)$ \eqref{projectionz} to $O(\lambda^2)$ 
\begin{align}
    v^2(z)\approx 1-\frac{1}{2\pi^2}\lambda^2 z^2 &&
    \Rightarrow && z(v^2)=\frac{\sqrt{2}\pi}{\lambda}\sqrt{1-v^2}\,,\label{zvhor}
\end{align}
which is a monotonically decreasing function in $z$ deviating from $v^2=1$ in BTZ. $z_*$ is given by the solution of $v^2(z)=0$, 
\begin{align}
    z_*=\frac{\sqrt{2}\pi}{\lambda}\,.\label{zStar}
\end{align}
We see that the relevant deformation \eqref{phiLambdaExp} of BTZ pulls $z_*$ from $\infty$ to a finite value. The entanglement dynamics in AdS$_3$-Kasner geometry is thus reminiscent of those in planar AdS$_{d+1}$-Schwarzschild black branes in $d\geq 3$ dimensions, in the sense that at late time, the HRT surface computing entanglement entropy stays at a finite $z_*$ instead of moving towards $z=\infty$. Unlike $z_{*}\in O(1)$ in AdS$_{d+1}$-Schwarzschild, however, $z_*$ in AdS$_3$-Kasner~\eqref{zStar} is large for small perturbations. When $\lambda\to 0$, $z_*\to\infty$, and the entanglement dynamics returns to that in BTZ. As $v_E$ \eqref{vE} is determined by the interior geometry at $z=z_*$, one can then proceed to obtain the small $\lambda$ expansion of $v_E$ 
\begin{align}
    v_E\approx 1+\frac{1}{2\pi^2}\lambda^2\log{\lambda}+\frac{1+\pi -7 \log 2 -2 \log \pi }{8 \pi ^3}\lambda^2\,.\label{vEexpansion}
\end{align}
Notice that the leading non-trivial term in $\lambda$ is at $O(\lambda^2\log{\lambda})$. A plot of \eqref{vEexpansion} with respect to the dimensionless parameter $\frac{\phi_0}{\sqrt{T}}$ can be found in Fig.~\ref{fig:vEvB}.\footnote{In plotting $v_E$ as a function of the dimensionless quantity $\frac{\phi_0}{\sqrt{T}}$, we also need to include the correction of temperature $T$ \eqref{temperature} due to the deformation \eqref{phiLambdaExp}: 
\begin{align}
    T\approx\frac{1}{2\pi}\Big(1+\frac{\lambda^2}{4\pi}\Big)\label{TwithLambda}
\end{align}
Notice that the $O(\lambda^2)$ correction drops out if we only keep the expansion of $v_E$ to $O(\lambda^2)$. The same applies to $v_B$. } For small $\phi_0$, \eqref{vEexpansion} matches well with numerical result. 

As we have already seen from the numerical solutions, the membrane tension is no longer degenerate in AdS$_3$-Kasner geometry. By plugging the above results into~\eqref{AreaFunct2}, we obtain the small $\lambda$ expansion of $\mathcal{E}(v)$:
\begin{align}
    \mathcal{E}(v)\approx v_E-\frac{\lambda^2}{4 \pi ^2}\log{(1-v^2)}\label{EvNearSing}
\end{align}
This formula was obtained
from the large $z$ expansion of the metric functions, and this limits its regime of validity in $v$. From \eqref{zvhor}, we see that unless $1-v=O(\lam^2)$, we stay in the large $z$ regime, hence  \eqref{EvNearSing} is widely applicable.
Indeed, this formula obeys the constraints $\mathcal{E}(0)=v_E$, $\mathcal{E}'(0)=0$ $\mathcal{E}'(v)\geq 0$, $\mathcal{E}''(v)\geq 0$ from \eqref{constraints}. A plot of the expansion \eqref{EvNearSing} can be found in Fig.~\ref{fig:EvAnalytic}. For small $\lambda$, \eqref{EvNearSing} matches well with numerical results. 

\subsubsection{Expansion near the horizon}
We expect that the butterfly velocity will move from its 2d CFT value $v_B=1$ to  $v_B=1-O(\lam^2)$. Plugging this into \eqref{EvNearSing} generates a $O(\lam^2\log(\lam))$ term that threatens the validity of the expansion. Indeed $v_B$ is determined by the geometry at the horizon $z=1$, as in  \eqref{vB}, whereas \eqref{EvNearSing} was obtained from a large $z$ expansion. Thus we examine the behaviour of $\mathcal{E}(v)$ near the butterfly velocity more carefully.

\begin{figure}[htbp]
\centering
\includegraphics[width=.66\textwidth]{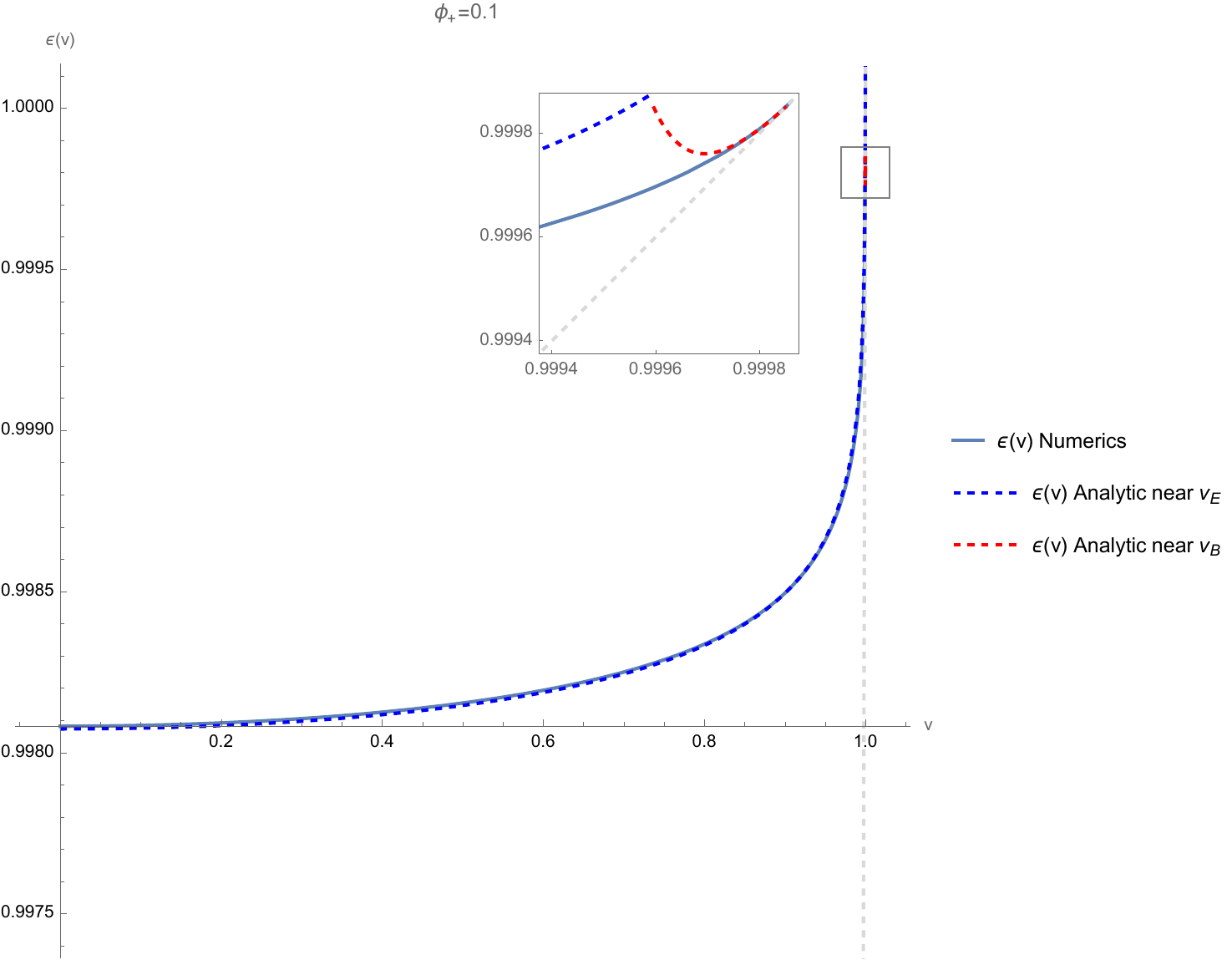}
\qquad
\caption{Plot of the membrane tension $\mathcal{E}(v)$ when $\phi_+=0.1$. For small $\phi_+$, $\mathcal{E}(v)$ is close to 1. The solid line is obtained numerically from shooting solutions, whereas the blue and red dashed lines are perturbative expansions around the deepest point $z_*$ \eqref{EvNearSing} and the horizon $z=1$ \eqref{vzEz2}, respectively. The dashed gray line has unit slope, greatly exaggerated here by the scale of the vertical axis.  Notice that the near-$v_B$ expansion of $\mathcal{E}(v)$ is valid only in a small interval near $v=v_B$, which is difficult to visualise and hence is plotted in the inset. \label{fig:EvAnalytic}}
\end{figure}

It is a straightforward exercise to obtain a series representation in $\de z\equiv z-1$ of the different metric components near the horizon. From these using \eqref{vB}, \eqref{projectionz}, and \eqref{AreaFunct2}  we can form:\footnote{To obtain these expansions we need to evaluate the following definite integral:
\begin{align}
    \chi_1(1)&=\int_0^{1}dz' \frac{\left((z'+1) K\left(\frac{1-z'}{2}\right)-2 z'
   E\left(\frac{1-z'}{2}\right)\right)^2}{\pi ^2 \left(z'^2-1\right)^2}=\frac{3}{8}-\frac{1}{2\pi}\label{chi1(1)}\,.
\end{align}}
\es{vzEz}{
{v_B-1\ov \lam^2}&=-{3\ov 32}+{1\ov 4\pi}<0\,,\\
{v^2-1\ov \lam^2}&={c(z)-1\ov \lam^2}=\le(-{3\ov 16}+{1\ov 2\pi}\ri)-{9\ov 128}\de z-{99\ov 2048}\de z^2+\dots\,,\\
{\mathcal{E}(v)-1\ov \lam^2}&={\sqrt{\frac{-a'(z)}{2z}}-1\ov \lam^2}=\le(-{3\ov 32}+{1\ov 4\pi}\ri)-{9\ov 256}\de z+{45\ov 4096}\de z^2+\dots\,,
}
where we have explicitly computed terms up to $O(\de z^4)$. By expressing $\de z$ as a function of 
\es{nudef}{
\nu\equiv {v^2-v_B^2\ov \lam^2}\ll1\,,
}
we can express $\mathcal{E}(v)$ in a $v$ expansion near the butterfly velocity (with a radius of validity in $v$ very small, $O(\lam^2)$) as follows:
\es{vzEz2}{
\mathcal{E}(v)=1+\lam^2\le(\le(-{3\ov 32}+{1\ov 4\pi}\ri)+{\nu\ov 2}+{64\nu^2\ov 9}+\frac{35840 \nu ^3}{243}+\frac{7813120 \nu ^4}{2187}+\dots\ri)\,.
}
This expression satisfies the constraints $\mathcal{E}(v_B)=v_B,\, \mathcal{E}'(v_B)=1$, $\mathcal{E}'(v)\geq 0$, $\mathcal{E}''(v)\geq 0$ from  \eqref{constraints}, as required. In Fig.~\ref{fig:EvAnalytic}, we plot the expansion \eqref{vzEz2}. It is found that \eqref{vzEz2} agrees with numerical results when $v$ is close to $v_B$.

It is instructive to convert our holographic results~\eqref{vzEz} to those in conformal perturbation theory~\cite{Davison:2024msq} 
\begin{align}
    S=S_{{\rm CFT}}+\sqrt{c}\,\mu \int d^2 x O
\end{align}
where $\mu$ is the relevant coupling that deforms the CFT, and the scalar primary operator $O$ is normalised such that its vacuum two-point function is $|x-x'|^{-2\Delta}$. This normalisation is related to our convention via $\phi_0=-4\sqrt{3}\pi\mu$.\footnote{The conversion between holographic and CFT normalisations   is~\cite{Davison:2024msq}
\begin{align}
    \phi=\frac{2\sqrt{3}}{1-\Delta}\mu z^{2-\Delta}+...\label{CFTNorm}
\end{align}
Comparing~\eqref{CFTNorm} with our falloff~\eqref{falloffs}, we have $\phi_0=\frac{2\sqrt{3}}{1-\Delta}\mu$. } The butterfly velocity in~\eqref{vzEz} can then be rewritten in terms of the dimensionless coupling $\frac{\mu}{\sqrt{T}}$ as 
\begin{align}
    v_B=1-\frac{3 \pi  (3 \pi -8) \Gamma \left(\frac{3}{4}\right)^2}{2 \Gamma
   \left(\frac{1}{4}\right)^2}\Big(\frac{\mu}{\sqrt{T}}\Big)^2\,.
\end{align}

\subsection{Adding $\phi^4$-interactions}

An interesting extension of the above story is to study an RG flow that interpolates between a UV and an IR CFT. We expect that ${\cal E}(v)$ would be degenerate both in the UV and IR, but nontrivial in between. We could use the effective temperature of the state or operator whose entanglement dynamics we are studying as the RG scale. We can holographically model such an RG flow by
 adding a $\frac{\mu}{4!}\phi^4$ interaction to the bulk scalar fields in \eqref{action}:
\begin{align}
    \mathcal{L}=\frac{1}{2}(R+2)-\frac{1}{2}(g^{ab}\partial_a\phi \partial_b\phi+m^2\phi^2+\frac{\mu}{12}\phi^4)\label{actionphi4}
\end{align}
The potential term in \eqref{actionphi4}, now containing both the mass term and the $\phi^4$ interaction, takes the form $V(\phi)=-\frac{3}{8}\phi^2+\frac{\mu}{24}\phi^4$. At zero temperature we would get a domain wall scalar profile that interpolates between $\phi=0$ at the boundary and the bottom of the potential $\phi_*={3\ov \sqrt{2\mu}}$, representing the IR fixed point \cite{Freedman:1999gp,Girardello:1998pd,Girardello:1999bd}.

The Einstein-scalar equations of motion in the presence of the $\phi^4$ interaction are 
\begin{align}
    \phi''+\Big(\frac{f'}{f}-\frac{1}{z}-\frac{\chi'}{2}\Big)\phi'+\frac{3}{4}\frac{\phi}{z^2f}-\frac{\mu}{6}\frac{\phi^3}{z^2f}&=0\,,\label{EoM1phi4}\\
    \chi'-2\frac{f'}{f}-\frac{3}{4}\frac{\phi^2}{zf}-\frac{4}{zf}+\frac{4}{z}+\frac{\mu}{12}\frac{\phi^4}{zf}&=0\,,\label{EoM2phi4}\\
    \chi'-z\phi'^2&=0\,.\label{EoM3phi4}
\end{align}
From \eqref{EoM1phi4}--\eqref{EoM3phi4}, we can obtain the shooting solution for $\phi^4$-interacting bulk scalars. A curious feature of these is that the scalar rolls through the minimum of the potential $\phi_*$ in the black brane interior and exhibits the same near-singularity behaviour in the presence of $\phi^4$  interactions as in their absence. We can understand this as follows: The terms in \eqref{EoM1phi4}--\eqref{EoM3phi4} that originate from $V(\phi)$ are 
\begin{align}
    \frac{3}{4}\frac{\phi}{z^2f}\,, && \frac{\mu}{6}\frac{\phi^3}{z^2f}\,, && \frac{3}{4}\frac{\phi^2}{zf}\,, && \frac{\mu}{12}\frac{\phi^4}{zf}\,,\label{VphiTemrs}
\end{align}
which upon substituting the asymptotic behaviour \eqref{Kasner}, all go to $0$ in the large $z$ limit. This is because $\phi$ and hence any polynomial of it only diverges $\emph{logarithmically}$, and the Kasner behaviour \eqref{Kasner} originates from the $\emph{kinetic}$ terms in the action \eqref{action} or \eqref{actionphi4}. To change \eqref{Kasner} near the singularity, one needs to introduce interactions that are (super-)exponential in $\phi$~\cite{Hartnoll:2022rdv}.  
This may make us doubt if the IR CFT leaves any imprint on entanglement dynamics that is determined by the interior. 

\begin{figure}[htbp]
\centering
\includegraphics[width=.49\textwidth]{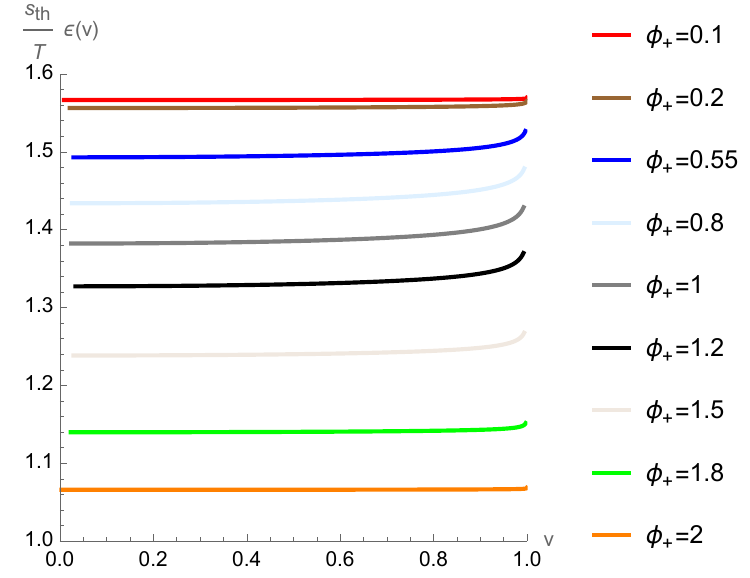}
\includegraphics[width=.49\textwidth]{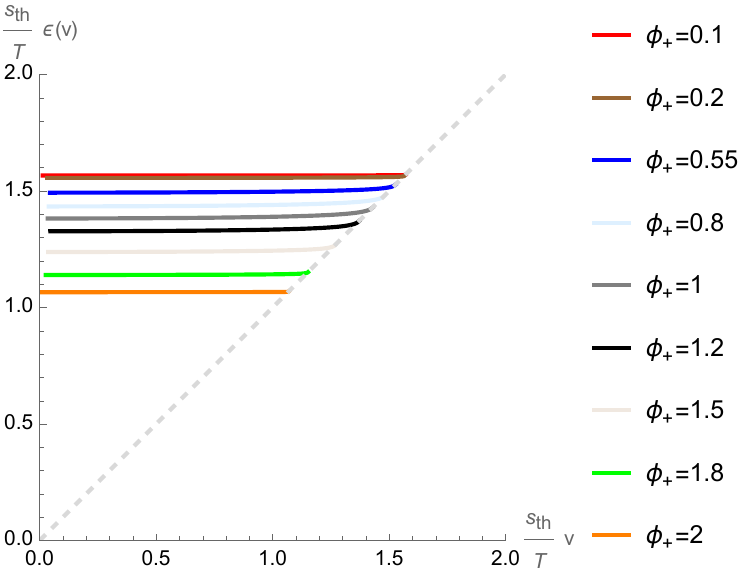}
\qquad
\caption{Plots of the membrane tension $\mathcal{E}(v)$ for a $\phi^4$-interacting bulk scalar with $\mu=1$ for the choices $\phi_+=0.1, 0.2, 0.55, 0.8, 1, 1.2, 1.5, 1.8$ and $2$. Note that the minimum of the potential that represents the IR fixed point is at $\phi_*=2.12$.\label{fig:EvPhi4}}
\end{figure}

Nevertheless, by tuning $\phi_+$, the horizon value of the scalar field, we can explore the RG flow, since  for  $\phi_+\to\phi_*$, i.e.~approaching the IR CFT,  can tune the Kasner exponent close to the BTZ value appropriate for the IR CFT. Our findings are fully consistent with RG intuition and we summarise them on two plots in Fig.~\ref{fig:EvPhi4}. Both show ${\cal E}(v)$ scaled differently. On the left we multiplied ${\cal E}(v)$ by the dimensionless  entropy density factor $s_\text{th}/ T$ that plays the role of the number of degrees of freedom. Recall that it is the combination $s_\text{th} {\cal E}(v)$ that shows up in the membrane action \eqref{MinMemb}. This combination indeed gives a monotonically decreasing  function  under RG flow, which is a satisfactory result on physical grounds. On the right we have also rescaled the horizontal axis by the same factor $s_\text{th}/ T$ to demonstrate that ${\cal E}(v)$ still is tangent to the $45^\circ$ line all along the RG flow. We note that we have found similar behaviour in the free scalar case as $\phi_+\to \infty$, but of course there was no IR CFT there.

\subsection{The transition between the generalised and the ordinary membrane theory}\label{The transition between the generalised and the ordinary membrane theory}

We are now prepared to understand the transition between the generalised and the ordinary membrane theories. Let us fix $\lam$ to be small and change the boundary data $X,T$. The relevant region of the bulk is for large $z$. In section~\ref{Expansion near the singularity} we have determined the large $z$ behaviour of the metric functions $a(z)$ and $b(z)$ to leading order in $\lam$, and repeating the derivation presented in \eqref{AreaFunct3} leads to the result 
\es{AreaFunct4}{
S &=s_\text{th} \int du \le[1+{d(e^{-\xi})\ov du}+ \mathcal{L}_\text{eff}+\dots \ri]\,\\
\mathcal{L}_\text{eff}&=\frac12\,e^{-2\xi}(\dot x^2-1-\dot \xi^2)+{\lam^2\ov 2\pi^2}\le(\xi-\frac\pi2\ri)\,,
}
where $s_\text{th}=\big(1+\frac{\lambda^2}{4\pi}\big)s_\text{th}^\text{(conf)}$ is the thermal entropy density to $O(\lambda^2)$ in the presence of the relevant deformation \eqref{phiLambdaExp}. This correction to the entropy density then leads to the shift of the linear in $\xi$ term in $\mathcal{L}_\text{eff}$. The only change in $\mathcal{L}_\text{eff}$ from the 2d CFT case is the last term. The $\xi$ equation of motion is then
\es{xiEOM2}{
\ddot \xi=-1+\dot \xi^2+p^2 e^{4\xi}+{\lam^2\ov 2\pi^2}e^{2\xi}\,.
}

Let us examine the last two terms. For $\lam=0$, we saw that $p^2=\exp(-4\xi_p)=\exp\le(-4(T-X)\ri)$, where $\xi_p$ is the deepest point in the bulk that the HRT surface explores. If $\lam\ll \exp\le(-2(T-X))\ri)$, the last term never becomes important. However, if $\lam$ is larger than this, it changes the behaviour of the solutions. Let us define 
\es{xilam}{
{\lam^2\ov 2\pi^2}\equiv e^{-2\xi_*}\,,
}
according to \eqref{zStar}, with which \eqref{xiEOM2} can be rewritten as
\es{xiEOM3}{
\ddot \xi=-1+\dot \xi^2+e^{4(\xi-\xi_p)}+e^{2(\xi-\xi_*)}\,.
}

\begin{figure}[htbp]
\centering
\includegraphics[width=.25\textwidth]{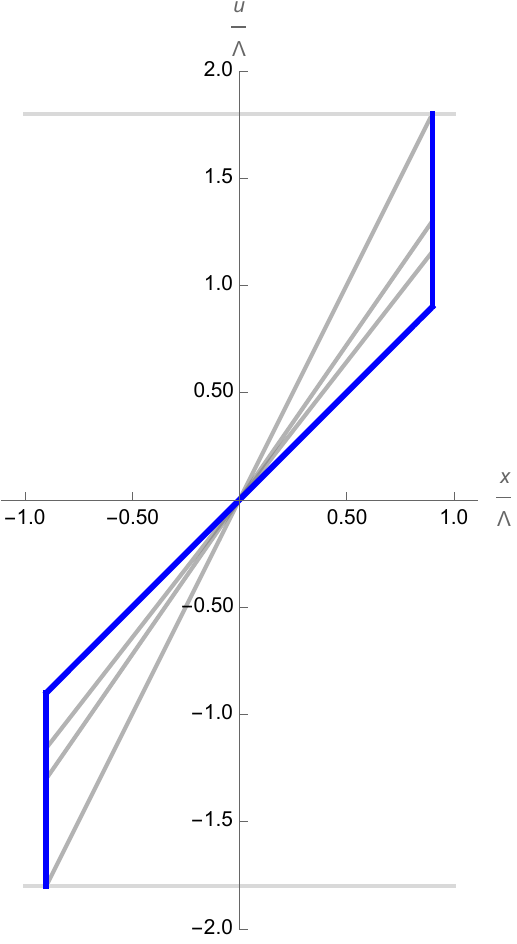}\hspace{0.5cm}
\includegraphics[width=.25\textwidth]{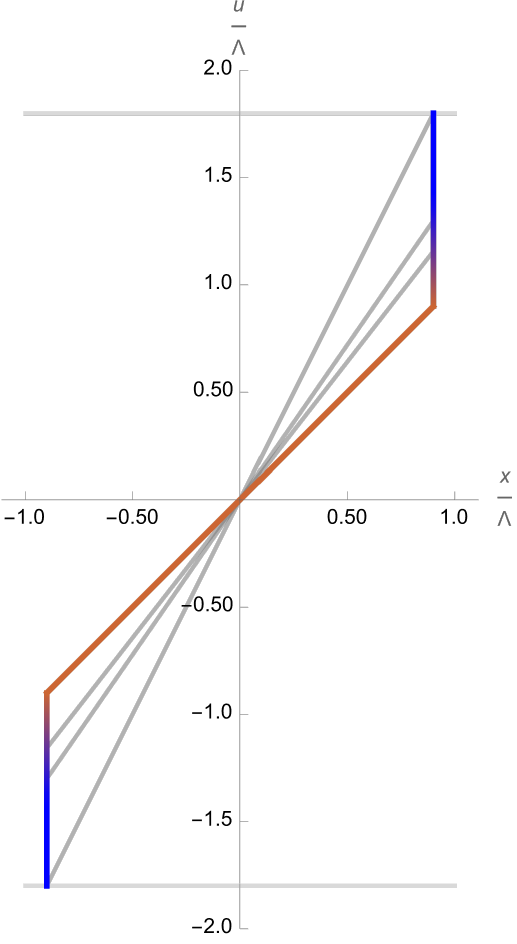}\hspace{0.5cm}
\includegraphics[width=.25\textwidth]{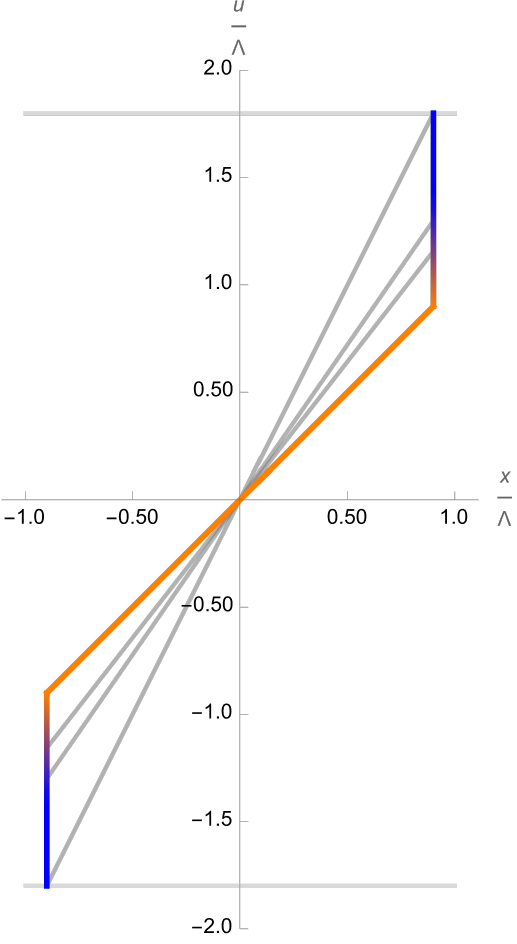}
\includegraphics[width=.25\textwidth]{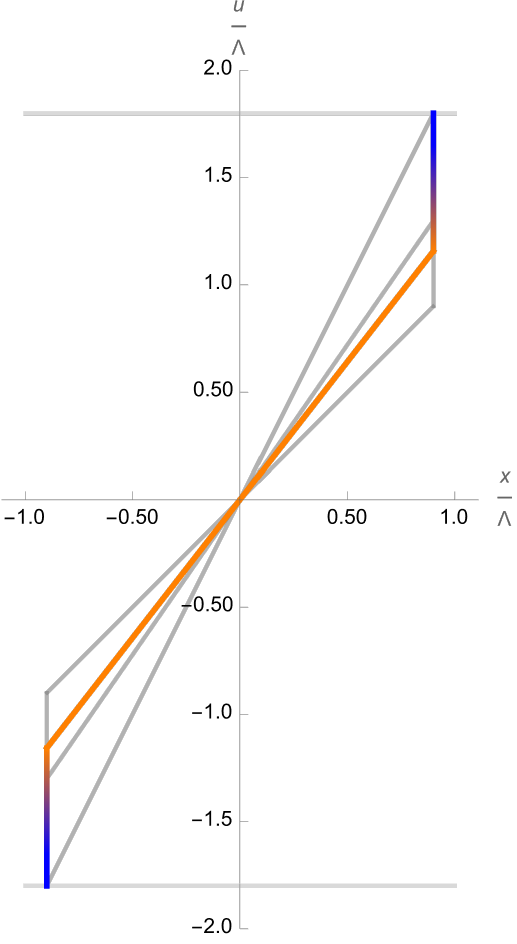}\hspace{0.5cm}
\includegraphics[width=.25\textwidth]{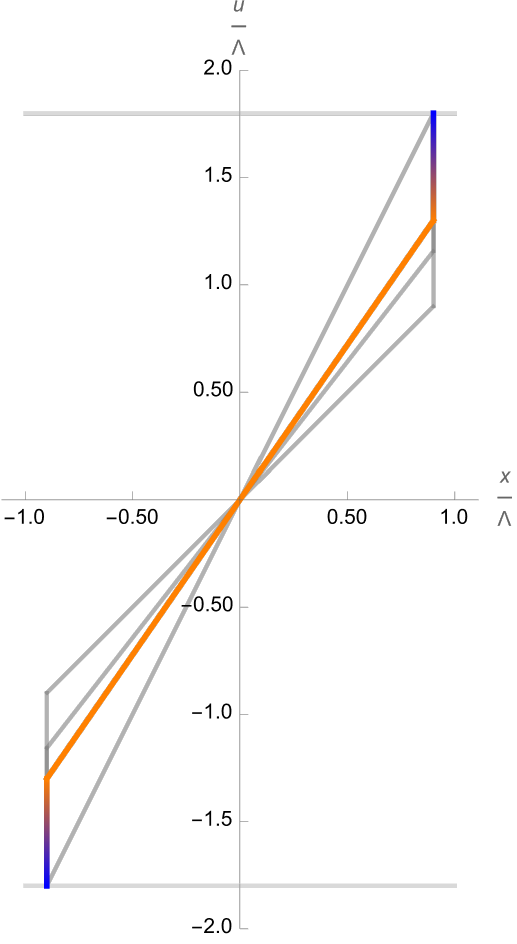}\hspace{0.5cm}
\includegraphics[width=.25\textwidth]{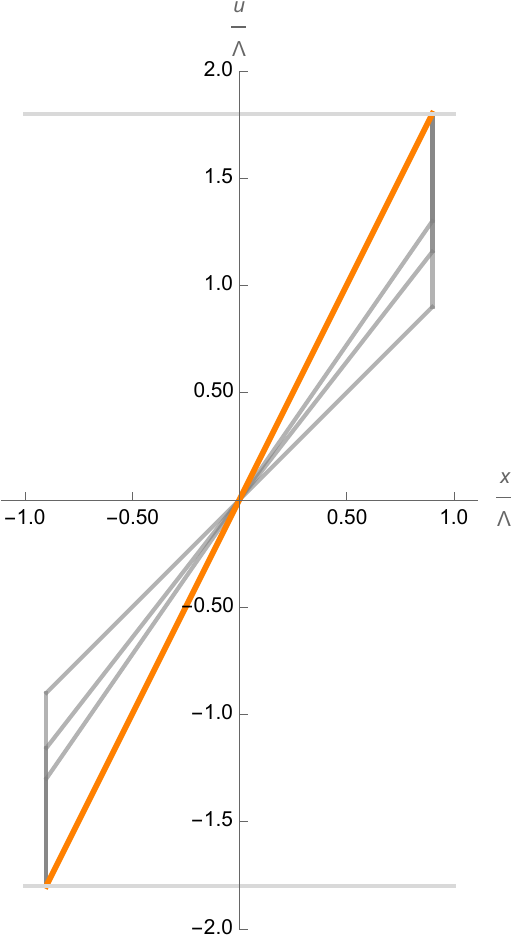}
\qquad
\caption{Plots of $x(u)$ in \eqref{FinalXi2} in the scaling limit, with blue $\to$ orange denoting the increase in the bulk depth $\xi(u)$ direction \eqref{FinalXi2}. Here, we fix $\xi_*=5 \Lam$ and $v=\frac{1}{2}$, and plot for $\frac{T}{\Lambda}=2,6,10,14,18$ and $\infty$. On each plot we indicate with gray where the other lines corresponding to the other choices of $T$ go. When $\frac{T}{\Lambda}=2,6,10$ we have the 2d CFT result (Fig.~\ref{fig:zxApproxLagrangian} left) as $\frac{\xi_*}{T}\leq 1-v$. The generalised membrane reaches deeper in $\xi$ as $T$ evolves. When $\frac{T}{\Lambda}=14,18$, the slope of the generalised membrane starts to increase continuously, transitioning to that of the ordinary membrane value $v$ (Fig.~\ref{fig:zxApproxLagrangian} right) when $T=\infty$. 
\label{fig:BTZKasnerTran}}
\end{figure}

The solutions that stay at large $\xi$ for a long infalling time $u$ still have a plateau at some $\xi=\xi_0$. This is where the majority of the motion in $x$ comes from, and it scales as $X=\exp(2(\xi_0-\xi_p))\, u_\text{plateau}$. Clearly, we need $\xi_p\sim \xi_0$ to get $X\sim T$. From these ingredients we can assemble together the solution to  \eqref{xiEOM3} in the regime ${\xi_*\ov T}<1-v$. (In the complementary regime we get the 2d CFT result.) We define:
\es{xis}{
\xi_p&\equiv\xi_*+a_p\,,\\
\xi_0&\equiv\xi_*+a_0\,,
}
where $\xi_*$ is large and $a_0, a_p=O(1)$. By neglecting $O(1)$ shifts, the evolution of $\xi$ is as follows 
\es{FinalXi2}{
(x(u),\xi(u))=\begin{cases}
\le(-X,T+u\ri)\qquad &-T<u<-T+\xi_*\,,\\
\le({v\ov 1-\xi_* /T}\,u,\xi_* \ri)\qquad &-T+\xi_*<u<T-\xi_*\,,\\
\le(X,T-u\ri)\qquad &T-\xi_*<u<T\,.
\end{cases}
}
See Fig.~\ref{fig:BTZKasnerTran}. We highlight two insightful limits of this formula. At the boundary of its applicability  ${\xi_*\ov T}=1-v$, we have
\es{limit1}{
\le({v\ov 1-\xi_* /T}\,u,\xi_* \ri)=\le(u,(1-v)\,T \ri)\,,
}
which is exactly the 2d CFT result given in \eqref{FinalXi2}; we conclude that the interpolation between the generalised and ordinary membrane is continuous. The other limit is ${\xi_*\ov T}\to 0$, in which case the first and third parts of the curve are negligible compared to $T$ and we get the membrane
 \es{limit2}{
\le(vu,\xi_* \ri)\,.
}

Since we have neglected $O(1)$ factors, we do not see in this derivation that the deepest point the HRT surface goes behind the horizon is $v$-dependent. This dependence can be restored by keeping $O(1)$ terms.\footnote{From the motion in the $x$ direction we get the following equation:
\es{xmotion}{
v=e^{2(a_0-a_p)}\le(1-{\xi_* \ov T}\ri)\,.
}
From the condition that the plateau is a solution, we get another equation
\es{plateau_cond}{
1=e^{4(a_0-a_p)}+e^{2 a_0}\,.
}
These two equations determine $a_0$ and $a_p$ to indeed be $O(1)$ numbers.
}
The result is then consistent with the treatment in the section~\ref{Expansion near the singularity}, where we found that  $\xi=\xi_*+\log\sqrt{1-v^2}$.

It would also be interesting to explore the interpolation between the generalised and ordinary membrane results for the reflected entropy discussed in section~\ref{sec:diffs}. For reflected entropy it is not just the membrane shapes that differ, but also the value of the entropy at late times. We leave this computation for the future.


\section{Information velocity}

An interesting probe of the speed of information spreading called the information velocity was proposed and analysed in~\cite{Couch:2019zni}. Consider a reference qubit that gets locally maximally entangled with an initial state that has a fraction $f$ of volume law entanglement of the thermal state. As time evolves the region entangled with the reference grows, and at every time there is a minimal subregion (of fixed shape), from which the entanglement with the reference can be recovered. The speed at which the minimal subregion grows is the shape dependent information velocity $v_I$.

 Since this quantity is defined from the entanglement dynamics of large regions, it should be possible to determine it from the membrane description: this is indeed the case, just like in the case of the entanglement wedge cross section. There are both conceptual and computational advantages to the membrane approach to the problem. We include a brief discussion here to illustrate the versatility of the membrane effective theory.\footnote{Some parts of this section have already appeared in~\cite{Eccles:2021zum} based on the unpublished notes of MM.}
 
 In the membrane theory the setup used to define $v_I$ is captured by introducing a penalty factor for the lower end of the spacetime slab~\cite{Jonay:2018yei}:
\es{MembraneAction}{
S[A(T)]=s_\text{th}\le(\int d^{d-1}\xi \ \sqrt{\abs{\ga}}\, {{\cal E}(v)\ov \sqrt{1-v^2}}+f \vol(A'(0))\ri)\,,
}
where the membrane is anchored on $\p A(t)$ and $\p A'(0)$ on the upper and lower boundaries respectively. The extra term in \eqref{MembraneAction} encodes the entanglement structure of the initial state, which has $f\, s_\text{th} \vol(A)$ entanglement. 

\begin{figure}[!h]
\begin{center}
\includegraphics[scale=0.4]{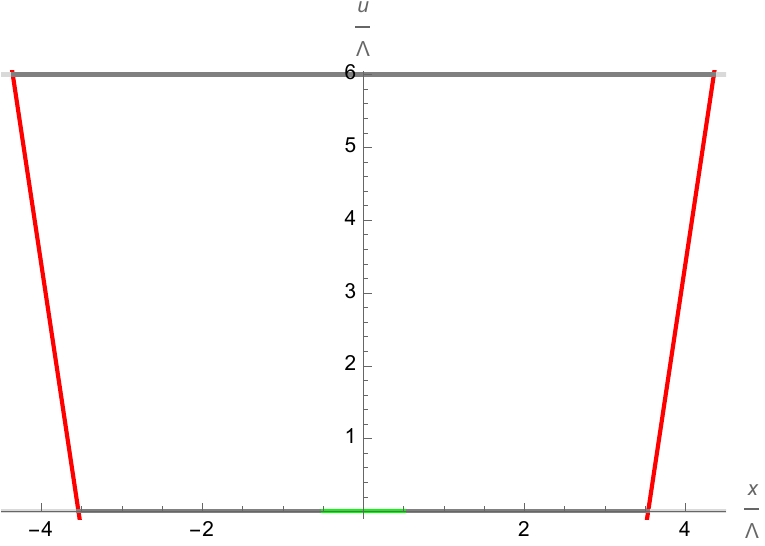}\hspace{1cm}
\includegraphics[scale=0.5]{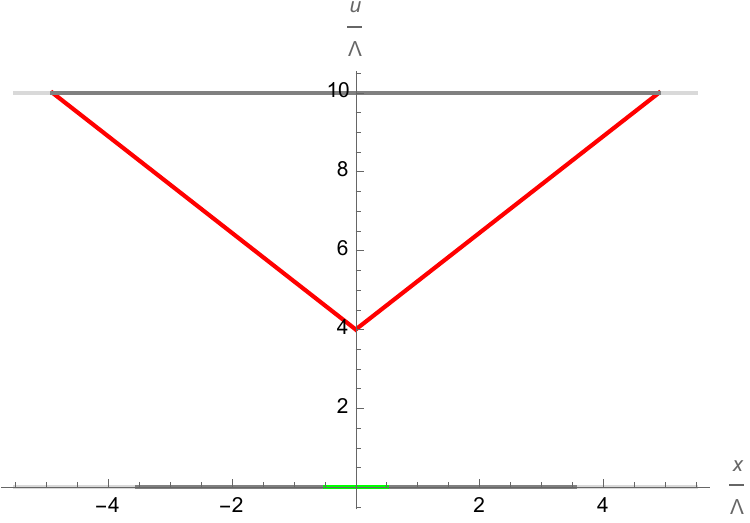}
\caption{A demonstration of information velocity $v_I$ from the entanglement membrane perspective in the scaling limit. Before saturation (left), the time-dependent membranes (red) reach down to the $t=0$ slice (gray). The density matrix of the subregion therefore contains the reference qubit (green). After saturation, however, the membrane forms a butterfly cone (red), which excludes the reference qubit on the $t=0$ slice. \label{fig:InfoVeloDemo}}
\end{center}
\end{figure}

It is argued in~\cite{Couch:2019zni} that $v_I$ is equal to the ratio of the size of the reconstruction region, as defined by the radius of the largest sphere inscribed in the region, divided by the saturation time of the entanglement entropy
\es{viRes}{
v_I&={R\ov t_\text{sat}}\,.
}
This result is very intuitive from the membrane perspective. The saturated membrane is a butterfly cone over the subsystem. It does not reach down to the time slice representing the initial state, to which the reference qubit was entangled. If we consider the membrane to be the minimal cut through the quantum circuit, the reference qubit lies outside the cut and cannot be reconstructed. However, just before saturation, the cut reaches down to this time slice and the reference qubit is contained in the density matrix of the subregion. See Fig.~\ref{fig:InfoVeloDemo}. 


\subsection{Strip}

It is clear from \eqref{MembraneAction} that the bulk equations of motion for the membrane remain unchanged, and the boundary term only affects boundary conditions (and the value of the entropy on shell). We also note for $f=0$ we recover the global quench setup.
Thus for strips of width $2R$ we get membranes with constant $v$, and \eqref{MembraneAction} becomes:
\es{MembraneActionStip}{
S[A(T)]=s_\text{th}\, \text{area}(\p A) \le({\cal E}(v)\, t +f (R-v\,t)\ri)\,,
}
whose minimum in $v$ is at
\es{Minimum}{
f= {\cal E}'(v)\,,
}
independently of $T$.

 The entropy saturates when $S[A(T_\text{sat})]=s\, \text{area}(\p A)R$, which gives
\es{vi}{
v_I&={R\ov T_\text{sat}}={{\cal E}(v)- {\cal E}'(v) v\ov 1-f}\Bigg\vert_{{\cal E}'(v)=f}\\
&\equiv{\Gamma(f)\ov 1-f}\,,
}
where in the second line we defined $\Gamma(f)$ to be the Legendre transform of ${\cal E}(v)$. It has the meaning of entanglement growth rate as discussed by~\cite{Jonay:2018yei} with $f$ playing the role of $f={1\ov s}\, {\p S\ov \p x}$.

\begin{figure}[!h]
\begin{center}
\includegraphics[scale=0.5]{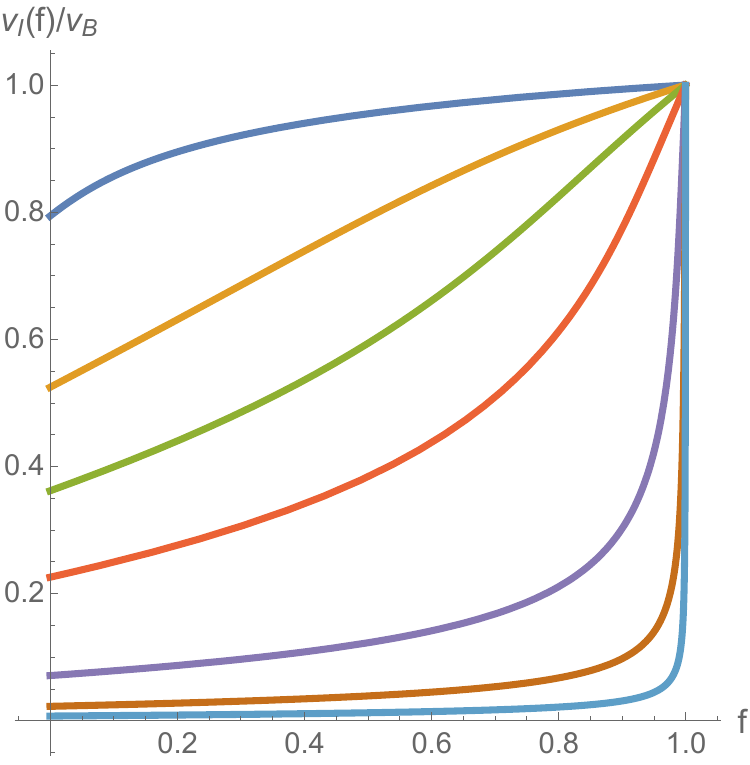}\hspace{1cm}
\includegraphics[scale=0.5]{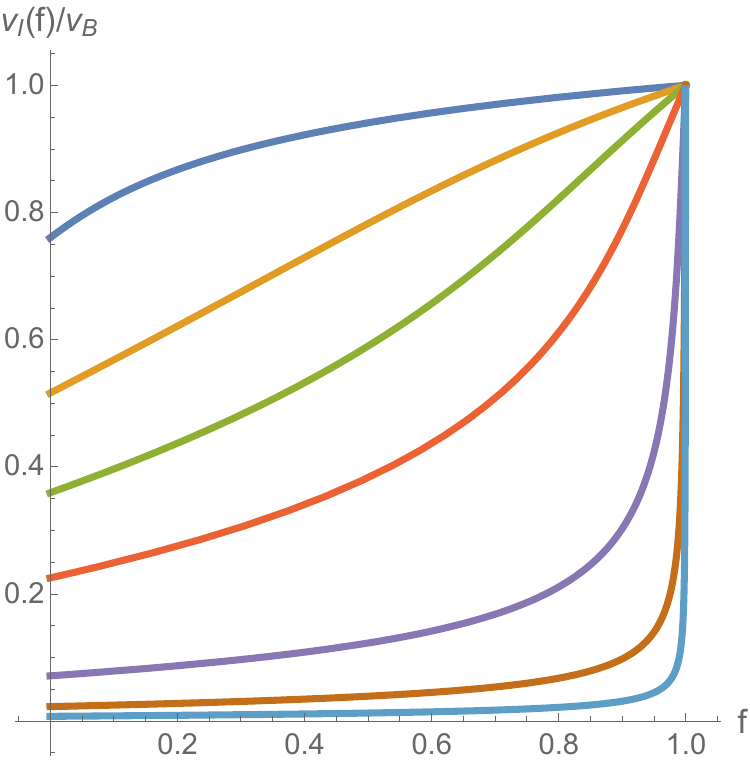}
\caption{$v_I(f)/v_B$ for strips in $d=3$ (left) and $d=4$ (right) for different values of the chemical potential of the equilibrium ensemble. The top curves are for zero chemical potential and in the extremal limit the graph becomes a step function centred on $f=1$.  \label{fig:strips}}
\end{center}
\end{figure}

From the general properties of ${\cal E}(v)$ it follows that $v_I(f=0)=v_E$ and $v_I(f=1)=v_B$. We note that $\Gamma(f)$ can also be thought of as $v_E(f)$, and will play a role for other shapes as well. On Fig.~\ref{fig:strips} we show some examples of $v_I(f)$ in $d=3,4$ for quenches that thermalise to the grand canonical ensemble, which has a charged black hole gravity dual. These analytic results are consistent with the numerical results of~\cite{Couch:2019zni}.

\subsection{Sphere and other shapes}

To determine $v_I$, the challenge is to determine the saturation time $t_\text{sat}$ for subregions more complicated than the strip. This problem was extensively studied in prior literature in quenches with $f=0$, and there are a couple of beautiful general results about it. For the subsequent discussion let us define the butterfly time as $t_B=R/v_B$ for arbitrary regions.  $t_B$ is the separation in time to the tip of the butterfly cone over the region. It follows from very general considerations that $t_\text{sat}\geq t_B$~\cite{Mezei:2016wfz}. 
Remarkably, spheres saturate this bound for the ${\cal E}(v)$ applicable to charge neutral states in four and higher dimensional holographic field theories~\cite{Liu:2013qca,Liu:2013iza,Mezei:2016wfz,Mezei:2016zxg,Mezei:2020knv}.\footnote{In three dimensions the bound is not saturated, but the violation is numerically very small, ${t_\text{sat}-t_B\ov t_B}\sim 10^{-6}$~\cite{Mezei:2016zxg}.} In~\cite{Mezei:2020knv} the saturation was found to extend to a large class of other shapes, but notably not the strip.

Now let us explore what these results imply for $v_I$. From $t_\text{sat}\geq t_B$ it follows immediately that $v_I\leq v_B$. 
It is easy to see that the boundary term in \eqref{MembraneAction} results in the boundary condition that is the local version of \eqref{Minimum}. In polar coordinates $(t,r,\Om)$:
\es{BC}{
f= {\cal E}'(v(t=0,\Om))\,,
}
which for $f=0$ becomes the familiar boundary condition that the membrane has to end perpendicular to the lower boundary of the slab. 

 In the setups where  $t_\text{sat}=t_B$ for $f=0$, the membrane is a $v_B$ light sheet, and $A'(0)$  is a point. Hence it does not matter what the value of $f$ is, they give $S=S_\text{sat}$. Since they were globally minimal already for $f=0$, they remain the global minimum also for $f>0$. We conclude that for these shapes $v_I=v_B$ irrespective of the value of $f$.

On the other hand, by turning on the chemical potential, we find that $t_\text{sat}>t_B$ at $f=0$~\cite{Mezei:2016zxg},\footnote{This is easiest to conclude from the analytic treatment of for a sphere~\cite{Mezei:2016zxg} or cylinder subregion~\cite{Mezei:2020knv}, but it is expected to hold for other shapes as well that require numerical treatment as in~\cite{Mezei:2020knv}.} and we will get a nontrivial $v_I(f)$.  To determine it, we have to solve a nontrivial membrane problem. Let us now review how we solve the membrane minimisation problem for spheres. We use spherical symmetry in \eqref{MembraneAction} to reduce it to a one dimensional classical mechanics problem, which we solve using energy conservation:
\es{Hcons}{
r^{d-2}\le({\cal E}(v)- {\cal E}'(v) v\ri)=\text{const}\,, \quad \text{with} \quad v=\dot{r}\,.
} 
The problem simplifies further because the scaling transformation $r\to \lam\, r ,\, t\to \lam\, t$ and time translation $t\to t+t_0$ take solutions to solutions. As a result, we only need the master solution $\rho(\tau)$ which obeys
\es{rhos}{
\rho(0)=1\,, \quad \dot{\rho}(0)=0\,, \quad \rho(\infty)= \infty\,, \quad \dot{\rho}(\infty)=v_B
}
which after scaling and a shift determines the problem we set out to solve:
\es{rt}{
r(t)&=\lam \,\rho\le({t\ov \lam}+\tau_0\ri)\,,
}
with $\tau_0$ and $\lam$ determined by the boundary conditions
\es{tau0lam}{
\dot{r}(0)&= \dot{\rho}(\tau_0)=\le({\cal E}'\ri)^{-1}(f)\,,\\
R&=r(T)=\lam \,\rho\le({T\ov \lam}+\tau_0\ri)\,.
}
Solving these we obtain $r(t)$ and can determine 
\es{ST}{
S(T)=\lam^{d-1}\le[s\le({T\ov \lam}+\tau_0\ri)-s\le(\tau_0\ri)+f \,v_{d-1}\, \rho(\tau_0)^{d-1}\ri]\,,
}
where $v_{d-1}$ is the volume of the unit $(d-1)$-ball and $s(\tau)$ is the on shell action of $\rho(\tau')$ (integrated from $0$ to $\tau$). Finally, $v_I$ is determined by finding $T=t_\text{sat}$ such that $S(t_\text{sat})=S_\text{sat}$. There is one subtlety: \eqref{tau0lam} may have multiple solutions for $\lam$, so in practice it is better to make parametric plots and to choose the branch of the multi-valued function that gives minimal entropy. The best parametrisation that we found was
\es{param}{
\{T, S(T)\}=\le\{{  R(\tau-\tau_0)\ov \rho(\tau)},\, \le({R\ov \rho(\tau)}\ri)^{(d-1)}\le[s\le(\tau\ri)-s\le(\tau_0\ri)+f \,v_{d-1}\, \rho(\tau_0)^{d-1}\ri] \ri\}\,,
}
which avoids having to solve for $\lam$ entirely. 

\begin{figure}[!h]
\begin{center}
\includegraphics[scale=0.57]{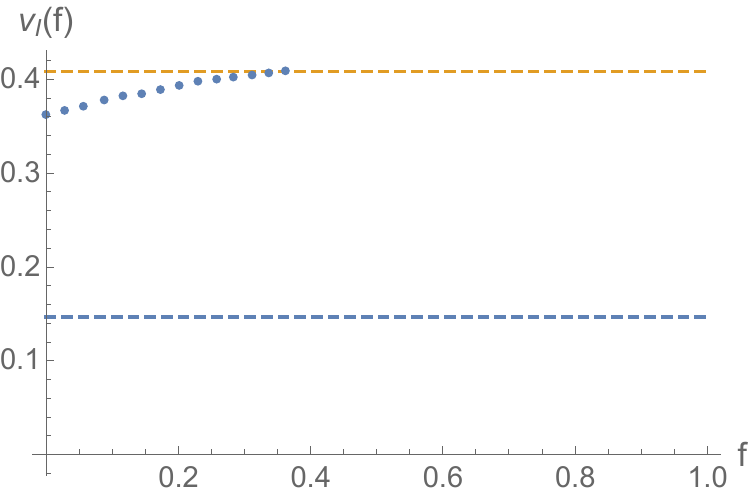}\hspace{0.5cm}
\includegraphics[scale=0.57]{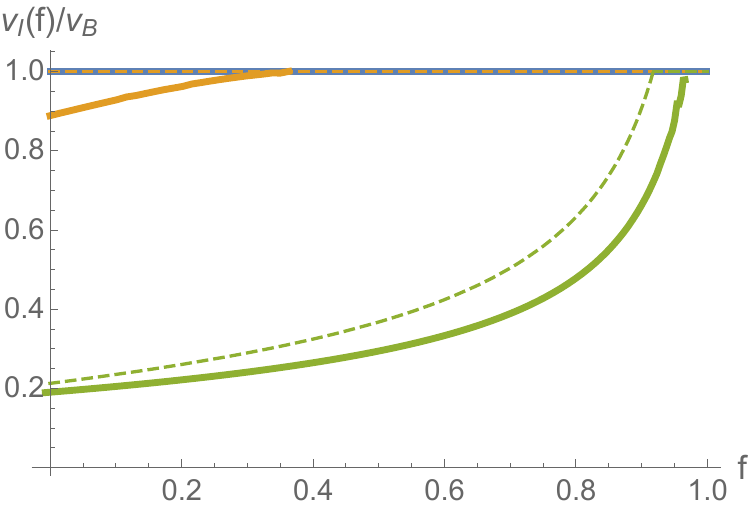}
\caption{\emph{Left:} $v_I(f)$ for a sphere in $d=4$ for  $\mu=0.75\, \mu_c$ with $\mu_c$ the extremal value of the chemical potential. The two dashed lines are $v_E$ and $v_B$ respectively.  \emph{Right:} The solid lines are $v_I(f)/v_B$ computed numerically from \eqref{param} for a sphere in $d=4$ for chemical potentials approaching extremality $\mu/\mu_c=0,\, 0.75,\,0.99$ (blue, orange, green). The dashed lines (blue, orange, green) are \eqref{vEformula}. The blue and orange lines are the constant $1$ function, only the green dashed line is nontrivial and follows the trend of the real curve.  \label{fig:sphere}}
\end{center}
\end{figure}

On Fig.~\ref{fig:sphere} we summarise our findings. As expected, for $f=0$, we have $v_I >v_E$, but we find that $v_I(f)$ saturates to $v_B$ at a critical $f_c$, and from then on, $v_I(f\geq f_c)=v_B$. Again, these semi-analytic results are consistent with the numerical results of~\cite{Couch:2019zni}. They are significantly easier to obtain and are significantly more precise.

For other shapes, such as those studied in~\cite{Mezei:2020knv}, one has to numerically solve the time evolution problem for a given $f$ and from there determine $T_\text{sat}$ to obtain $v_I(f)$. Recall that in the sphere case, we only had to determine the function $\rho(\tau)$ that is independent of $f$, and from it we could extract $v_I(f)$. We expect that we would again find an $f_c$, above which $v_I(f\geq f_c)=v_B$.

We end with some further attempts at qualitative understanding the curves in Fig.~\ref{fig:sphere}.
The existence of $f_c$ can be understood from the following semi-analytic argument: Let us analyse the small and large $\tau$ behaviour of \eqref{param}. For $\tau=\tau_0$, $S(T)<S_\text{sat}$. For large $\tau$ the terms in \eqref{param} behave as follows:
\es{Largetau}{
\le({R\ov \rho(\tau)}\ri)^{(d-1)}\,s\le(\tau\ri)&=S_\text{sat}+{c_1\ov \tau^{(d-1)}}+\dots\,,\\
\le({R\ov \rho(\tau)}\ri)^{(d-1)}\le[-s\le(\tau_0\ri)+f \,v_{d-1}\, \rho(\tau_0)^{d-1}\ri]&=-{c_2(f)\ov \tau^{(d-1)}}+\dots\,.
}
We found that $c_2(f)$ is a monotonically increasing function of $f$, hence for $c_1=c_2(f_c)$ we get that $\tau_\text{sat}=\infty$. For $f\geq f_c$ it remains true that $\tau_\text{sat}=\infty$, which using \eqref{rhos}  gives $t_\text{sat}=t_B$.  For $f<f_c$ instead we have a finite $\tau_\text{sat}$, and $t_\text{sat}>t_B$. 

It may also be of interest to get a gross approximation to the curves. To this end, we note that if the $S(T)$ curves were precisely linear, we would have
\es{vEformula0}{
v_I^\text{lin. app.}={(d-1)v_E(f)\ov 1-f}\,,
}
where $v_E(f)=\Gamma(f)$ is the initial growth rate of the entropy.
This would contradict the general bound $v_I\leq v_B$ derived above, hence we correct it:
\es{vEformula}{
v_I^\text{lin. app.}=\min\le[{(d-1)v_E(f)\ov 1-f},v_B\ri]\,.
}
For comparison we also plot \eqref{vEformula} on Fig.~\ref{fig:sphere}. We note that \eqref{vEformula} does not become exact even in the extremal limit.


\section{Conclusion and outlook}
In this paper, we addressed the status of the entanglement membrane in 2d CFT. Using holography, we found that one needs to introduce an additional degree of freedom on the membrane, which we depicted in colour in our figures. Beyond EE, this generalised membrane theory also captures the reflected entropy computed holographically by the entanglement wedge cross sections. To study the interpolation between generalised and ordinary membrane theory, we introduced a relevant deformation to the 2d CFT, which is holographically dual to a hairy planar BTZ black hole with an interior Kasner universe. Under such deformation, the conventional non-degenerate membrane tension emerges, which we computed both numerically and from conformal perturbation theory. We also studied reflected entropy in higher dimensional CFTs, and showed that their dynamics in the scaling limit are described by ordinary membrane theory. We found that due to the additional degree of freedom in generalised membrane theory for 2d CFT, the dynamics of reflected entropy show different behaviours in $d=2$ and $d\geq3$. Lastly, we showed that membrane theory sheds new light on and provides a streamlined computation of the information spreading velocity introduced by~\cite{Couch:2019zni}. 

We discuss some interesting future directions opened by this work:
\begin{itemize}
\item In generalised membrane theory, the additional degree of freedom $\xi(u)$ labels the depth of the bulk $z=e^{\xi}$ direction. While we believe that $\xi(u)$ reflects the infinite-dimensional conformal symmetry of 2d CFT, a direct map of $\xi(u)$ to field theoretic quantities remains lacking. 

\item It would be interesting to reproduce our results for $v_B$~\eqref{vzEz} and $v_E$~\eqref{vEexpansion} in (some partial resummation of) conformal perturbation theory in 2d CFT. The latter contains a logarithmic enhancement $\lam^2\log \lam$, possibly indicating an IR divergence in perturbation theory.   

\item Unlike in the case of the ordinary entanglement membrane theory of holographic gauge theories~\cite{Mezei:2018jco}, there is currently no random unitary circuit interpretations of generalised membrane theory. As a starting point, one can investigate dual unitary circuits~\cite{Bertini:2018fbz,Piroli:2019umh}, which also have a degenerate tension function $\mathcal{E}(v)=1$~\cite{Zhou:2019pob}. 

\item In section~\ref{sec:2dEWCS} and~\ref{sec:EWCS}, we studied reflected entropy and EWCS in (generalised) entanglement membranes, implying that the concept of entanglement wedge~\cite{Czech:2012bh,Headrick:2014cta,Wall:2012uf} is meaningful beyond holography. However, our approach in computing reflected entropy from EWCS is still holographic. It would be interesting to calculate reflected entropy from a random unitary circuit starting point, and in particular, to explore its relations with the ``entanglement wedge cross section" in entanglement membrane (see e.g. Fig.~\ref{fig:NeqEWCSProj}, \ref{fig:EqEWCSProj} and \ref{fig:MembraneEWCS}). Related works in this direction include~\cite{Akers:2021pvd,Akers:2022zxr,Akers:2023obn,Akers:2024pgq}. One can attempt to generalise their methods to the \emph{boundary} tensor network, which is relevant for the entanglement membrane. 

\end{itemize}


\section*{Acknowledgments}

We are grateful to Richard Davison, Luca Delacr\'etaz, Christopher Herzog, Jonah Kudler-Flam, and Ayan Patra for discussions. HJ is partially supported by Lady Margaret Hall, University of Oxford. MM is supported in part by the STFC grant ST/X000761/1. JV is partially funded by the DFG grant No. 406116891 within the Research Training Group RTG 2522/1.

\appendix

\section{Detailed analysis of solutions in the generalised membrane theory}\label{ApproxLagD}
\subsection{Displaced half-space}\label{ApproxLagDispHalf}
In this appendix, we solve the equations of motion arising from the effective action $\mathcal{L}_{{\rm eff}}$ \eqref{Leff} exactly. Our solution is still an approximation to the full solution presented in Appendix~\ref{Displaced half spaces}, since the effective action $\mathcal{L}_{{\rm eff}}$ \eqref{Leff} is derived under the scaling limit \eqref{AreaFunct3}. 

\begin{figure}[htbp]
\centering
\includegraphics[width=.4\textwidth]{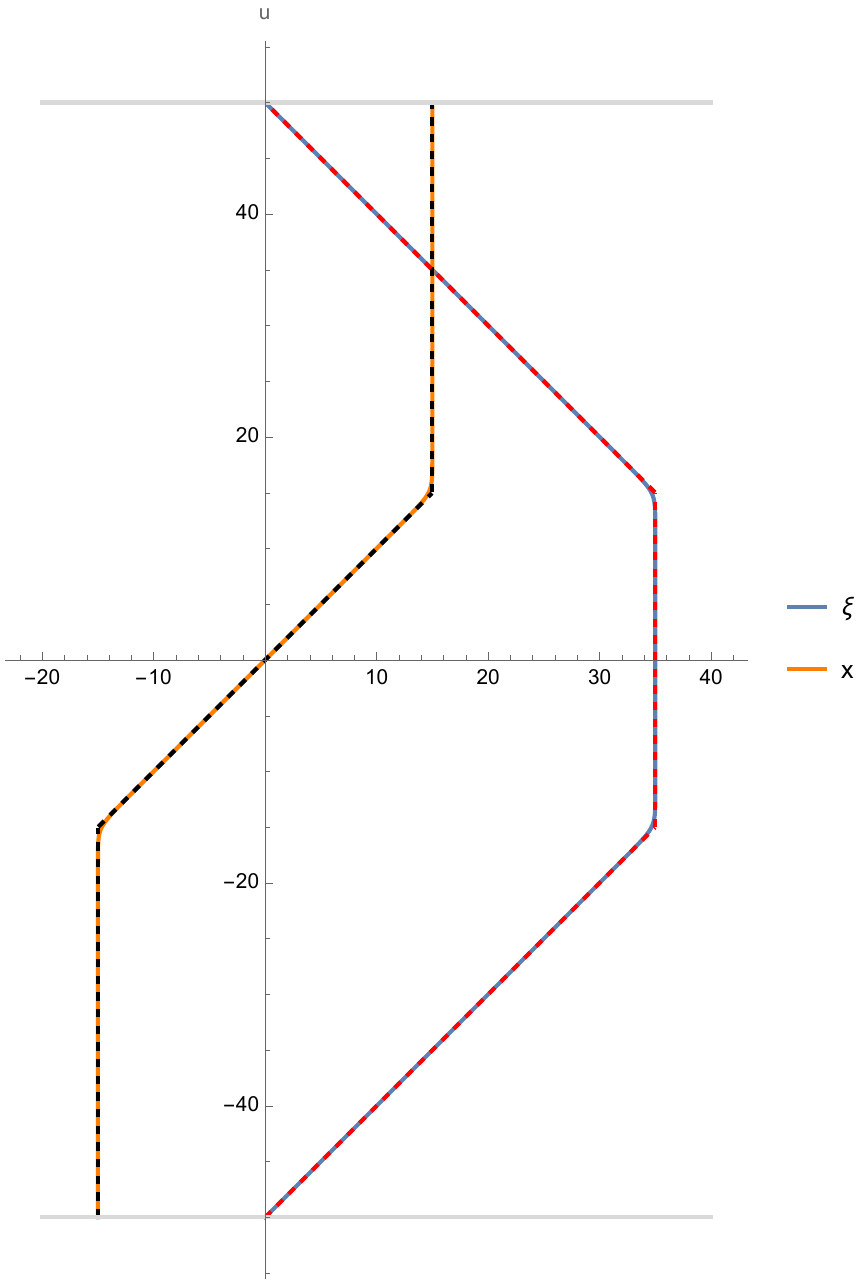}
\qquad
\caption{Plot of $\xi(u)$ (blue) \eqref{z(t)2} and $x(u)$ (orange) \eqref{x(t)2} when $T=50$ and $X=15$. The scaling approximation of $\xi(u)$ and $x(u)$ \eqref{FinalXi} are plotted as dashed in red and black, respectively. \label{fig:xiXApproxLag}}
\end{figure}

As \eqref{Leff} does not contain $u$ or $x$, it has two conserved quantities, the energy $E$ and momentum $p$, 
\begin{align}
    E&=\dot{\xi}\frac{\partial \mathcal{L}}{\partial \dot{\xi}}+\dot{x}\frac{\partial \mathcal{L}}{\partial \dot{x}}-\mathcal{L}=\frac{1}{2} e^{-2 \xi} \left(1+\dot{x}^2-\dot{\xi}^2\right)\label{simplifiedEBTZ}\\
    p&=\frac{\partial \mathcal{L}}{\partial \dot{x}}=e^{-2\xi}\dot{x}\label{simplifiedPBTZ}
\end{align}
The generalised velocity $\dot{\xi}$ vanishes at the furthest point $\xi_0$ the geodesic can reach in the interior, $\dot{\xi}|_{\xi=\xi_0}=0$. Therefore, we can evaluate $E$ \eqref{simplifiedEBTZ} at $\xi_0$, which then allows us to solve for $\dot{\xi}$
\begin{align}
    e^{-2 \xi} \left(1+p^2e^{4\xi}-\dot{\xi}^2\right)&=e^{-2 \xi_0} \left(1+p^2e^{4\xi_0}\right)\\
    \Rightarrow \dot{\xi}^2&=(e^{2\xi_0}-e^{2\xi})(e^{-2\xi_0}-p^2e^{2\xi})\label{zdot}
\end{align}
where we have plugged in \eqref{simplifiedPBTZ}. We see that by making use of conserved quantity $E$ \eqref{simplifiedEBTZ}, we reduced the problem from a second order ODE \eqref{xiEOM} to a first order ODE \eqref{zdot}. Solving \eqref{zdot}, we find
\begin{align}
    \xi(u)=\log \left(\frac{e^{\text{$\xi $}_0} \text{sech}u}{\sqrt{1-e^{4 \text{$\xi $}_0} p^2
   \tanh ^2u}}\right)\label{z(t)}
\end{align}
Then, plug \eqref{z(t)} into \eqref{simplifiedPBTZ} and integrate, we can further obtain $x(u)$
\begin{align}
    x(u)={\rm arctanh}(p e^{2\xi_0}\tanh u)\label{x(t)}
\end{align}
From \eqref{x(t)}, we can find the total displacement $2X$ in terms of $p$ and $z_0$ by taking $|u|$ large, resulting in 
\begin{align}
    X={\rm arctanh}(Pe^{2\xi_0}) && \Rightarrow && p=e^{-2\xi_0}\tanh X\label{PwX}
\end{align}
\eqref{PwX} indicates that $p\to e^{-2\xi_0}$ for large $X$. Plugging \eqref{PwX} into \eqref{z(t)}, we have 
\begin{align}
    \xi(u)=\log \left(\frac{e^{\text{$\xi $}_0} \text{sech}u}{\sqrt{1-\tanh
   ^2X \tanh ^2u }}\right)\label{z(t)3/2}
\end{align}
This allows us to solve for $\xi_0$ in terms of $T$ and $X$ from \eqref{z(t)3/2}
\begin{align}
    \xi_0=\log \left(\sqrt{\frac{\cosh 2 T+\cosh 2 X}{\cosh 2 X +1}}\right)\label{xi0X}
\end{align}
which matches \eqref{z0X} obtained from the BTZ-AdS$_3$ mapping method. Plug \eqref{xi0X} and \eqref{PwX} into \eqref{z(t)} and \eqref{x(t)} we can express the geodesic as a parametric curve parameterised by the infalling time $u$
\begin{align}
    \xi(u)&=\frac{1}{2} \log \left(\frac{\cosh 2 T+\cosh 2 X}{\cosh 2 u+\cosh 2
   X}\right)\label{z(t)2}\\
    x(u)&={\rm arctanh}\ (\tanh X \tanh u)\label{x(t)2}
\end{align}
A plot of \eqref{z(t)2} and \eqref{x(t)2} can be found in Fig.~\ref{fig:xiXApproxLag}. In the large $T$, large $X$ ($T>X$) limit, one can check that $\xi(u)$ \eqref{z(t)2} and $x(u)$ \eqref{x(t)2} match \eqref{FinalXi}.

 \subsection{Saturated EWCS}\label{Plateau+Null geodesic}
With the change of variable from $u$ to $x$ in \eqref{xiEOMx}, $E$ \eqref{simplifiedEBTZ} becomes 
\begin{align}
    E=\frac{1}{2} e^{-2 \xi } \left(1-e^{4 (\xi -\xi_p)} \left(\xi
   '^2-1\right)\right)\label{xixDiffEqn}
\end{align}
where we have plugged in $u'$ \eqref{xiEOMx}. Again, the conserved $E$ allows us to reduce the problem from a second-order ODE \eqref{xiEOMx} to a first-order ODE \eqref{xixDiffEqn}. Solving \eqref{xixDiffEqn} we find
\begin{align}
    \xi(x)=\frac{1}{2}\log\left(\frac{e^{4\xi_p}\left(1-E^2 e^{4\xi_p}\right) e^{-2 x} -e^{2
   x}+2 E e^{4\xi_p}}{2}\right)\label{xixv1}
\end{align}
Notice that $\xi(-x)$ is also a solution to \eqref{xixDiffEqn}. This allows us to flip the sign of $u'$ \eqref{xiEOMx}.\footnote{It turned out that the effective theory allows both $\pm$ signs. To match with the exact solution, however, we choose $-$ sign here. } Plugging \eqref{xixv1} into \eqref{xiEOMx} and integrating, we have
\begin{align}
    u(x)=u_p+\ell_p-{\rm arctanh}(E e^{2\xi_p}-e^{2(x-\xi_p)})\label{txv1}
\end{align}
where for later convenience we write the integration constant as $u_p+\ell_p$. It can be easily observed from \eqref{txv1} that as $x\to-\infty$, $u(x)$ asymptotes a constant value. Denoting this value as $u_p$, we can solve for $E$ 
\begin{align}
    E=e^{-2\xi_p}\tanh\ell_p\label{ENullGeod}
\end{align}
Notice the similarity between \eqref{ENullGeod} and \eqref{PwX}. Plugging \eqref{ENullGeod} into \eqref{xixv1}, one finds
\begin{align}
    \xi(x)=\frac{1}{2}\log\frac{1}{2}\Big(-e^{2x}+e^{-2(x-2\xi_p)}{\rm sech}^2\ell_p+2e^{2\xi_p}\tanh\ell_p\Big)\label{xixexact}
\end{align}
which, in the large $\ell_p$ limit, simplifies to 
\begin{equation}
\label{xixscaling}
    \xi(x)=
    \left\{
    \begin{aligned}
        &-x-\ell_p+2\xi_p+\frac{1}{2}\log2,&& x\in(-\infty,-\ell_p+\xi_p+\frac{1}{2}\log2]\\
        &\xi_p,&& x\in[-\ell_p+\xi_p+\frac{1}{2}\log2,\xi_p+\frac{1}{2}\log2)
    \end{aligned}
    \right.
\end{equation}
When $x\in[-\ell_p+\xi_p+\frac{1}{2}\log2,\xi_p+\frac{1}{2}\log2)$, the geodesic is on a $\emph{plateau}$ $\xi=\xi_p$. $\ell_p$ is therefore identified as the (horizontal) $\emph{length}$ of this plateau. When $x\in(-\infty,-\ell_p+\xi_p+\frac{1}{2}\log2]$, the geodesic shoots into the deep interior exponentially in $x$. Plugging \eqref{ENullGeod} into \eqref{txv1}, one finds 
\begin{align}
    u(x)=u_p+\ell_p+{\rm arctanh}(\tanh\ell_p-e^{2(x-\xi_p)})\label{txexact}
\end{align}
which, in the large $\ell_p$ limit, simplifies to 
\begin{equation}
\label{txscaling}
    u(x)=
    \left\{
    \begin{aligned}
        &u_p,&& x\in(-\infty,-\ell_p+\xi_p+\frac{1}{2}\log2]\\
        &x+u_p+\ell_p-\xi_p-\frac{1}{2}\log2,&& x\in[-\ell_p+\xi_p+\frac{1}{2}\log2,\xi_p+\frac{1}{2}\log2)
    \end{aligned}
    \right.
\end{equation}
Notice that when $(-\infty,-\ell_p+\xi_p+\frac{1}{2}\log2]$, the geodesic is on the $u=u_p$ slice. 

In general, one can shift $x\to x-x_c$ in \eqref{xixexact} and \eqref{txexact}. In the scaling limit, the shifted \eqref{xixexact} and \eqref{txexact} match the approximate solution \eqref{XiEWCS} in the main text with $x_0=x_c-\ell_p+\xi_p$. Therefore, one cannot fix all four parameters $(u_p,\ell_p,\xi_p,x_c)$ even solving the approximate equations of motion \eqref{xiEOMx} exactly. 

\begin{figure}[htbp]
\centering
\includegraphics[width=.6\textwidth]{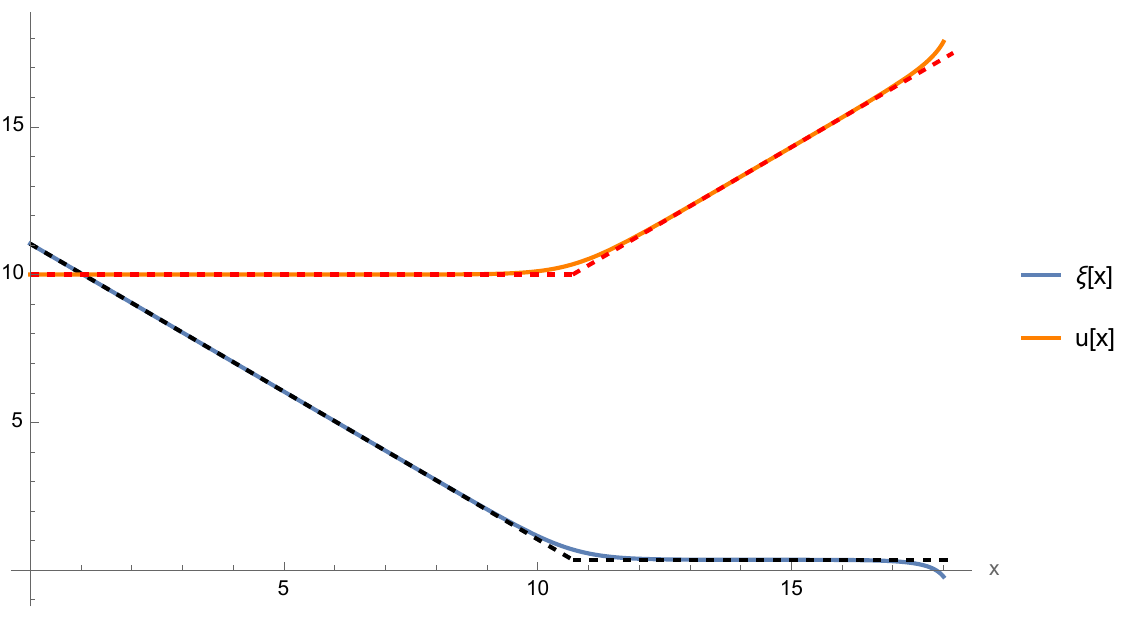}
\qquad
\caption{Plot of $\xi(x)$ \eqref{xixscaling} and $u(x)$ \eqref{txscaling} in \eqref{XiEWCS} when $u_p=10$, $\ell_p=7.5$, $x_c=17.5$ and $\xi_p=\log\sqrt{2}$ which is negligible in the scaling limit. The exact \eqref{xixexact} and scaling approximation \eqref{XiEWCS} of $\xi(x)$ are plotted in blue and dashed black, respectively;  the exact \eqref{txexact} and scaling approximation \eqref{XiEWCS} of $u(x)$ are plotted in orange and dashed red, respectively. The scaling approximations \eqref{xixscaling} and \eqref{txscaling} match \eqref{XiEWCS}. \label{fig:ApproxLagNullGeod}}
\end{figure}

In the exact solution in Appendix~\ref{SatEWCS}, we have $\xi_p=\frac{1}{2}\log2$ \eqref{zsqrt2} which is negligible, and $\ell_p=\frac{\ell}{2}$ \eqref{SatEWCSLen}, $u_p=T-\frac{1}{2}(\ell+\frac{D}{2})$ \eqref{BProjCoord}. Adapting to the coordinate system in Fig.~\ref{fig:ABProj} via a shift $x_c=\ell-\frac{D}{4}$, one finds that $\eqref{txscaling}$ becomes 
\begin{align}
    u(x)=x+T-\ell-\frac{D}{2}
\end{align}
which matches the exact solution \eqref{x(t)SatEWCS} in the scaling limit. Notice that the effective theory in this section is technically much simpler than the exact solution in subsection~\ref{SatEWCS}. A plot of \eqref{xixexact}-\eqref{txscaling} with shift $x_c=\ell-\frac{D}{4}$ can be found in Fig.~\ref{fig:ApproxLagNullGeod}.

\section{Exact solutions of entanglement entropy in BTZ black hole} \label{app:AdStoBTZ}

In the main text as well as Appendix~\ref{ApproxLagD}, we have solved the equation of motions for various geodesics in planar BTZ black hole $\emph{approximately}$ in the scaling limit. These geodesic are in fact solvable $\emph{exactly}$, by virtue of the map between planar BTZ black hole and Poincare AdS$_3$~\cite{Hartman:2013qma}. We will see that all approximate solutions found earlier match with exact results detailed below.  

The planar BTZ black brane metric is given by
\begin{align}
    ds^2=\frac{1}{z^2}\Big(-f(z)dt^2+\frac{dz^2}{f(z)}+dx^2\Big)\label{BTZmetric}
\end{align}
where $f(z)=1-z^2$ for planar BTZ, and the BTZ horizon is at $z=1$. Unlike in the main text or Appendix~\ref{ApproxLagD}, for technical reasons we will work in Schwarzschild time $t$ in this appendix. The subtlety one needs to bear in mind is that in the interior, $t$ becomes complex~\cite{Fidkowski:2003nf}. We will use $\Tilde{t}=t+\frac{i\pi}{2}$ to denote the real part of the Schwarzschild time in the interior. As we will see in subsection~\ref{SchVSInfallingSec}, the differences between Schwarzschild time $t$ and infalling time $u$ turn out to be immaterial in the region $z\gg1$ that we are mostly interested in. 

\subsection{Difference between Schwarzschild and infalling time}\label{SchVSInfallingSec}
The infalling time $u$ is given by
\begin{align}
    du=dt-\frac{dz}{f(z)}\label{SchInf}
\end{align}
where $f(z)=1-z^2$. Outside the horizon $z<1$, one can simply integrate \eqref{SchInf} to obtain \eqref{u(z)2dExpansion}. In the interior $z>1$, notice that \eqref{SchInf} contains a simple pole at the horizon $z=1$. The resulting residue cancels the imaginary part of the interior Schwarzschild time $t$, and we have
\begin{align}
    u=\Tilde{t}-{\rm arccoth}z\approx\Tilde{t}-\frac{1}{z}\to\Tilde{t}\label{SchApproxInf}
\end{align}
for large $z$. Therefore, in the deep interior $z\gg1$, adopting Schwarzschild time $u$ or infalling time $t$ yields asymptotically the same result. 

\subsection{Map between BTZ and Poincare AdS$_3$}\label{BTZAdSMap}
\begin{figure}[htbp]
\centering
\includegraphics[width=.45\textwidth]{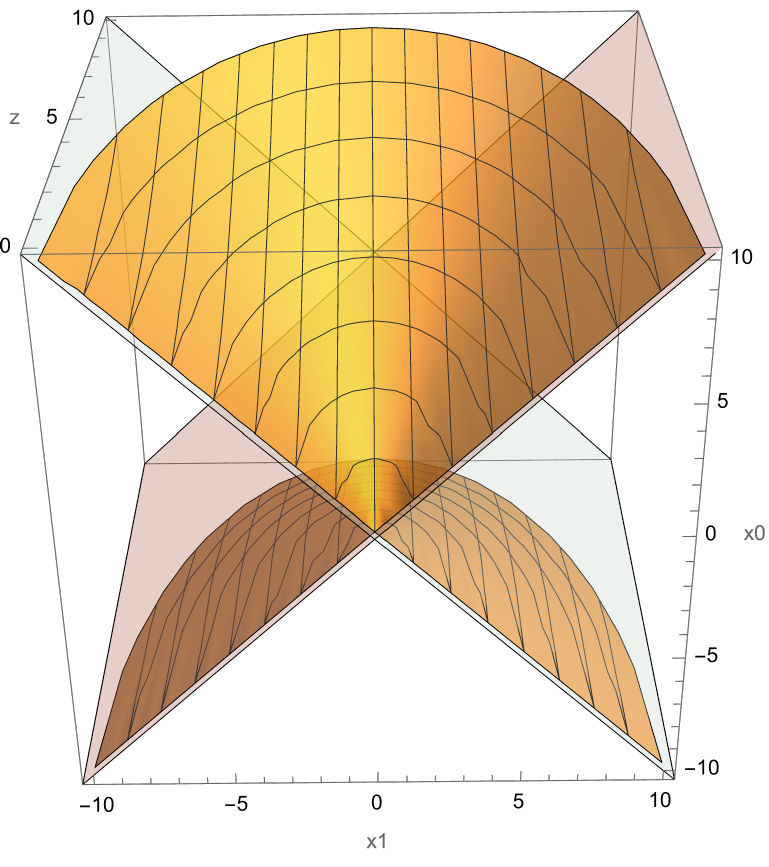}
\qquad
\caption{Map from BTZ black hole to Poincare AdS$_3$. The two boundaries on each side of the BTZ black hole becomes the two Rindler wedges in Poincare AdS$_3$. The BTZ black hole horizon is mapped to $|x_0|=|x_1|$, which are plotted in light blue. The exterior and interior regions are mapped to $x_0^2-x_1^2\leq 0$ and $0\leq x_0^2-x_1^2\leq Z^2$, respectively. In the interior, the hyperboloid $Z^2=x_0^2-x_1^2$ corresponds to $z\to\infty$ in BTZ. 
\label{fig:BTZAdS3Map}}
\end{figure}

BTZ black hole is a quotient of AdS$_3$~\cite{Banados:1992wn,Banados:1992gq}. We can therefore unwrap the BTZ angle and obtain Poincare AdS$_3$ as in~\cite{Hartman:2013qma}. In AdS$_3$ there is only one boundary, and the two boundaries on each side of the BTZ black hole become the two Rindler wedges in Poincare AdS$_3$ (See Fig.~\ref{fig:BTZAdS3Map})~\cite{Hartman:2013qma}. Notice that these two Rindler wedges do not cover the full Poincare AdS$_3$ boundary; the complementary regions are known as the two Milne patches. We would like to obtain from \eqref{BTZmetric} the Poincare AdS$_3$ metric 
\begin{align}
    ds^2=\frac{1}{Z^2}(-dx_0^2+dx_1^2+dZ^2)\label{AdSMetric}
\end{align}
It is helpful to discuss the exterior and interior regions of the BTZ black brane metric \eqref{BTZmetric}. To this end, we follow~\cite{Hartman:2013qma} and massage the BTZ radial coordinate $z$ via a coordinate transformation 
\begin{align}
    z={\rm sech}\rho\label{zRho}
\end{align}
where $\rho>0$ so that $z\in[0,1]$. The $\emph{exterior}$ part of the BTZ metric \eqref{BTZmetric} thus becomes
\begin{align}
    ds^2=- \sinh ^2\rho dt^2+d\rho ^2+\cosh ^2\rho dx^2\label{ExteriorMetric}
\end{align}
The BTZ horizon is now at $\rho=0$. The coordinate time $t$ in \eqref{ExteriorMetric} diverges at the horizon $\rho=0$. The BTZ interior coordinate can be obtained via an analytic continuation on $\rho$ and adding an imaginary part in $t$~\cite{Hartman:2013qma}
\begin{align}
    \rho=i\alpha\ \Rightarrow\ z={\rm sec}\alpha\in[1,+\infty) && \Tilde{t}=t+\frac{i\pi}{2} \label{zAlpha}
\end{align}
The $\emph{interior}$ part of the BTZ metric \eqref{BTZmetric} is then given by 
\begin{align}
    ds^2=\sin ^2\alpha d\Tilde{t}^2-d\alpha ^2+\cos ^2\alpha dx^2  \label{InteriorMetric}
\end{align}
The interior time $\Tilde{t}$ in \eqref{InteriorMetric} diverges at the horizon $\alpha=0$. Notice that in the interior, $\alpha$ is timelike, and $\Tilde{t}$ is spacelike. 

Using the embedding coordinate in~\cite{Hartman:2013qma}, the coordinate transformation between BTZ exterior and Poincare AdS$_3$ are found to be
\begin{align}
    Z&=e^x{\rm sech}\rho\label{coordtransfExtZ}\\
    x_0&=e^x{\rm sinh}t\ {\rm tanh}\rho\label{coordtransfExtx0}\\
    x_1&=e^x{\rm cosh}t\ {\rm tanh}\rho\label{coordtransfExtx1}
\end{align}
For the convenient of the readers, we also list the inverse of \eqref{coordtransfExtZ}-\eqref{coordtransfExtx1}
\begin{align}
    \rho(\lambda)&={\rm arcsinh}\frac{\sqrt{x_1(\lambda)^2-x_0(\lambda)^2}}{Z(\lambda)}\label{coordtransfExtZInv}\\
    t(\lambda)&={\rm arcsinh}\frac{x_0(\lambda)}{\sqrt{x_1(\lambda)^2-x_0(\lambda)^2}}\label{coordtransfExtx0Inv}\\
    x(\lambda)&=\frac{1}{2}\log\big(Z(\lambda)^2-x_0(\lambda)^2+x_1(\lambda)^2\big)\label{coordtransfExtx1Inv}
\end{align}
The coordinate transformation between BTZ interior and Poincare AdS$_3$ are
\begin{align}
    Z&=e^x{\rm sec}\alpha\label{coordtransfIntZ}\\
    x_0&=e^x{\rm cosh}\Tilde{t}\ {\rm tan}\alpha\label{coordtransfIntx0}\\
    x_1&=e^x{\rm sinh}\Tilde{t}\ {\rm tan}\alpha\label{coordtransfIntx1}
\end{align}
the inverses of \eqref{coordtransfIntZ}-\eqref{coordtransfIntx1}
\begin{align}
    \alpha(\lambda)&={\rm arcsin}\frac{\sqrt{x_0(\lambda)^2-x_1(\lambda)^2}}{Z(\lambda)}\label{coordtransfIntZInv}\\
    \Tilde{t}(\lambda)&={\rm arcsinh}\frac{x_1(\lambda)}{\sqrt{x_0(\lambda)^2-x_1(\lambda)^2}}\label{coordtransfIntx0Inv}\\
    x(\lambda)&=\frac{1}{2}{\rm log}\big(Z(\lambda)^2-x_0(\lambda)^2+x_1(\lambda)^2\big)\label{coordtransfIntx1Inv}
\end{align}
Notice that the expression of $x(\lambda)$ is the same in the exterior and interior. 

Under the above coordinate transformations, the BTZ black hole horizon $\rho=\alpha=0$ is mapped to $|x_0|=|x_1|$; the exterior region is mapped to $x_0^2-x_1^2\leq 0$; the BTZ black hole interior is mapped to the region $0\leq x_0^2-x_1^2\leq Z^2$ enclosed by the two planes $|x_0|=|x_1|$ and the hyperboloid $Z^2=x_0^2-x_1^2$. Notice that $z\to\infty$ in BTZ \eqref{BTZmetric} is mapped to the hyperboloid $Z^2=x_0^2-x_1^2$ in Poincare AdS$_3$ \eqref{AdSMetric}. See Fig.~\ref{fig:BTZAdS3Map} for a demonstration. 

The above BTZ$\to$AdS     $_3$ mapping allows us to map the problem of interest in BTZ to that in AdS$_3$, where geodesic equations avoid the complexity due to the presence of horizons. In the next subsection, we will find the most general geodesic equations in Poincare AdS$_3$. 

\subsection{Spacelike Geodesics in Poincare AdS$_3$}\label{AdSGeod}
The Poincare AdS$_3$ metric \eqref{AdSMetric} indicates that there are two conserved quantities, the Energy and Momentum of the geodesic 
\begin{align}
    E=\frac{1}{Z^2}\frac{dx_0}{d\lambda} && P=\frac{1}{Z^2}\frac{dx_1}{d\lambda}\label{conservedQuantities}
\end{align}
where $\lambda$ is the affine parameter along the geodesic. For spacelike geodesics, the norm of the tangent vector is 1, so we have
\begin{align}
    -\frac{1}{Z^2}\Big(\frac{dx_0}{d\lambda}\Big)^2+\frac{1}{Z^2}\Big(\frac{dZ}{d\lambda}\Big)^2+\frac{1}{Z^2}\Big(\frac{dx_1}{d\lambda}\Big)^2=1\label{normTan}
\end{align}
Plug \eqref{conservedQuantities} into \eqref{normTan} then yields the differential equation 
\begin{align}
    \Big(\frac{dZ}{d\lambda}\Big)^2=Z^2\big(1+(E^2-P^2)Z^2\big)\label{ZLambda0}
\end{align}
We are interested in spacelike geodesics that reach some deepest points $Z_0$, where $\frac{dZ}{d\lambda}=0$. If $E^2-P^2>0$, then $\frac{dZ}{d\lambda}\neq 0$ for $\forall\ Z$; if instead $E^2-P^2<0$, then 
\begin{align}
    Z_0=\frac{1}{\sqrt{P^2-E^2}} && Z\in [0,Z_0]
\end{align}
And we can solve the equation \eqref{ZLambda0}
\begin{align}
    \lambda(Z)=\pm{\rm arctanh}\sqrt{1-(P^2-E^2)Z^2}    \label{ZLambda2}
\end{align}
At the AdS$_3$ boundary $Z=1$, $\lambda=\pm\infty$. We choose conventions such that $\lambda=+\infty$ and $\lambda=-\infty$ are at the right and left Rindler wedge boundary points, respectively. At the furthest point $Z_0$, $\lambda=0$. Inverting \eqref{ZLambda2} then gives 
\begin{align}
    Z(\lambda)=\frac{1}{\sqrt{P^2-E^2}}{\rm sech}\lambda\label{zLambda}
\end{align}
Plug \eqref{zLambda} into \eqref{conservedQuantities} and integrate, we have
\begin{align}
    x_0(\lambda)&=\frac{E}{P^2-E^2}{\rm tanh}\lambda+C_0\label{x0Lambda}\\
    x_1(\lambda)&=\frac{P}{P^2-E^2}{\rm tanh}\lambda+C_1\label{x1Lambda}
\end{align}
where $C_0$ and $C_1$ are integration constants that can be determined by the geodesic endpoints on the AdS$_3$ boundary. 

The separations in $x_0$ and $x_1$ determine $E$ and $P$ via 
\begin{align}
    \Delta x_0&=2\int_0^\infty\frac{dx_0}{d\lambda}d\lambda=2E\int_0^\infty Z^2 d\lambda=2E\int_0^{Z_0}\frac{ZdZ}{\sqrt{1-(P^2-E^2)Z^2}}=\frac{2E}{P^2-E^2}\label{Deltax0}\\
    \Delta x_1&=2\int_0^\infty\frac{dx_1}{d\lambda}d\lambda=2P\int_0^\infty Z^2 d\lambda=2P\int_0^{Z_0}\frac{ZdZ}{\sqrt{1-(P^2-E^2)Z^2}}=\frac{2P}{P^2-E^2}\label{Deltax1}
\end{align}
Where in both equations, in the first step the 2 comes from $\lambda\to -\lambda$ symmetry, in the second step we plugged in \eqref{conservedQuantities}, and in the third step we change the integration variable from $\lambda$ to $Z$ using \eqref{ZLambda0}. From \eqref{Deltax0} and \eqref{Deltax1}, one can then solve for $E$ and $P$ in terms of $\Delta x_0$ and $\Delta x_1$
\begin{align}
    E=\frac{2\Delta x_0}{\Delta x_1^2-\Delta x_0^2} && P=\frac{2\Delta x_1}{\Delta x_1^2-\Delta x_0^2}\label{EPDelta}
\end{align}
The two points $P_1$ and $P_2$ where the geodesic is anchored on the boundary therefore determine the value of $E$ and $P$. 

\subsection{Displaced half spaces}\label{Displaced half spaces}
In this subsection, we find the exact solution to the displaced half space geodesic considered in subsection~\ref{ApproxApproxLag} and Appendix~\ref{ApproxLagDispHalf}. 
\subsubsection{In Poincare AdS$_3$}
\begin{figure}[htbp]
\centering
\includegraphics[width=.5\textwidth]{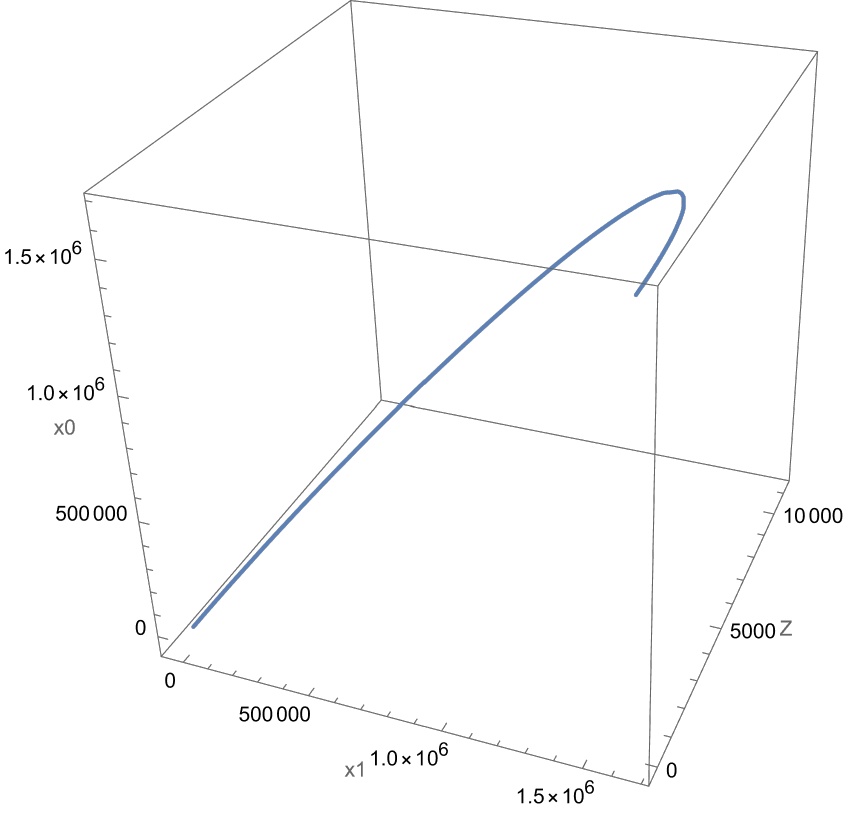}
\qquad
\caption{Spacelike geodesic in Poincare AdS$_3$, with coordinates $(Z,x_0,x_1)$. Here we plot for $T=10$ and $X=5$. \label{fig:PoincareAdS3Geodesic}}
\end{figure}

We have a spacelike geodesic anchoring on the two boundary points in AdS$_3$
\begin{align}
    P_1=(e^X{\rm sinh}T,\ e^X{\rm cosh}T) && P_2=(e^{-X}{\rm sinh}T,\ -e^{-X}{\rm cosh}T)\label{P1P2}
\end{align}
therefore, 
\begin{align}
    \Delta x_0=2\sinh X \sinh T && \Delta x_1=2\cosh X \cosh T \label{deltax0x1}
\end{align}
The explicit values of $E$ and $P$ can be determined by plugging \eqref{deltax0x1} into \eqref{EPDelta}. Further substituting these $E$ and $P$ into \eqref{zLambda}, \eqref{x0Lambda} and \eqref{x1Lambda}, fixes $Z(\lambda)$ and leaves $x_0(\lambda)$ and $x_1(\lambda)$ with $C_0$ and $C_1$ unknown. $C_0$ and $C_1$ can be evaluated by noting that the geodesic is anchored on the point $P_1$ on the AdS$_3$ boundary, 
\begin{align}
    x_0(+\infty)={\rm sinh}T && x_1(+\infty)={\rm cosh}T
\end{align}
Put everything together, we have the following parametric curves
\begin{align}
    Z(\lambda)&=\sqrt{\frac{\cosh 2T+\cosh 2X}{2}}\ {\rm sech}\lambda\label{Z(lambda)}\\
    x_0(\lambda)&=\sinh T(\sinh X\tanh\lambda+\cosh X)\label{x0(lambda)}\\
    x_1(\lambda)&=\cosh T(\cosh X\tanh\lambda+\sinh X)\label{x1(lambda)}
\end{align}
where $\lambda$ is the affine parameter along the geodesic. At $\lambda=0$, the geodesic reaches its deepest point $Z_0$; at $P_1$ and $P_2$, $\lambda=\pm\infty$, respectively. When the geodesic is at the horizon, $x_0=\pm x_1$, one finds $\lambda_{{\rm hor}}=\pm T-X$. Therefore, the interior range of the geodesic correspond to $\lambda\in(-T-X,T-X)$, and the exterior range $\lambda\in(-\infty,-T-X)\cup (T-X,+\infty)$. A plot of the parametric curve \eqref{Z(lambda)}, \eqref{x0(lambda)}, \eqref{x1(lambda)} in Poincare AdS$_3$ can be found in Fig.~\ref{fig:PoincareAdS3Geodesic}. 

When $X=0$, the above parametric curve \eqref{Z(lambda)}, \eqref{x0(lambda)} and \eqref{x1(lambda)}  becomes 
\begin{align}
    Z(\lambda)=\cosh T\ {\rm sech}\lambda && x_0(\lambda)=\sinh T && x_1(\lambda)=\cosh T\ \tanh \lambda \label{semicircleHM0}
\end{align}
which is the semi-circle 
\begin{align}
    Z^2+x_1^2=\cosh^2T && x_0=\sinh T\label{semicircleHM}
\end{align}
studied in~\cite{Hartman:2013qma}. The furthest point this semi-circle can reach in Poincare AdS$_3$ is $Z_0=\cosh T$. 




\subsubsection{In planar BTZ}
By applying the inverse coordinate transformation \eqref{coordtransfIntZInv}-\eqref{coordtransfIntx1Inv}, one can map the spacelike geodesic \eqref{Z(lambda)}, \eqref{x0(lambda)} and \eqref{x1(lambda)} into the BTZ interior \footnote{\eqref{zBTZLambda} is obtained from 
\begin{align}
    \alpha(\lambda)&=\arcsin\left(\sqrt{\frac{\cosh 2 T-\cosh (2 (\lambda +X))}{\cosh 2 T+\cosh 2 X}}\right)\label{alphaLambda}
\end{align} using \eqref{zAlpha}.}
\begin{align}
    z(\lambda)&=\sqrt{\frac{\cosh 2 T+\cosh 2 X}{\cosh (2 (\lambda +X))+\cosh 2 X}}\label{zBTZLambda}\\
    \Tilde{t}(\lambda)&={\rm arcsinh} \left(\frac{\sqrt{2} \cosh T \sinh (\lambda +X)}{\sqrt{\cosh 2T-\cosh (2 (\lambda +X))}}\right)\label{TildeTLambda}\\
   x(\lambda)&=\frac{1}{2} \log (\text{sech}\lambda \cosh (\lambda +2 X))\label{xLambda}
\end{align}
Notice that $\lambda$ is still the affine parameter in Poincare AdS$_3$, although the geodesic is mapped to BTZ. $z(\lambda)$ \eqref{zBTZLambda} peaks at $\lambda=-X$,
\begin{align}
    z_0=z(-X)=\sqrt{\frac{\cosh 2 T+\cosh 2 X}{1+\cosh 2 X}}\approx e^{T-X}\label{z0X}
\end{align}
where $\Tilde{t}=0$. Like in the radial $X=0$ case, $z_0$ grows exponentially with $T$ for large $T$. When $X=0$, \eqref{z0X} becomes $z_0={\rm cosh}T$, which is the same as $Z_0$ in AdS$_3$. From \eqref{xLambda} one have $x(+\infty)=X$ and $x(-\infty)=-X$. Notice that $x(\lambda)$ does not depend on $T$. 

The complicated parametric curves \eqref{zBTZLambda}, \eqref{TildeTLambda} and \eqref{xLambda} in the interior regions $-T-X<\lambda<T-X$  simplify greatly in the large $T$, large $X$ ($T>X$) limit. one can show that $\Tilde{t}$ is $\emph{linear}$ in $\lambda$, 
\begin{align}
    \Tilde{t}(\lambda)=\lambda+X\label{TildeTApprox}
\end{align}
and $z(\lambda)$ behaves as 
\begin{equation}
\label{zBTZApprox}
    z(\lambda)=
    \left\{
    \begin{aligned}
        &e^{T-X-\lambda},&&\lambda\in (0,T-X)\\
        &e^{T-X},&&\lambda\in (-2X,0)\\
        &e^{T+X+\lambda},&&\lambda\in (-T-X,-2X)
    \end{aligned}
    \right.
\end{equation}
which is at a $\emph{constant}$ value $z_0=e^{T-X}$ for $\lambda$ between $(-2X,0)$. This value is the $\emph{maximum}$ value of $z(\lambda)$, i.e. the deepest point $z_0$ \eqref{z0X} the geodesic can reach in the BTZ interior. Moreover, $x(\lambda)$ is $\emph{piecewise\ linear}$ in $\lambda$
\begin{equation}
\label{xApprox}
    x(\lambda)=
    \left\{
    \begin{aligned}
        &X,&&\lambda\in (0,T-X)\\
        &\lambda+X,&&\lambda\in (-2X,0)\\
        &-X,&&\lambda\in (-T-X,-2X)
    \end{aligned}
    \right.
\end{equation}
In other words, almost all the displacements in the $x$ direction occurred in the region $\lambda\in[-2X,0]$ linearly with $\lambda$, whereas in the complementary region of $\lambda$ (which includes all the exterior) there are nearly no displacements. 

From \eqref{zBTZApprox} and \eqref{xApprox}, it can be observed that when $\lambda\in(-2X,0)$, $z(\lambda)$ stays at $z_0$ while $x(\lambda)$ varies linearly; in the complementary interior region $\lambda\in(-T-X,-2X)\cup (0,T-X)$, $z(\lambda)$ varies non-trivially while $x(\lambda)$ stays at constant values. Combining with the fact that $\Tilde{t}(\lambda)=\lambda+X$ is linear in $\lambda$ for all interior values $\lambda\in(-T-X,T-X)$, one can eliminate $\lambda$ and write down the expression $z(\Tilde{t})$ and $x(\Tilde{t})$ in the scaling limit. Noticing that for large $z$ there is asymptotically no difference between Schwarzschild and infalling time \eqref{SchApproxInf}, and making the change of variable $z=e^{\xi}$, one arrives at \eqref{FinalXi} in the main text. 

By applying the inverse coordinate transformation \eqref{coordtransfExtZInv}-\eqref{coordtransfExtx1Inv}, one can map the spacelike geodesic \eqref{Z(lambda)}, \eqref{x0(lambda)} and \eqref{x1(lambda)} into the BTZ exterior 
\footnote{\eqref{zBTZLambdaExt} is obtained from 
\begin{align}
    \rho&={\rm arccosh}\left(\sqrt{\frac{\cosh (2 (\lambda +X))+\cosh 2 X}{\cosh 2 T+\cosh 2 X}}\right)
\end{align} using \eqref{zRho}.}
\begin{align}
    z(\lambda)&=\sqrt{\frac{\cosh 2 T+\cosh 2 X}{\cosh (2 (\lambda +X))+\cosh 2 X}}\label{zBTZLambdaExt}\\
    t(\lambda)&={\rm arcsinh} \left(\frac{\sqrt{2} \sinh T \cosh (\lambda +X)}{\sqrt{\cosh (2 (\lambda +X))-\cosh 2 T}}\right)\label{tLambdaExt}\\
    x(\lambda)&=\frac{1}{2} \log (\text{sech}\lambda \cosh (\lambda +2 X))\label{xLambdaExt}
\end{align}
Notice that the expression of $z(\lambda)$ \eqref{zBTZLambdaExt} and $x(\lambda)$ \eqref{xLambdaExt} are the same as those in the BTZ black hole interior \eqref{zBTZLambda} and \eqref{xLambda}, respectively. It is just that the range of $\lambda$ are different. 

Under the scaling limit, the parametric curves \eqref{zBTZLambdaExt}, \eqref{tLambdaExt} and \eqref{xLambdaExt} in the exterior regions $\lambda\in(-\infty,-T-X)\cup (T-X,+\infty)$ also simplify greatly.  
\begin{equation}
\label{TApproxExt}
    t(\lambda)=
    \left\{
    \begin{aligned}
        &T,&&\lambda\in (T-X,+\infty)\\
        &-T,&&\lambda\in (-\infty,-T-X)
    \end{aligned}
    \right.
\end{equation}
where we have used the fact that for geodesics in the thermal double, i.e. when $\lambda\in(-\infty,-T-X)$, ${\rm Im}\ t=-i\pi$~\cite{Klebanov:1999tb}; we can therefore let $t\to t+i\pi$ in the thermal double, like introducing $\Tilde{t}=t+\frac{i\pi}{2}$ \eqref{zAlpha} in the interior. \eqref{TApproxExt} indicates that in the exterior regions, the HRT surface is on a $\emph{constant}$-$t$ $\emph{slice}$ $t=T$ ($t=-T$ in the thermal double) apart from the near-horizon divergence. For $z(\lambda)$ and $x(\lambda)$, we have 
\begin{equation}
\label{zBTZApproxExt}
    z(\lambda)=
    \left\{
    \begin{aligned}
        &e^{T-X-\lambda},&&\lambda\in (T-X,+\infty)\\
        &e^{T+X+\lambda},&&\lambda\in (-\infty,-T-X)
    \end{aligned}
    \right.
\end{equation}
\begin{equation}
\label{xApproxExt}
    x(\lambda)=
    \left\{
    \begin{aligned}
        &X,&&\lambda\in (T-X,+\infty)\\
        &-X,&&\lambda\in (-\infty,-T-X)
    \end{aligned}
    \right.
\end{equation}
which are the same as $z(\lambda)$ \eqref{zBTZApprox} and $x(\lambda)$ \eqref{xApprox} when $\lambda\in(0,T-X)$ and $\lambda\in(-T-X,-2X)$ in the interior, respectively.


When $T<X$. The expressions of $z(\lambda)$, $\Tilde{t}(\lambda)$, $t(\lambda)$ and $x(\lambda)$  are the same as \eqref{zBTZLambda}, \eqref{TildeTLambda}, \eqref{tLambdaExt}, and \eqref{xLambda}, respectively, but the scaling of $z(\lambda)$ is different. 
\begin{equation}
\label{zBTZApproxTleqX}
    z(\lambda)=
    \left\{
    \begin{aligned}
        &e^{-\lambda},&&\lambda\in (0,\infty)\\
        &1,&&\lambda\in (-2X,0)\\
        &e^{T+X+\lambda},&&\lambda\in (-\infty,-2X)
    \end{aligned}
    \right.
\end{equation}
\eqref{zBTZApproxTleqX} then indicates that $z$ reaches just to the horizon instead of going deeper into the interior. The geodesic moves transversely in $x$ when at the plateau $z=1$. 

\begin{figure}[htbp]
\centering
\includegraphics[width=.4\textwidth]{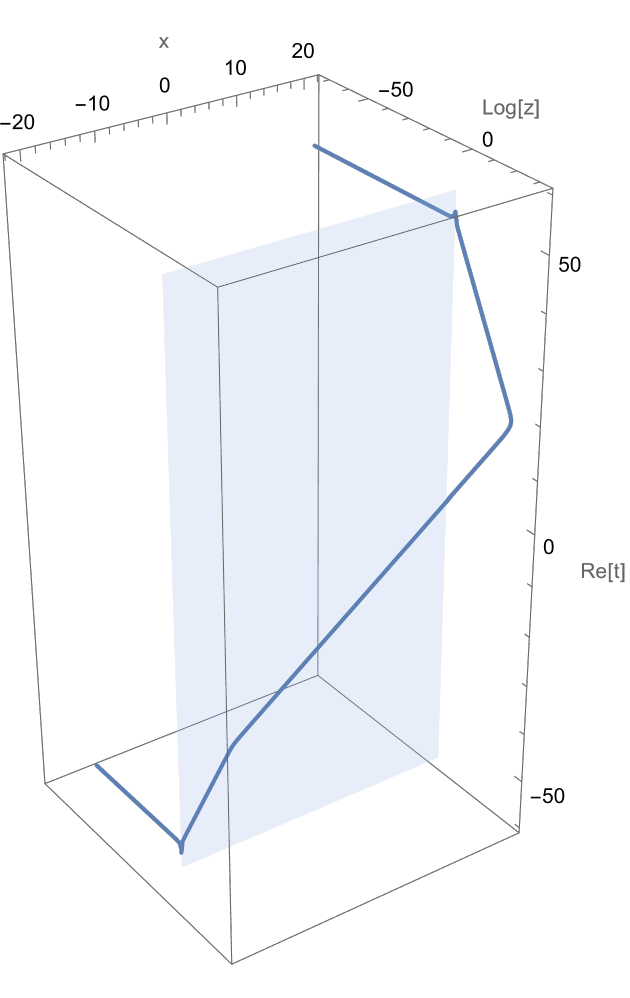}
\includegraphics[width=.42\textwidth]{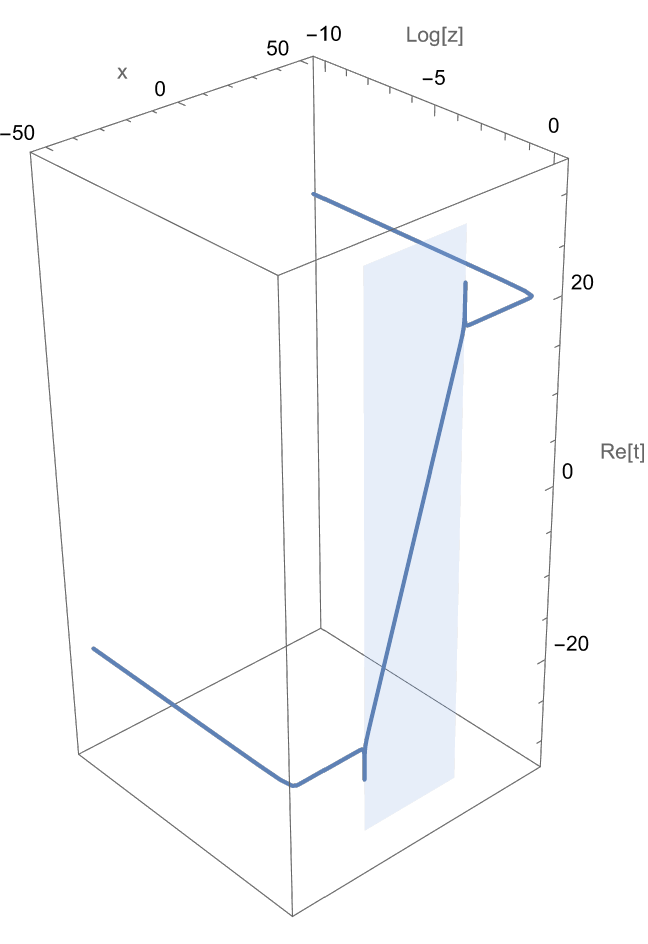}
\qquad
\caption{Spacelike geodesic in BTZ in the coordinates $(\log z,\Tilde{t},x)$. Here we plot for $T=50$ and $X=20$ in the left figure and $T=20$ and $X=50$ in the right figure. The light blue surface is the black hole horizon. \label{fig:BTZGeodesic}}
\end{figure}

\begin{figure}[htbp]
\centering
\includegraphics[width=.45\textwidth]{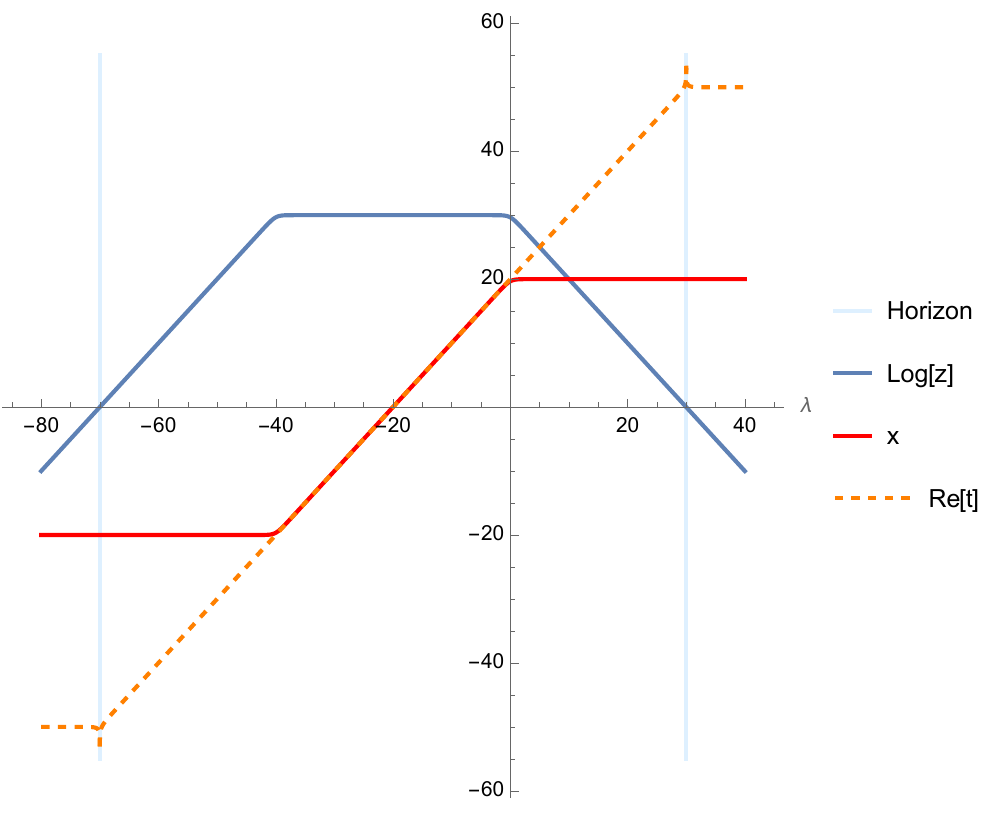}
\includegraphics[width=.45\textwidth]{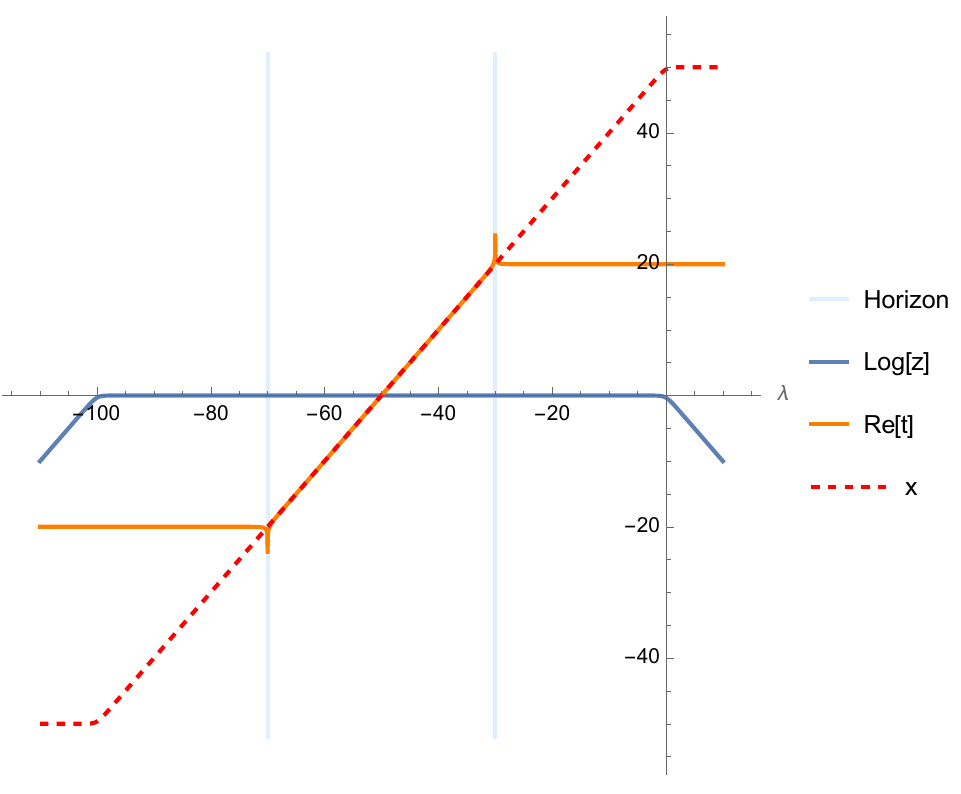}
\qquad
\caption{$\emph{Left}$: Plot of $\log z(\lambda)$ (blue), $\Tilde{t}(\lambda)$ or $t(\lambda)$ (dashed orange) and $x(\lambda)$ (red) when $T=50$ and $X=20$. $\emph{Right}$: Plot of $\log z(\lambda)$ (blue), $\Tilde{t}(\lambda)$ or $t(\lambda)$ (orange) and $x(\lambda)$ (dashed red) when $T=20$ and $X=50$. The light blue lines are the black hole horizons. \label{fig:ztx}}
\end{figure}

Plots of the parametric curves \eqref{zBTZLambda}-\eqref{xLambda}, \eqref{zBTZLambdaExt}-\eqref{xLambdaExt} and \eqref{zBTZApproxTleqX} can be found in Fig.~\ref{fig:BTZGeodesic} and~\ref{fig:ztx}. 

\subsection{Static geodesic}\label{Static geodesic}
To study the static geodesic only, it suffices to work on the $t=0$ slice in BTZ; we will, however, work on the $t=T$ slice for the need of subsection~\ref{SatEWCS}. 

\subsubsection{In Poincare AdS$_3$}
To find the expression of the spacelike geodesic with endpoints $P_2$ and $P_3$, with 
\begin{align}
    P_2=\Big(-\frac{D}{2},T\Big) && P_3=\Big(\frac{D}{2},T\Big)\label{P2P3}
\end{align}
 as a parametric curve in BTZ, we again map it to Poincare AdS$_3$ using the transformations in Appendix~\ref{BTZAdSMap}. The general expression of the spacelike geodesic in AdS$_3$ is the parametric curve \eqref{zLambda}-\eqref{x1Lambda}, but the boundary conditions are now given instead by $P_2$ and $P_3$ \eqref{P2P3} mapped to Poincare AdS$_3$ 
\begin{align}
    P_2=(e^{-\frac{D}{2}}\sinh T,\ e^{-\frac{D}{2}}\cosh T) && P_3=(e^{\frac{D}{2}}\sinh T,\ e^{\frac{D}{2}}\cosh T)\label{P2P3AdS}
\end{align}
which yields 
\begin{align}
    \Delta x_0=2\sinh \frac{D}{2} \sinh T && \Delta x_1=2\sinh \frac{D}{2} \cosh T \label{deltax0x1Static}
\end{align}
Plugging \eqref{deltax0x1Static} into \eqref{EPDelta}, and using the boundary condition \eqref{P2P3AdS}, one finds 
\begin{align}
    Z(\lambda)&=\sinh \frac{D}{2}\ {\rm sech}\lambda\label{Z(lambda)Static}\\
    x_0(\lambda)&=\sinh T\Big(\sinh \frac{D}{2}\tanh\lambda+\cosh \frac{D}{2}\Big)\label{x0(lambda)Static}\\
    x_1(\lambda)&=\cosh T\Big(\sinh \frac{D}{2}\tanh\lambda+\cosh \frac{D}{2}\Big)\label{x1(lambda)Static}
\end{align}
Notice that $x_0(\lambda)< x_1(\lambda)$ due to $\sinh T<\cosh T$, indicating that the geodesic is always in the exterior region. 

\subsubsection{In planar BTZ}
Mapping the geodesic \eqref{Z(lambda)Static}-\eqref{x1(lambda)Static} to BTZ using \eqref{coordtransfExtZInv}-\eqref{coordtransfExtx1Inv}, one finds 
\begin{align}
    z(\lambda)&=\frac{\sinh \frac{D}{2}}{\sqrt{\cosh \lambda \cosh (D+\lambda )}}\label{zLambdaStatic}\\
    x(\lambda)&=\frac{1}{2} \log (\text{sech}\lambda  \cosh (D+\lambda ))\label{xLambdaStatic}
\end{align}
and $t=T$. In \eqref{zLambdaStatic} and \eqref{xLambdaStatic} $\lambda\in(-\infty,+\infty)$. $z(\lambda)$ \eqref{zLambdaStatic} peaks at $\lambda=-\frac{D}{2}$: $z(-\frac{D}{2})=\tanh \frac{D}{2}$. As $D$ gets large, $z(-\frac{D}{2})\to 1$. Notice that $x(\lambda)$ takes the same form as \eqref{xLambda} and \eqref{xLambdaExt} with $2 X\to D$. A plot of the full expression \eqref{zLambdaStatic}-\eqref{xLambdaStatic} can be found in Fig.~\ref{fig:StaticGeod}. 

\begin{figure}[htbp]
\centering
\includegraphics[width=.48\textwidth]{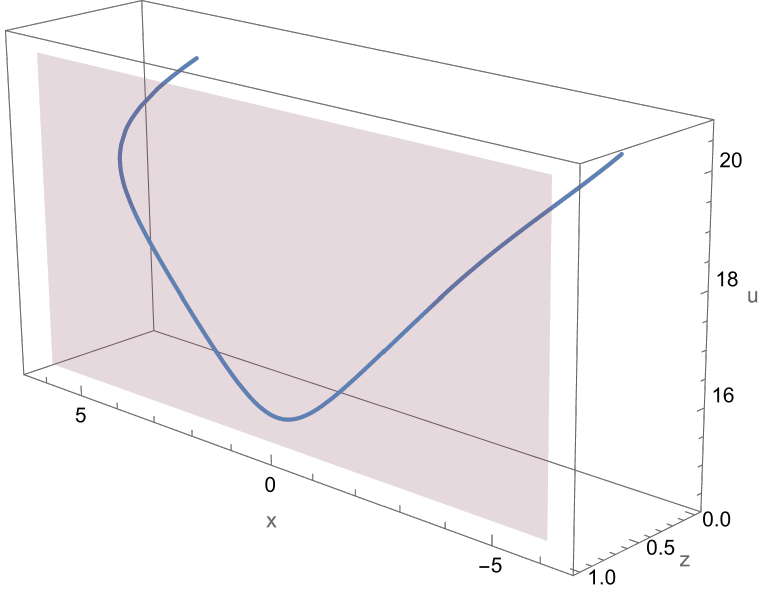}
\qquad
\caption{A 3d plot of the static geodesic given by \eqref{zLambdaStatic}-\eqref{xLambdaStatic} and shown in Fig.~\ref{fig:StaticGeod} in the $(u,z,x)$ coordinate, , for $T=20$ and $D=10$. The light blue surface denotes the black hole horizon $z=1$. The plateau part of the geodesic lies close to the $z=1$ plane.  \label{fig:RTInfalling}}
\end{figure}

As is explained in the main text, there is a finite difference between Schwarzschild and infalling time in the plateau regions due to \eqref{u(z)2dExpansion}. For example, at the deepest point the RT surfaces reach, we have $u(-\frac{D}{2})=T-\frac{D}{2}$. The plot of the static RT surface \eqref{zLambdaStatic}-\eqref{xLambdaStatic} in infalling coordinate can be found in Fig.~\ref{fig:RTInfalling}.

In the scaling limit, 
\begin{equation}
\label{zBTZStaticApprox}
    z(\lambda)=
    \left\{
    \begin{aligned}
        &e^{-\lambda},&&\lambda\in (0,+\infty)\\
        &1-\frac{1}{2}(e^{2\lambda}+e^{-2(\lambda+D)})\approx 1,&&\lambda\in (-D,0)\\
        &e^{\lambda+D},&&\lambda\in (-\infty,-D)
    \end{aligned}
    \right.
\end{equation}
One can then obtain \eqref{vBcone2dApprox} by plugging \eqref{zBTZStaticApprox} into \eqref{u(z)2dExpansion}, and using \eqref{xApprox} with $2 X\to D$ to eliminate $\lambda$. 

According to the BTZ-AdS$_3$ map \eqref{coordtransfExtZ}-\eqref{coordtransfExtx1}, a shift of $\ell+\frac{D}{2}$ in $x$ corresponds to a boost of $e^{\ell+\frac{D}{2}}$ to \eqref{Z(lambda)Static}-\eqref{x1(lambda)Static}, which are \eqref{AcoordZ}-\eqref{Acoordx1} in subsection~\ref{SatEWCS} of the main text. 

\subsection{Entanglement entropy of displaced half space}
In this appendix, we calculate the entanglement entropy computed by the displaced half space geodesic solution found in Appendix~\ref{Displaced half spaces}, using both holographic and CFT approaches. We will see that results from these two methods match each other. 
\subsubsection{Gravity calculation}\label{EEGravCalc}
The entanglement entropy is computed by the regulated length $A_{12}$ of the spacelike geodesic connecting $P_1$ and $P_2$. 
\begin{align}
    A_{12}=\int_{\lambda_2}^{\lambda_1} d\lambda = \lambda_1- \lambda_2
\end{align}
where $\lambda_i$ ($i=1,2$) are the value of $\lambda$ corresponding to the near boundary cutoffs, $Z(\lambda_1)$ and $Z(\lambda_2)$. $\lambda_1\to +\infty$ and $\lambda_2\to -\infty$. In BTZ, the cutoff is at some large radius $\rho=\rho_{{\rm max}}$. Following~\cite{Hartman:2013qma} we introduce the notion $\epsilon=e^{-\rho_{{\rm max}}}$.  When mapped to Poincare AdS$_3$ using \eqref{coordtransfExtZ}, the near-boundary cutoffs $\rho=\rho_{{\rm max}}$ on the two sides are mapped to 
\begin{align}
    Z(\lambda_1)=e^{x(\lambda_1)}\ {\rm sech}{\rho_{{\rm max}}}= 2\epsilon e^{X} && Z(\lambda_2)=e^{x(\lambda_2)}\ {\rm sech}{\rho_{{\rm max}}}=2\epsilon e^{-X}
\end{align}
where we have used $x(+\infty)=X$ and $x(-\infty)=-X$ from \eqref{xLambdaExt}. Therefore, from \eqref{Z(lambda)}, one finds 
\begin{align}
    \lambda_1&=-X+\log \Big(\frac{1}{\sqrt{2}\epsilon}\sqrt{{\rm cosh}2T+{\rm cosh}2X}\Big)\\
    -\lambda_2&= X+\log \Big(\frac{1}{\sqrt{2}\epsilon}\sqrt{{\rm cosh}2T+{\rm cosh}2X}\Big)
\end{align}
and the regulated geodesic length $A_{12}$ is given by 
\begin{align}
    A_{12}=\lambda_1-\lambda_2={\rm log}\Big(\frac{1}{2\epsilon^2}({\rm cosh}2T+{\rm cosh}2X)\Big)\label{A12}
\end{align}
The entanglement entropy is thus 
\begin{align}
    S(T)=\frac{1}{4G_N}A_{12}=\frac{c}{6}\ {\rm log}\Big(\frac{1}{2\epsilon^2}({\rm cosh}2T+{\rm cosh}2X)\Big)\label{HEE}
\end{align}
where we have used the Brown-Henneaux central charge~\cite{Brown:1986nw}. Notice that when $X=0$, $S(T)=\frac{c}{3}\ {\rm log}\big(\frac{{\rm cosh}T}{\epsilon}\big)$, which matches the result in~\cite{Hartman:2013qma}. In the scaling limit, 
\begin{equation}
\label{TApproxExt2}
    S(T)-S_{{\rm div}}=
    \left\{
    \begin{aligned}
        &\frac{c}{3}T,&& T>X\\
        &\frac{c}{3}X,&& T<X
    \end{aligned}
    \right.
\end{equation}
where $S_{{\rm div}}=\frac{c}{3}\log \frac{1}{\epsilon}$. \eqref{TApproxExt2} matches the operator entanglement entropy \eqref{OpEntFinal} ($T>X$) and \eqref{OpEntFinal3} ($T<X$) in the main text. For small $T$, 
\begin{align}
    S(T)\approx \frac{c}{3} \log \cosh X+\frac{c}{3 (\cosh 2X+1)}T^2+S_{{\rm div}}
\end{align}
indicating that the entanglement entropy grows quadratically with $T$, with a coefficient that depends on $X$. At $T=0$, one finds
\begin{align}
    S(T)\approx \frac{c}{3}X+S_{{\rm div}}
\end{align}
for large $X$. 


\subsubsection{CFT calculation}
\begin{figure}[htbp]
\centering
\includegraphics[width=.35\textwidth]{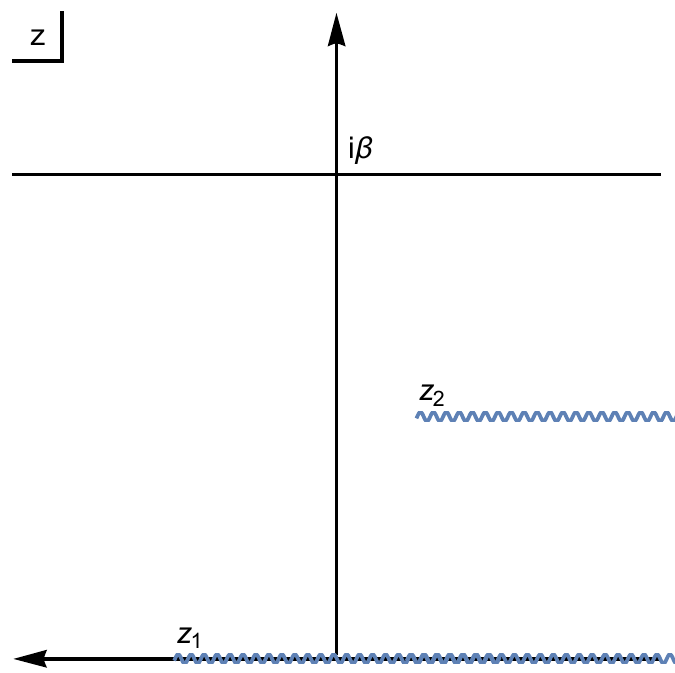}
\qquad
\caption{Insertion of twist operators at $z_1$ and $z_2$ on the Euclidean cylinder. \label{fig:TwistOp}}
\end{figure}

The CFT entanglement entropy is computed by the two-point function of twist operators~\cite{Calabrese:2004eu}
\begin{align}
    \langle \Phi_+(z_1,\overline{z}_1)\Phi_-(z_2,\overline{z}_2)\rangle\label{twopt}
\end{align}
with dimension $\Delta_n=\frac{c}{12}\Big(n-\frac{1}{n}\Big)$. The twist operators are inserted at 
\begin{align}
    z_1=\overline{z}_1=X && z_2=2T-X+\frac{i\beta}{2} && \overline{z}_2=-2T-X-\frac{i\beta}{2}
\end{align}
see Fig.~\ref{fig:TwistOp} for illustrations. 
Mapping to the complex plane, $w=e^\frac{2\pi z}{\beta}$, $\overline{w}=e^\frac{2\pi \overline{z}}{\beta}$, one finds the holomorphic and anti-holomorphic part of \eqref{twopt} are given by 
\begin{align}
    \langle \Phi_+(z_1)\Phi_-(z_2)\rangle&=\Big(\frac{\partial z_1}{\partial w_1}\Big)^{-\frac{\Delta_n}{2}}\Big(\frac{\partial z_2}{\partial w_2}\Big)^{-\frac{\Delta_n}{2}}\frac{1}{|w_1-w_2|}\nonumber\\
    &=\Big(\frac{\beta}{4\pi}\cosh\Big(\frac{2\pi}{\beta}(T-X)\Big)\Big)^{-\Delta_n}\label{holomorphic}\\
    \langle \overline{\Phi}_+(\overline{z}_1)\overline{\Phi}_-(\overline{z}_2)\rangle&=\Big(\frac{\partial \overline{z}_1}{\partial \overline{w}_1}\Big)^{-\frac{\Delta_n}{2}}\Big(\frac{\partial \overline{z}_2}{\partial \overline{w}_2}\Big)^{-\frac{\Delta_n}{2}}\frac{1}{|\overline{w}_1-\overline{w}_2|}\nonumber\\
    &=\Big(\frac{\beta}{4\pi}\cosh\Big(\frac{2\pi}{\beta}(T+X)\Big)\Big)^{-\Delta_n}\label{antiholomorphic}
\end{align}
Taking the product of \eqref{holomorphic} and \eqref{antiholomorphic}, setting $\beta=2\pi$ and introducing the cutoff, one finds 
\begin{align}
    \langle \Phi_+(z_1,\overline{z}_1)\Phi_-(z_2,\overline{z}_2)\rangle=\Big(\frac{1}{2\epsilon^2}\big(\cosh 2T+\cosh 2X\big)\Big)^{-\Delta_n}
\end{align}
The entanglement entropy is then obtained by $S(t)=-\partial_n I_n|_{n=1}$, which is precisely \eqref{HEE}.

\section{Exact solutions of EWCS in BTZ black hole}\label{ExactEWCS}
In this Appendix, we find exact solutions to the dynamics of entanglement wedge cross sections in $d=2$, adopting the setup in section~\ref{sec:2dEWCS} in the main text. 
\subsection{Non-equilibrium EWCS}
\begin{figure}[htbp]
\centering
\includegraphics[width=.4\textwidth]{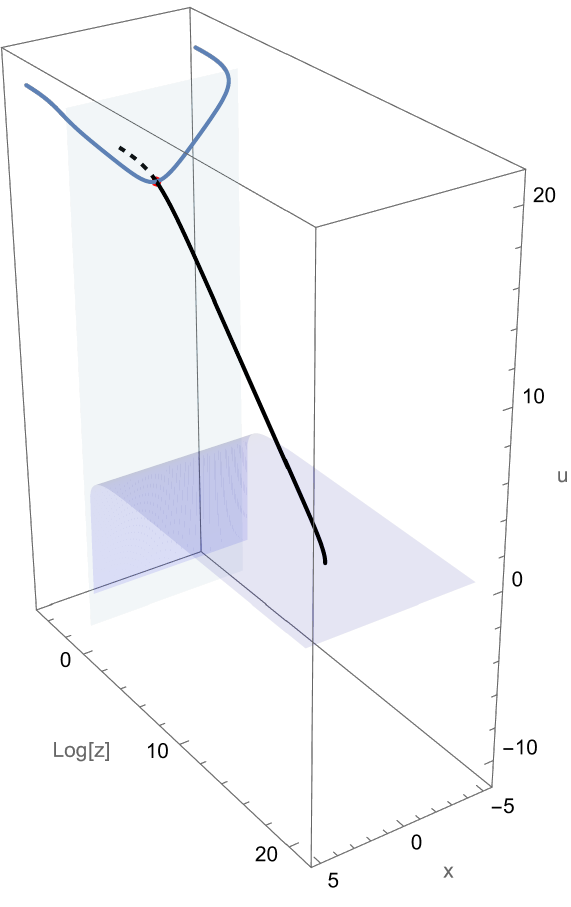}
\qquad
\caption{Plot of non-equilibrium EWCS in $d=2$ with BTZ coordinates $(u,\log(z),x)$, for $T=20$ and $D=10$. The static RT surface (blue) stays close to the horizon $z=1$ (vertical light green surface) before shooting exponentially towards the boundary $z=0$ (didn't plot). The saturated EWCS (black) starts from the static RT surface at the midpoint (red dot) of the latter, and then enters the interior where its length grows linearly according to \eqref{lEWCSNeqApprox}. The non-equilibrium EWCS ends perpendicularly on the EoW brane (light blue surface), whose location is given by $\tilde{t}=0$ or equivalently $u=-{\rm arccoth}z\to 0$ for large $z$. Notice that if one continues the EWCS to the boundary (black dashed line), it anchors on the $T-\frac{D}{2}$ line, instead of the $T$ line that the static RT surface ends. This $-\frac{D}{2}$ difference ensures that the EWCS growth is continuous in time in the scaling limit. \label{fig:NeqEWCS}}
\end{figure}

The time depending EWCS starts from the static RT surface $P_2P_3$ and ends on the end of the world brane. By symmetry, it starts from the mid-point $(t,x)=(T,\tanh\frac{D}{2})$ of $P_2P_3$. As all spacelike geodesics in AdS$_3$ reach the conformal boundary~\cite{Hubeny:2012ry}, we can imagine the EWCS as the part of a spacelike geodesic between $\big(T,\tanh\frac{D}{2}\big)$ and the EoW brane. Let us suppose that this fictitious geodesic starts from the $t=T'$ line on the boundary. As $P_1$ does not play any roles in this part of calculation, we can choose a coordinate such that this fictitious geodesic is at $x=0$. The exterior part of this fictitious geodesic is then given by \eqref{zBTZLambdaExt} and \eqref{xLambdaExt} with $X=0$
\begin{align}
    z(\lambda)&=\cosh T' \text{sech}\lambda\\
    \sinh t(\lambda)&=\frac{\sinh T'}{\sqrt{1-\cosh ^2T' \text{sech}^2\lambda  }}
\end{align}
As the fictitious geodesic passes the mid-point/furthest point of the static geodesic $\big(T,\tanh\frac{D}{2}\big)$, we have 2 equations
\begin{align}
    \tanh\frac{D}{2}&=\cosh T' \text{sech}\lambda\label{EqEWCS1}\\
    \sinh T&=\frac{\sinh T'}{\sqrt{1-\cosh ^2T' \text{sech}^2\lambda  }}\label{EqEWCS2}
\end{align}
which allows us to solve for $T'$ and $\lambda$ in terms of $T$ and $D$. Plugging \eqref{EqEWCS1} into \eqref{EqEWCS2}, we have 
\begin{align}
    T'&=\text{arcsinh}\Big(\sinh T \text{sech}\frac{D}{2}\Big)\label{tb'exact}\\
    &\approx T-\frac{D}{2}+\log 2\label{tb'}
\end{align}
where in getting \eqref{tb'}, we have taken the large $T$, large $D$ limit. This means that the fictitious geodesic emanates from the $t=T-\frac{D}{2}+\log 2$ line on the conformal boundary. Plugging \eqref{tb'exact} back into \eqref{EqEWCS1}, we can solve for $\lambda$. The EWCS intersects the EoW brane at the point of time reflection symmetry $\Tilde{t}=0$. According to \eqref{TildeTLambda} with $X=0$, this is at $\lambda=0$. Therefore, the length of EWCS is simply given by $\lambda$

\begin{align}
    \ell(\text{EWCS})&=\text{arccosh}\left(\coth \frac{D}{2} \sqrt{\frac{\cosh D+\cosh 2T}{\cosh D+1}}\right)\label{lEWCSNeq}\\
    &\approx T-\frac{D}{2}+\log 2\label{lEWCSNeqApprox}
\end{align}
which grows linearly with $T$ with slope $v_E=1$, see Fig.~\ref{fig:EWCSDyn}. Notice that the EWCS starts to be non-zero at $t={\rm arccosh}\sinh\frac{D}{2}\approx\frac{D}{2}$. Therefore, apart from the small ${\rm arccosh}\sqrt{2}$ factor that is negligible in the scaling limit, the EWCS growth is actually $\emph{continuous}$ in time. This is because the fictitious geodesic starts on the $t=T-\frac{D}{2}$ line \eqref{tb'} instead of $t=T$ line on the boundary. See Fig.~\ref{fig:NeqEWCS} for a plot of the non-equilibrium EWCS. 

As a consistency check, taking the $D\to 0$ limit of \eqref{lEWCSNeq}, one finds 
\begin{align}
    \ell(\text{EWCS})\approx\log\Big(\frac{4}{D}\cosh T\Big)
\end{align}
which reduces to the length of the spacelike geodesic computing the time-depending holographic entanglement entropy~\cite{Hartman:2013qma}. Here, the small $D$ can be interpreted as proportional to the UV cutoff $\epsilon$. 

\subsection{Saturated Entanglement Wedge Cross Section}\label{SatEWCS}
The saturated EWCS starts from the point $A$ on the static geodesic $P_2P_3$ and ends on the point $B$ on the radial geodesic ending on the EoW brane. Again, it is most straightforward to map this calculation to Poincare AdS$_3$. Notice that the static geodesic $P_2P_3$ is just the parametric curve \eqref{zLambdaStatic}-\eqref{xLambdaStatic} with a $\emph{shift}$ of $\ell+\frac{D}{2}$ in the $x$ coordinate \eqref{xLambdaStatic}; in AdS$_3$, this shift of $\ell+\frac{D}{2}$ in $x$ corresponds to a $\emph{boost}$ of $e^{\ell+\frac{D}{2}}$ to $\big(Z(\lambda),x_0(\lambda),x_1(\lambda)\big)$, see \eqref{coordtransfExtZ}-\eqref{coordtransfExtx1}. Therefore, in Poincare AdS$_3$, the coordinates of $A$ on the static geodesic $P_2P_3$ are given by (see Appendix~\ref{Static geodesic})
\begin{align}
    Z(\lambda)&=e^{\ell+\frac{D}{2}} \sinh \frac{D}{2}\  \text{sech}\lambda\label{AcoordZ}\\
    x_0(\lambda)&=e^{\ell+\frac{D}{2}} \sinh T\ \text{sech}\lambda \cosh \Big(\lambda +\frac{D}{2}\Big)\label{Acoordx0}\\
    x_1(\lambda)&=e^{\ell+\frac{D}{2}} \cosh T\ \text{sech}\lambda \cosh \Big(\lambda +\frac{D}{2}\Big)\label{Acoordx1}
\end{align}
As $B$ is on a radial geodesic starts from the boundary point $P_1$ and ends on the EoW brane, its coordinates are simply given by \eqref{semicircleHM0}
\begin{align}
    Z(\mu)&=\cosh T\ \text{sech}\mu\label{BcoordZ}\\
    x_0(\mu)&=\sinh T\label{Bcoordx0}\\
    x_1(\mu)&=\cosh T\tanh\mu\label{Bcoordx1}
\end{align}
where $\mu$ is the affine parameter along the geodesic emanating from $P_1$ and ends on the EoW brane. 
\subsubsection{Length of saturated EWCS}
\begin{figure}[htbp]
\centering
\includegraphics[width=.6\textwidth]{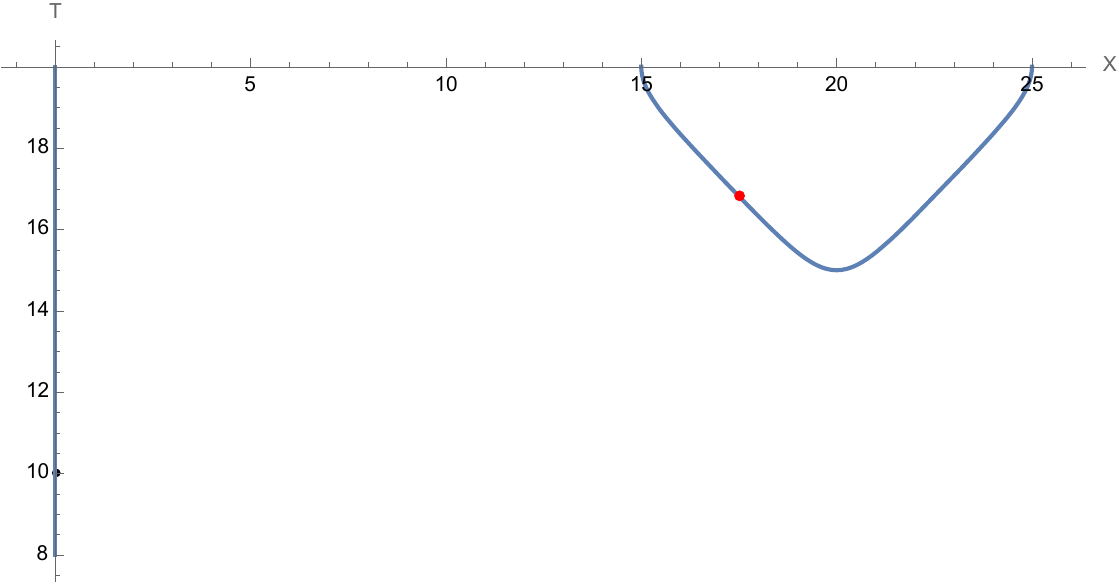}
\qquad
\caption{Locations of $A$ (red) and $B$ (black) upon projections to the boundary along constant infalling time, for $T=20$, $\ell=15$ and $D=10$. \label{fig:ABProj}}
\end{figure}

Now, we would like to calculate the geodesic distance between $A$ and $B$, and find its minimum with respect to the two variables $\lambda\in(-\infty,+\infty)$ and $\mu\in[0,+\infty)$.\footnote{This is because the radial geodesic $P_1B$ ends on the EoW brane, where $\mu=0$} In $2d$, one can find the geodesic distance between two arbitrary points using the geodesic distance formula in Poincare AdS$_3$ below, without knowing the explicit expressions of the geodesic connecting them
\begin{align}
    \cosh d(A,B)=T_1^AT_1^B+T_2^AT_2^B-X_1^AX_1^B-X_2^AX_2^B\label{geodDistAdS3}
\end{align}
where the embedding coordinates are given by 
\begin{align}
    T_1&=\frac{1+(Z^2-x_0^2+x_1^2)}{2 Z}\\
    T_2&=\frac{x_0}{Z}\\
    X_1&=\frac{x_1}{Z}\\
    X_2&=\frac{1-(Z^2-x_0^2+x_1^2)}{2 Z}
\end{align}
Plugging in the coordinates of $A$ \eqref{AcoordZ}-\eqref{Acoordx1} and $B$ \eqref{BcoordZ}-\eqref{Bcoordx1} into the geodesic distance formula \eqref{geodDistAdS3}, one can find the geodesic distance $d(A,B)$. The explicit expression appears to be lengthy and uninspiring, so we will only present results in the scaling limit, 
\begin{align}
    T\to \Lambda \mathfrak{T} && \ell\to \Lambda\mathfrak{l} && d\to\Lambda \mathfrak{D} && \lambda\to\Lambda s  && \mu\to\Lambda m\label{scalingSatEWCS}
\end{align}
where $\Lambda$ is taken to be large. The EWCS length in the scaling limit \eqref{scalingSatEWCS} is  
\begin{align}
    d(A,B)\approx \frac{\ell}{2}+\log 2\label{SatEWCSLen}
\end{align}
which takes place at 
\begin{align}
    \lambda_A\approx -\frac{3}{4}D && \mu_B\approx T-\frac{1}{2}\Big(\ell+\frac{D}{2}\Big)\label{lambdaAmuB}
\end{align}
here $\approx$ means up to exponentially small corrections. Henceforth, the ``equilibrium" EWCS is not truly in equilibrium, instead it asymptotes the extremal value \eqref{SatEWCSLen} exponentially slow. From \eqref{lambdaAmuB}, one finds that 
\begin{align}
    z(\lambda_A)\approx\frac{1}{\sqrt{1+e^{-\frac{D}{2}}}}, && u(\lambda_A)\approx T-\frac{D}{4}-\log 2, && x(\lambda_A)\approx \ell+\frac{D}{4} && \label{AProjCoord}
\end{align}
indicating that $A$ is approximately at the $\emph{middle}$ point of the $v_B$ cone upon projection; and\footnote{Notice that as $z(\mu_B)\gg1$, the difference between Schwarzschild and infalling time at $B$ is negligible, i.e. $u(\mu_B)\approx\Tilde{t}(\mu_B)$.  One then have $\Tilde{t}(\mu_B)\approx\mu_B$ combining \eqref{AProjCoord} and \eqref{BProjCoord}, consistent with \eqref{TildeTApprox} in the radial, i.e. $X=0$ case. }
\begin{align}
    z(\mu_B)\approx e^{\frac{1}{2}\big(\ell+\frac{D}{2}\big)}\gg1, && u(\mu_B)\approx T-\frac{1}{2}\Big(\ell+\frac{D}{2}\Big), && x(\mu_B)=0 \label{BProjCoord} 
\end{align}
implying that $B$ reaches deep into the interior. Notice that the location of $z(B)$ does not time evolve with $T$ to leading order. This is made possible by $B$ traveling towards the horizon since the geodesic goes deeper into the interior as time evolves, as can be seen from \eqref{lambdaAmuB}.\footnote{Notice that at where the geodesic meets the EoW brane, $\Tilde{t}=0$ and therefore $\mu=0$ according to \eqref{TildeTApprox} with $X=0$. Therefore, $\mu_B$ measures the distance between $B$ and the EoW brane along the geodesic. } A plot of the location of $A$ and $B$ upon projections can be found in Fig.~\ref{fig:ABProj}. 

\subsubsection{In Poincare AdS$_3$}
Having computed $l(EWCS)$ \eqref{SatEWCSLen} and the coordinates of $A$ and $B$ \eqref{lambdaAmuB}, we would like to now obtain the expression of the EWCS. Again, we imagine the EWCS as the part of a fictitious spacelike geodesic between $A$ and $B$. The general expression of a spacelike geodesic in AdS$_3$ is given as a parametric curve by \eqref{zLambda}-\eqref{x1Lambda}. Imposing the constraints that it passes point $A$ \eqref{AProjCoord} and $B$ \eqref{BProjCoord}, we have
\begin{align}
    Z(A)&=\frac{1}{\sqrt{\mathfrak{P}^2-\mathfrak{E}^2}}{\rm sech}\nu_A\\
    x_0(A)&=\frac{\mathfrak{E}}{\mathfrak{P}^2-\mathfrak{E}^2}{\rm tanh}\nu_A+\mathfrak{C}_0\\
    x_1(A)&=\frac{\mathfrak{P}}{\mathfrak{P}^2-\mathfrak{E}^2}{\rm tanh}\nu_A+\mathfrak{C}_1\\
    Z(B)&=\frac{1}{\sqrt{\mathfrak{P}^2-\mathfrak{E}^2}}{\rm sech}\nu_B\\
    x_0(B)&=\frac{\mathfrak{E}}{\mathfrak{P}^2-\mathfrak{E}^2}{\rm tanh}\nu_B+\mathfrak{C}_0\\
    x_1(B)&=\frac{\mathfrak{P}}{\mathfrak{P}^2-\mathfrak{E}^2}{\rm tanh}\nu_B+\mathfrak{C}_1
\end{align}
where $\nu$ is the affine parameter along the fictitious geodesic. Now, there are 6 variables $\mathfrak{E}$, $\mathfrak{P}$, $\mathfrak{C}_0$, $\mathfrak{C}_1$, $\nu_A$, and $\nu_B$, as well as 6 constraints \eqref{AProjCoord} and \eqref{BProjCoord}. To solve for these variables, we can proceed by first finding $\tanh \nu_A$ and $\tanh \nu_B$ from the $Z$ equations above.\footnote{It turned out that the physical range of parameters corresponds to $\nu_B<\nu_A<0$.} Substituting them into the $x_0$ and $x_1$ equations, and taking their differences $x_0(\nu_A)-x_0(\nu_B)$ and $x_1(\nu_A)-x_1(\nu_B)$, leaving us with two equations of $\mathfrak{E}$ and $\mathfrak{P}$. We can then solve for all 6 variables. The full expressions are too lengthy to be included in the paper, so we will only discuss their scaling limits \eqref{scalingSatEWCS} and show plots of this fictitious geodesic. 

An interesting feature of this fictitious geodesic is that it hits the hyperboloid $Z^2=x_0^2-x_1^2$ corresponding to the BTZ orbifold singularity $z=\infty$ (see Fig.~\ref{fig:BTZAdS3Map}), and then ends on the Milne patch! Of course, only the portion of the geodesic below the hyperboloid is physical when mapped to BTZ. In BTZ, the fictitious geodesic emanates from a point on the conformal boundary, passes $A$ and $B$, and eventually hits the BTZ singularity. 

In the scaling limit \eqref{scalingSatEWCS}, we found
\begin{align}
    \ell(\text{EWCS})=\nu_A-\nu_B\approx \frac{\ell}{2}+\log 2
\end{align}
where $\nu$ is the affine parameter along the fictitious geodesic. This agrees with \eqref{SatEWCSLen}. 

\subsubsection{In planar BTZ}
\begin{figure}[htbp]
\centering
\includegraphics[width=.55\textwidth]{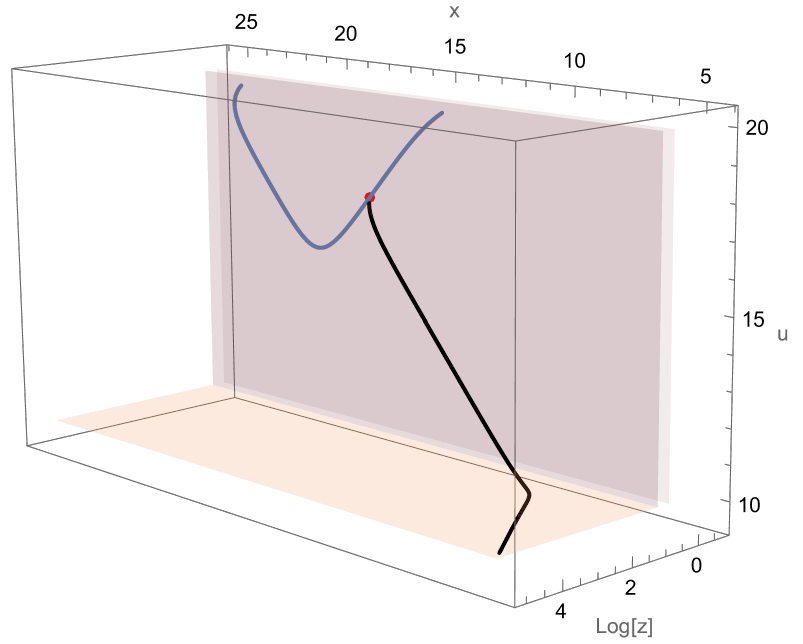}
\qquad
\caption{Plot of saturated EWCS in $d=2$ with BTZ coordinates $(u,\log(z),x)$, for $T=20$, $\ell=15$, and $d=10$. The static RT surface (blue) stays close to $z=1$ (vertical light green surface in the back) before shooting exponentially towards the boundary $z=0$ (didn't plot). The saturated EWCS (black) starts from the static RT surface at point $A$ (red dot), then enters the horizon and stays at $z=\sqrt{2}$ (vertical light blue surface in the front) before becoming almost null and shooting exponentially towards the BTZ singularity $z=\infty$ at constant interior Schwarzschild time $\Tilde{t}=T-\big(\ell+\frac{D}{2}\big)$ (horizontal light red surface). Notice that at large $z$ the difference between Schwarzschild and infalling time is negligible. \label{fig:SatEWCS}}
\end{figure}

When mapped to BTZ, we found that the saturated EWCS shows interesting behavior as depicted in Fig.~\ref{fig:SatEWCS}: it consists of a plateau in the interior plus an almost null part that shoots exponentially towards the interior on a constant $\Tilde{t}$ slice. 

\begin{figure}[htbp]
\centering
\includegraphics[width=.35\textwidth]{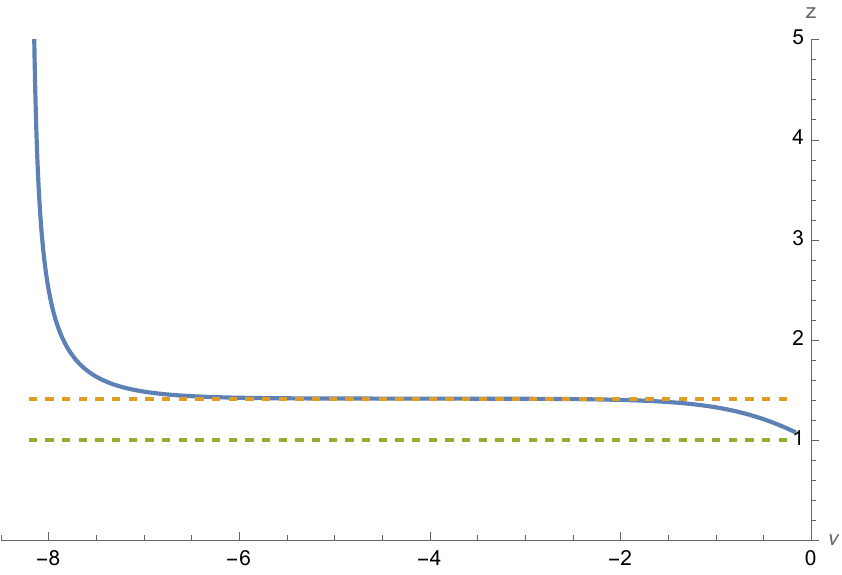}
\includegraphics[width=.28\textwidth]{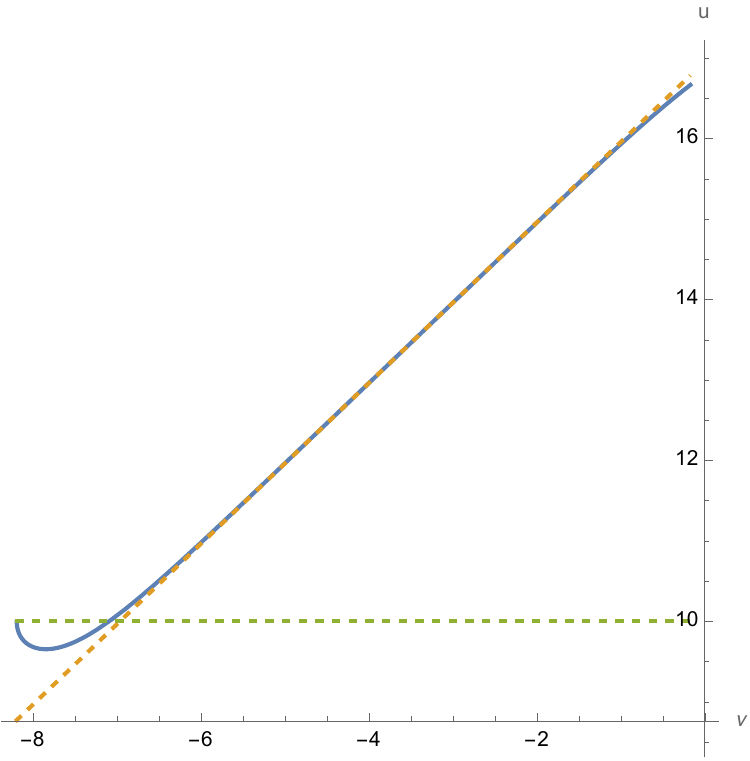}
\includegraphics[width=.28\textwidth]{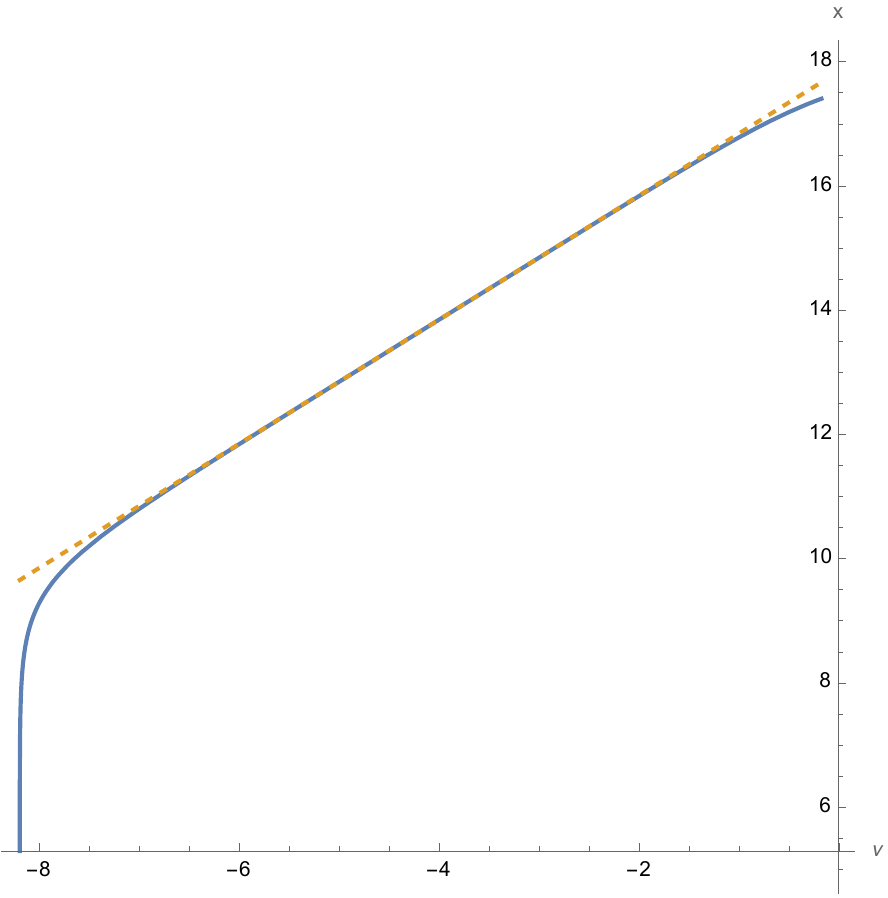}
\qquad
\caption{Plot of $z(\nu)$ (left), $u(\nu)$ (middle) and $x(\nu)$ (right) as well as their scaling limits (dashed), for $T=20$, $\ell=15$, and $D=10$. In the plot of $z(\nu)$, the two dashed lines represent the horizon $z=1$ and $z=\sqrt{2}$, respectively. In the plot of $u(\nu)$, the 45$^\circ$ and horizontal dashed lines stand for \eqref{unu} and the constant interior time slice $\Tilde{t}=T-\big(\ell+\frac{D}{2}\big)$ the null geodesic is on, respectively; the small lump near the horizontal dashed line is due to the difference between Schwarzschild and infalling time \eqref{SchInf} when $z$ is not yet large. In the plot of $x(\nu)$, the dashed line is \eqref{xnu}. \label{fig:zuxNu}}
\end{figure}

The plateau corresponds to $\nu\in(\nu_B,\nu_A]$, excluding a very small region near $\nu\to\nu_B^+$. In this regime we found 
\begin{align}
    z(\nu)&\approx\sqrt{2}\\
    \tilde{t}(\nu)&\approx \nu+T-\frac{D}{4}+\frac{1}{2}\log 2\\
    x(\nu)&\approx \nu+\ell+\frac{D}{4}+\frac{1}{2}\log 2\label{xnu}
\end{align}
and therefore 
\begin{align}
    u(\nu)\approx\tilde{t}(\nu)-{\rm arccoth}z(\nu)=\tilde{t}(\nu)-{\rm arccoth}\sqrt{2}\label{unu}
\end{align}
according to \eqref{SchApproxInf}, see Fig.~\ref{fig:zuxNu}. Putting them together yields two equations \begin{align}
    z&\approx\sqrt{2}\label{zsqrt2}\\
    u(x)&\approx x+T-\ell-\frac{D}{2}-{\rm arccoth}\sqrt{2}\label{x(t)SatEWCS}
\end{align}
describing saturated EWCS. As the range of $\nu$ for the plateau region is almost the entire $[\nu_B,\nu_A]$, the EWCS in this region has length 
\begin{align}
    \ell({\rm Plateau})\approx \nu_A-\nu_B\approx \frac{\ell}{2}+\log 2\approx \ell(\text{EWCS})
\end{align}
In other words, nearly $\emph{all}$ the length of EWCS \eqref{SatEWCSLen} comes from this plateau region. 

\begin{figure}[htbp]
\centering
\includegraphics[width=.5\textwidth]{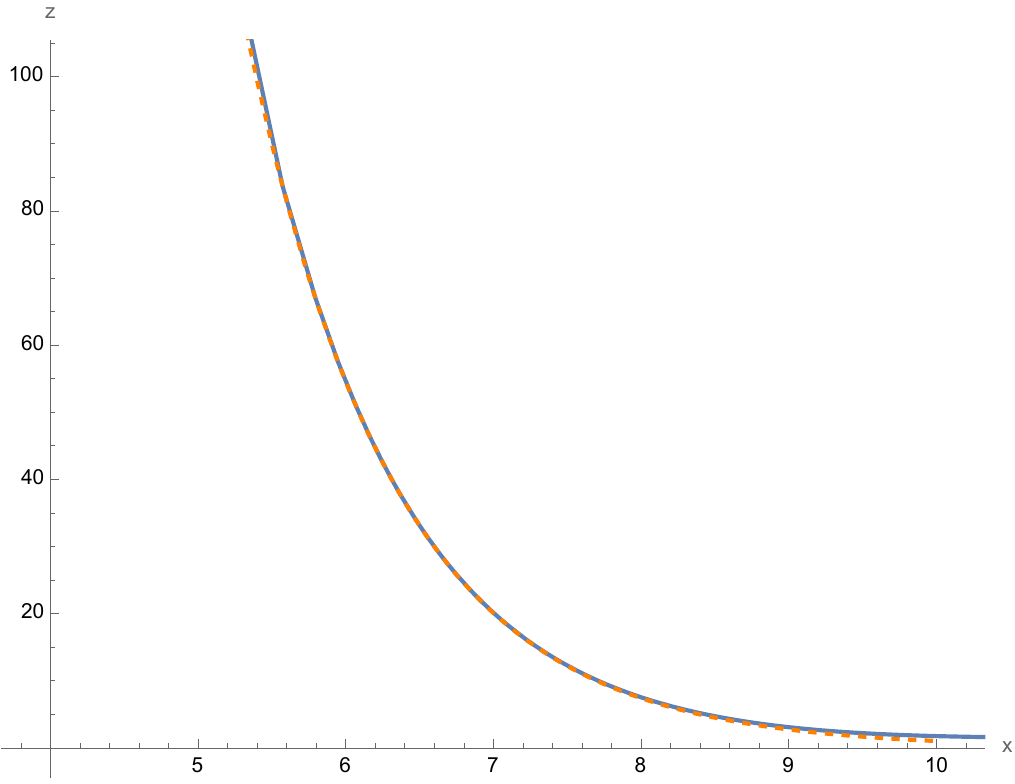}
\qquad
\caption{Exponential relation between $z$ and $|x|$ in the "null" part of EWCS, for $T=20$, $\ell=15$, and $D=10$. The dashed orange line represents $z\propto e^{-x}$. \label{fig:FlyingGeod}}
\end{figure}

The EWCS then transforms smoothly into a almost-$\emph{null}$ part that lies on the $\emph{constant}$ $\Tilde{t}\approx T-\frac{1}{2}\big(\ell+\frac{D}{2}\big)$ $\emph{slice}$ while shooting exponentially towards the BTZ singularity. In fact, the EWCS is a portion of a spacelike geodesic that $\emph{will}$ hit the BTZ singularity $z=\infty$! To understand the relation between $z$ and $x$ for the almost-null geodesic, we demand the metric \eqref{InteriorMetric} to be null on a constant-$\Tilde{t}$ slice, 
\begin{align}
    0=-d\alpha^2+\cos^2 \alpha dx^2 &&  \Rightarrow\ z=\cosh x \propto e^{|x|}\label{zxFlying}
\end{align}
where we have used \eqref{zAlpha}. The $\emph{exponential}$ relation between $z$ and $|x|$ (see Fig.~\ref{fig:FlyingGeod}) is consistent with the coordinates of $A$ and $B$, and the width of the plateau: we have $x(A)\approx \ell+\frac{D}{4}$ according to \eqref{AProjCoord}, and the plateau width is $\Delta x\approx \frac{\ell}{2}$. Therefore, at where the plateau ends and where the exponential growth of $z$ with respect to $x$ starts, we have $x\approx\frac{1}{2}\big(\ell+\frac{D}{2}\big)$. Thus, when the EWCS arrives at $B$, where $x=0$, one expects $z(B)\propto e^{\frac{1}{2}\big(\ell+\frac{D}{2}\big)}$ according to \eqref{zxFlying}. This is indeed the case as in \eqref{BProjCoord}.\footnote{Here we work in the scaling limit \eqref{scalingSatEWCS} and neglect small contributions such as factors of $\log 2$. } 

\begin{figure}[htbp]
\centering
\includegraphics[width=.6\textwidth]{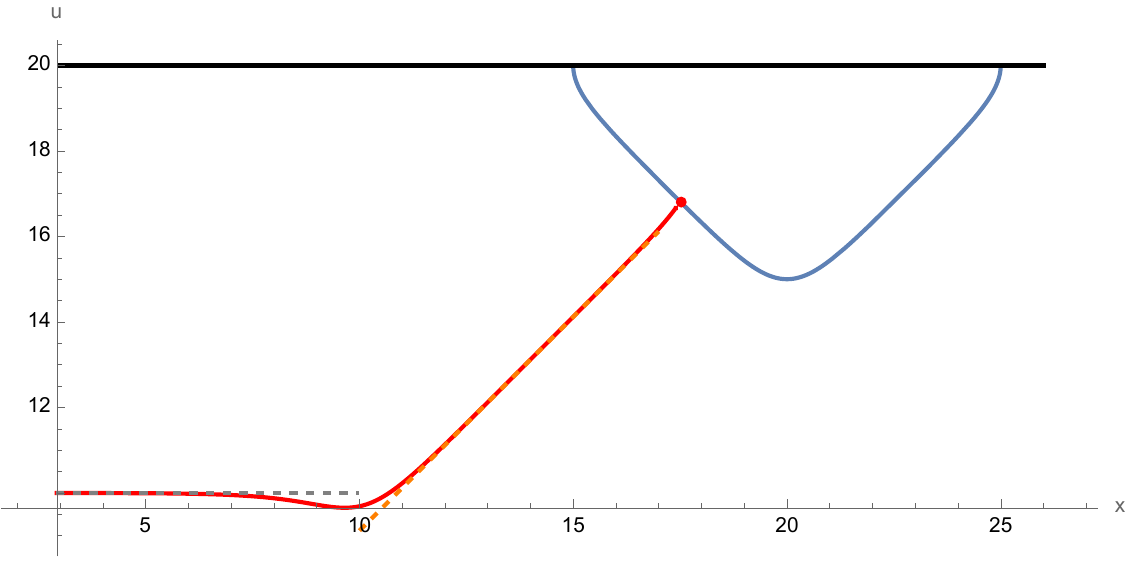}
\qquad
\caption{Projection of saturated EWCS (red) in $d=2$ along constant infalling time, for $T=20$, $\ell=15$, and $D=10$. The blue triangle is the projection of the static RT surface anchoring on $P_2$ and $P_3$ on the boundary. The dashed orange line is the plateau \eqref{x(t)SatEWCS}, which is responsible for almost all the length of the EWCS \eqref{SatEWCSLen}. The horizontal part of the projection is almost null on the 
 $\Tilde{t}=T-\frac{\ell}{2}\big(\ell+\frac{D}{2}\big)$ slice (dashed grey). The lump near the transition point is due to the difference between Schwarschild and infalling time \eqref{SchInf} when $z$ is not yet large.\label{fig:SatEWCSProj}}
\end{figure}

The projection of the saturated EWCS along constant infalling time can be found in Fig.~\ref{fig:SatEWCSProj}. Notice that $u(x)$ agrees with Fig.~\ref{fig:ApproxLagNullGeod} obtained from solving the approximate equations of motion in the scaling limit. 

\section{Scaling behavior of EWCS in $d>2$}\label{App:ScalingECS}
In this appendix we compare the direct numerical evaluation of the reflected entropy and the corresponding membrane prediction. As described in the main text, these two results should agree in the scaling limit $D\rightarrow\infty$. However, we are constrained in our capacity to evaluate the EWCS within this limit. This problem arises from the numerical solution of the intersection equations between the EWCS and the HRT surfaces. Even for the relatively small values of $D$ we do have access to, we can see that the membrane theory is a surprisingly accurate description, able to capture all the main features of the reflected entropy.

To further test the validity of the membrane prediction, we can take a closer look at some of these key features. First, we can consider the time at which the reflected entropy begins to grow linearly and the size of the gap between the linear growth and the initial vanishing value.
\begin{figure}[h!]
\centering
\includegraphics[scale=0.35]{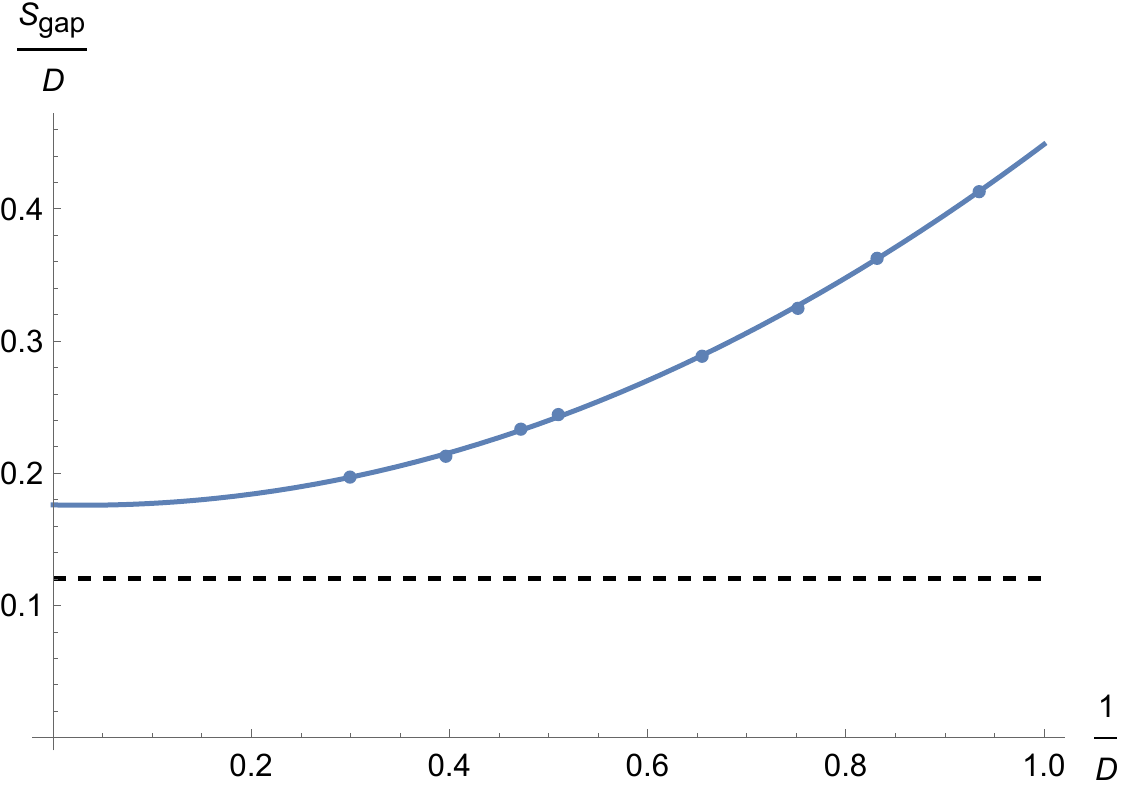}
\includegraphics[scale=0.35]{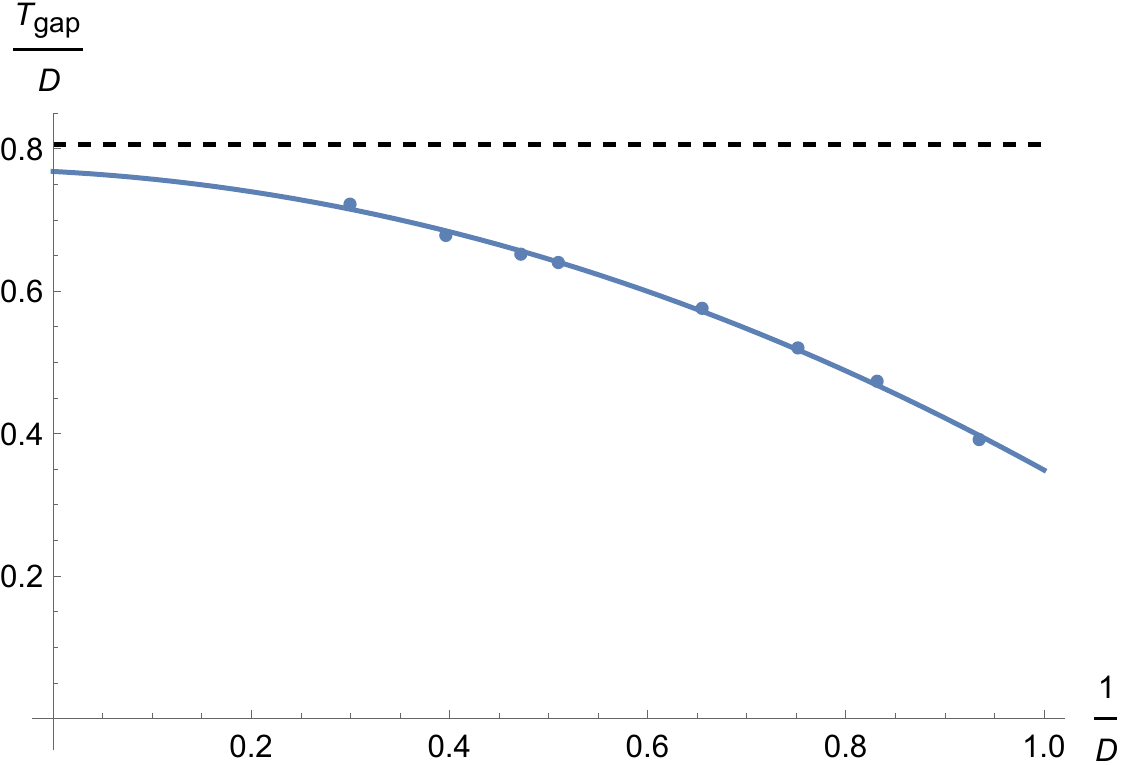}
\caption{Comparisson of the time of transition between vanishing EWCS and its regime of linear growth, as well as the size of the discontinuity, as functions of $1/D$. The membrane prediction is depicted in black.}
\end{figure}

While the discrepancy is somewhat worrying, we can provide an argument for this being simply a numerical issue. The transition between the vanishing and linearly growing regime is dictated not by the EWCS per se, but by the dynamics of entanglement entropy; namely, whether the entanglement wedge is connected or disconnected. This has been shown to be properly captured by the membrane prediction to a high degree of accuracy.

Another point we can compare is the transition between linear growth and the plateau. Once again we see a favorable trend in our evaluation, although precise numerical agreement remains elusive.
\begin{figure}[h!]
\centering
\includegraphics[scale=0.35]{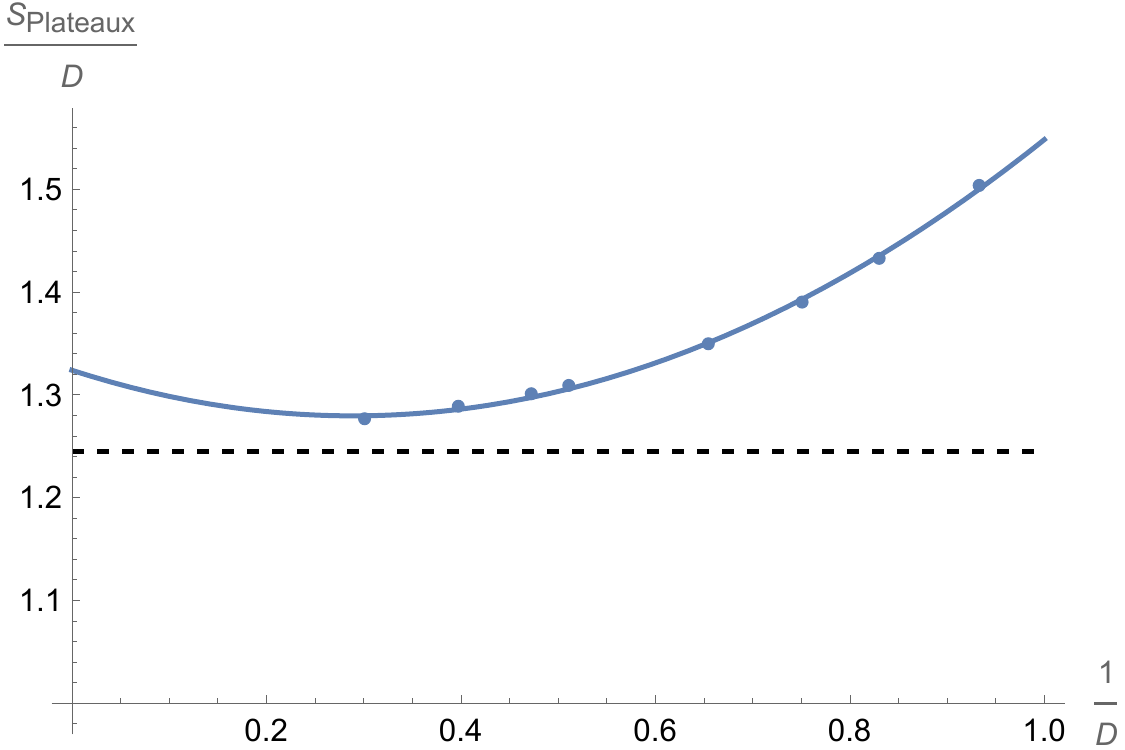}
\includegraphics[scale=0.35]{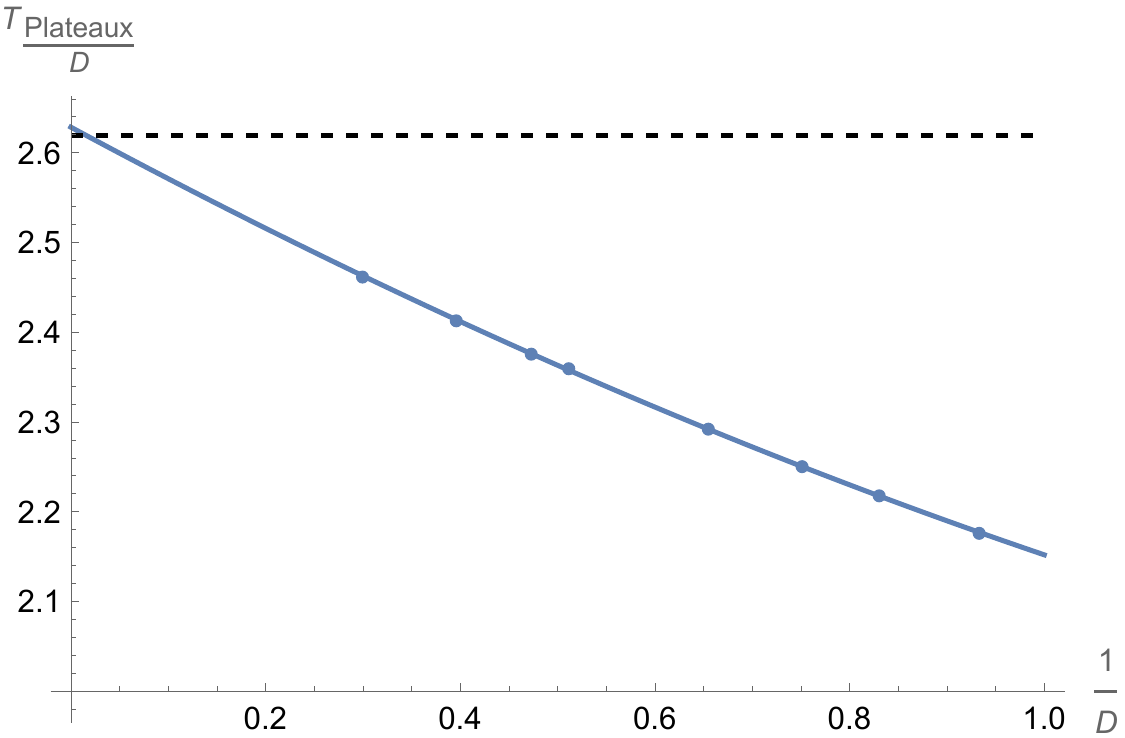}
\end{figure}

\bibliographystyle{JHEP}
\bibliography{biblio.bib}

\end{document}